\documentclass[namedreferences]{solarphysics}
\usepackage[hyperref,optionalrh,solaromanenum]{spr-sola-addons} 
\usepackage{graphicx}                    
\usepackage[usenames]{color}                       
\usepackage{paralist}

\usepackage{solaheader}

\ifx \urlurl    \undefined \def \urlurl#1{\href{http://#1}{\textsf{#1}}}\fi
\ifx \urlurls    \undefined \def \urlurls#1{\href{https://#1}{\textsf{#1}}}\fi

\newcommand{\reviewdel}[2]{\textcolor{blue}{{}}}

\newcommand{\adv}{    {\it Adv. Space Res.}}
\newcommand{\annG}{   {\it Ann. Geophys.}}
\newcommand{\aap}{    {\it Astron. Astrophys.}}

\newcommand{\apj}{    {\it Astrophys. J.}}
\newcommand{\apjl}{   {\it Astrophys. J. Lett.}}

\newcommand{\grl}{    {\it Geophys. Res. Lett.}}

\newcommand{\jastp}{  {\it J. Atmos. Solar-Terr. Phys.}}
\newcommand{\jgr}{    {\it J. Geophys. Res.}}

\newcommand{\solphys}{{\it Solar Phys.}}

\newcommand{\ssr}{    {\it Space Sci. Rev.}}
\chardef\us=`\_

\usepackage{graphicx}
\usepackage{natbib}
\usepackage{tabularx}
\usepackage{lscape}

\usepackage{hyperref} 

\usepackage{ulem}

\begin{document}
\begin{article}

\begin{opening}
	\title{Statistical Analysis of Solar Events Associated with Storm Sudden Commencements over One Year of Solar Maximum during Cycle 23: Propagation from the Sun to the Earth and Effects }
	
	\author[addressref={aff1},corref,email={karine.bocchialini@ias.u-psud.fr}]{\inits{K.}\fnm{K.}~\lnm{Bocchialini}}
	\author[addressref=aff7]{\inits{B.}\fnm{B.}~\lnm{Grison}}
	\author[addressref={aff2,aff3}]{\inits{M.}\fnm{M.}~\lnm{Menvielle}}
	\author[addressref=aff4]{\inits{A.}\fnm{A.}~\lnm{Chambodut}}
	\author[addressref={aff5,aff6}]{\inits{N.}\fnm{N.}~\lnm{Cornilleau-Wehrlin}}
	\author[addressref=aff6]{\inits{D.}\fnm{D.}~\lnm{Fontaine}}
	\author[addressref={aff10,aff12}]{\inits{A.}\fnm{A.}~\lnm{Marchaudon}}
	\author[addressref={aff5}]{\inits{M.}\fnm{M.}~\lnm{Pick}}
	\author[addressref={aff10,aff12}]{\inits{F.}\fnm{F.}~\lnm{Pitout}}
	\author[addressref={aff5}]{\inits{B.}\fnm{B.}~\lnm{Schmieder}}
	\author[addressref=aff11]{\inits{S.}\fnm{S.}~\lnm{R\'egnier}}
	\author[addressref=aff13]{\inits{I.}\fnm{I.}~\lnm{Zouganelis}}

	\address[id=aff1]{Institut d'Astrophysique Spatiale, Univ. Paris-Sud, CNRS, Universit\' e Paris-Saclay, B\^atiment 121, 91405 Orsay CEDEX, France}
	\address[id=aff7]{Institute of Atmospheric Physics CAS, Bocni II, 1401, 141 31 Prague 4, Czech Republic}
	\address[id=aff2]{Universit\'e Versailles Saint Quentin, CNRS, Laboratoire Atmosph\`eres, Milieux, Observations Spatiales, Guyancourt, France}
	\address[id=aff3]{Univ. Paris Sud, D\'epartement des Sciences de la Terre, 91405 Orsay CEDEX, France}
	\address[id=aff4]{Institut de Physique du Globe de Strasbourg, UMR7516; Universit\'e de Strasbourg/EOST, CNRS, 5 rue Ren\'e Descartes, 67084 Strasbourg CEDEX, France}
	\address[id=aff5]{Observatoire de Paris, LESIA, PSL Research University, 5 place Jules Janssen, 92195 Meudon CEDEX, France}
	\address[id=aff6]{LPP, CNRS, Ecole Polytechnique, UPMC Univ. Paris 06, Univ.  Paris Sud, Observatoire de Paris, Universit\'e Paris-Saclay, Sorbonne Universit\'es, PSL Research University, Ecole Polytechnique, 91128 Palaiseau CEDEX, France}
	\address[id=aff10]{Institut de Recherche en Astrophysique et Plan\'etologie, Universit\'e de Toulouse, Toulouse, France}
	\address[id=aff12]{CNRS, UMR 5277, 9 Av. du Colonel Roche, BP 44346, 31028 Toulouse CEDEX 4, France}
	\address[id=aff11]{Department of Mathematics, Physics and Electrical Engineering, Northumbria University, Newcastle upon Tyne, NE1 8ST, United Kingdom}
	\address[id=aff13]{European Space Agency, ESAC, Madrid, Spain}

	\runningauthor{K. Bocchialini \textit{et al.}}
	\runningtitle{Statistical Analysis of Solar Events Associated with SSCs over One Year}
	
	\begin{abstract}
		Taking the 32 storm sudden commencements (SSCs) listed by the International Service of Geomagnetic Indices (ISGI) of the Observatory de l'Ebre during 2002 (solar activity maximum in cycle 23) as a starting point, we performed a multi-criterion analysis based on observations (propagation time, velocity comparisons, sense of the magnetic field rotation, radio waves) to associate them with solar sources, identified their effects in the interplanetary medium, and looked at the response of the terrestrial ionized and neutral environment. We find that 28 SSCs can be related to 44 coronal mass ejections (CMEs), 15 with a unique CME and 13 with a series of multiple CMEs, among which 19 (68\,\%) involved halo CMEs. 
		Twelve of the 19 fastest CMEs with speeds greater than 1000 km\,s$^{-1}$ are halo CMEs. For the 44 CMEs, including 21 halo CMEs, the corresponding X-ray data classes are: 3 X-class, 19 M-class, and 22 C-class flares.
		The probability for an SSC to occur is 75\,\% if the CME is a halo CME. Among the 500, or even more, front-side, non-halo CMEs recorded in 2002, only 23 could be the source of an SSC, \textit{i.e.} 5\,\%. 
		The complex interactions between two (or more) CMEs and the modification of their trajectories have been examined  using joint white-light and multiple-wavelength radio observations. The detection of long-lasting type IV bursts observed at metric--hectometric wavelengths is a very useful criterion for the CME--SSC events association. The events associated with the most depressed Dst values are also associated with type IV radio bursts. The four SSCs associated with a single shock at L1 correspond to four radio events exhibiting characteristics different from type IV radio bursts.
		The solar-wind structures at L1 after the 32 SSCs are 12 magnetic clouds (MCs), 6 interplanetary coronal mass ejections (ICMEs) without an MC structure, 4 miscellaneous structures, which cannot unambiguously be classified as ICMEs, 5 corotating or stream interaction regions (CIRs/SIRs), and 4 Shock events; note than one CIR caused two SSCs.
		The 11 MCs listed in 3 or more MC catalogs covering the year 2002 are associated with SSCs.
		For the three most intense geomagnetic storms (based on Dst minima) related to MCs, we note two sudden  increases of the  Dst, at the arrival of the sheath and the arrival of the MC itself.
		In terms of geoeffectiveness, the relation between the CME speed and
		the magnetic-storm intensity, as characterized using the Dst magnetic
		index, is very complex but generally, CMEs with velocities at the Sun
		larger than 1000 km\,s$^{-1}$ have larger probabilities to trigger
		moderate or intense storms. The most geoeffective events are MCs,
		since 92\,\% of them trigger moderate or intense storms, followed
		by ICMEs (33\,\%). At best, CIRs/SIRs only cause weak storms. We show
		that these geoeffective events (ICMEs or MCs) trigger an increased and
		combined auroral kilometric radiation (AKR) and non-thermal continuum
		(NTC) wave activity in the magnetosphere, an enhanced convection in the
		ionosphere, and a stronger response in the thermosphere. However, this
		trend does not appear clearly in the coupling functions, which exhibit
		relatively weak correlations between the solar-wind energy input and
		the amplitude of various geomagnetic indices, whereas the role of the
		southward component of the solar-wind magnetic field is confirmed. Some
		saturation appears for Dst values $< -100$ nT on the integrated
		values of the polar and auroral indices.
	\end{abstract}
\end{opening}

\section{Introduction}
\label{section:Introduction}

Coronal mass ejections (CMEs) are known as events driving geomagnetic storms most
efficiently. The aim of this study is to investigate the link between
CMEs, geomagnetic storm intensities, and the occurrence of a
storm sudden commencement (SSC). An SSC is defined by a sudden increase of
the magnetic field strength at the Earth's surface, due to the impinging
of a shock on the magnetopause. Those shocks often lead to geomagnetic storms,
which are characterized by solar-wind and magnetosphere energy coupling
enhancement and the growth of the ring current, see \cite{Saiz13} and references therein.

Following \cite{Gonzalez1994}, geomagnetic storms are usually
classified by the Dst-index value defined by
\citet{Sugiura1964} and provided by the World Data Center for Geomagnetism in Kyoto \citep{Nose2015}.

Large-scale terrestrial magnetic disturbances have long been linked to
eruptive events originating at the Sun's atmosphere. Since
continuous Sun observations from space became available,
the solar sources of the geoeffective events could be identified with
more accuracy.  Intense flares are associated with solar energetic
particle (SEP) events and/or CMEs, which can reach the Earth and thus
be responsible for geoeffective events. The occurrence of CMEs depends
on the solar cycle, increasing with solar activity. CMEs
are statistically more likely to lead to  geomagnetic disturbances
when their solar sources are close to the central meridian and
are observed as fast halo CMEs with an apparent width near
$360^\circ$ \citep{Zhang07,Gopalswamy10a, Gopalswamy10b}, {\it i.e.} facing
the Earth \citep{Bothmer07,Bein11,Wimmer14}.
Before CME observations, \citet{Caroubalos1964} showed that
solar flares associated with type IV bursts radiating in the
microwave domain (with a radiated energy often greater than $10^{-17}$
J\,m$^{-2}$\,Hz$^{-1}$) and with a second long-duration phase at
metric wavelengths gave rise to SSCs. It is now well known that the
development, in the corona, of large flare and/or CME events is often associated
with strong non-thermal radio emission.

According to their speed and their interplanetary signatures, interplanetary coronal mass ejections (ICMEs)
may reach the Earth in one to five days after the eruptive event
\citep{Yashiro06,Gopalswamy09,Bein11}. Statistical analysis showed
that the interplanetary drivers of intense storms during
Solar Cycle 23 (1996\,--\,2008) were found to be associated with 
magnetic clouds (MCs) that drove fast shocks causing 24\,\%
of the storms, sheath fields (Sh) also causing 24\,\% of the storms,
combined sheath and MC fields (Sh+MC) causing 16\,\% of the storms,
and corotating interaction regions (CIRs) causing 13\,\% of the
storms. These four interplanetary structures are responsible for three
quarters of the intense magnetic storms, as found by \citet{Echer2008}.
For moderate storms, the statistics are different; the more important
drivers being the CIRs (48\,\%), followed by the MCs
or ICMEs ($21\,\%$), sheath
fields ($11\,\%$), or combinations of sheaths and ICMEs (10\,\%)
\citep{Echer13}. The more intense are the storms, the more they are
associated with ICMEs \citep{Gonzalez07,Zhang07}.

Despite these facts, achieving a definitive one-to-one correlation between
a CME, its interplanetary signature, and the subsequent terrestrial
response is not always an easy task \citep{Hanuise2006}. From statistics
on ICMEs properties observed during Cycle 23, \cite{Richardson2010}
manage to associate only about half of the ICMEs with a solar event. Among
them, about 30\,\% present characteristics of MCs, \textit{i.e.} with
a well-identified magnetic field structure described as a
flux rope \citep{Burlaga1981, Gosling1990}. \cite{Cid12} revisited
very carefully every link along the Sun--Earth chain
during Solar Cycle 23. They concluded that a CME originating from a solar source 
close to the limb cannot be really geoeffective (\textit{i.e.}
associated with at least a moderate geomagnetic storm) unless it belongs
to a complex series of events.

Thus, despite the important progress accomplished in recent years, in this article we have decided to develop 
an innovative approach by studying each Earth-directed event leading to an
SSC in the magnetosphere during 2002 (maximum activity
of Cycle 23). This unprecedented, comprehensive study consists in:
\begin{inparaenum}[i)]
\item starting from the list of SSCs\footnote{Built by the Service of Rapid
	Magnetic Variations, Observatori de l'Ebre (Spain) in the framework of
	the International Service of Geomagnetic Indices (ISGI) activities},
\item linking each SSC to a CME,
\item filling in the observational
gaps along the Sun--Earth chain as much as possible, and
\item identifying one or many solar sources,
with a clear signature on the Sun, in a temporal window determined
by considering two extreme propagation velocities (300 and 1500
km\,s$^{-1}$) \citep{Bein2012}.
\end{inparaenum} 
When there is no CME identified as the
source of the observed SSC, we seek what process in the solar wind may
have led to the SSC, with the help of observations at the Lagrangian point L1. Regarding the
terrestrial consequences, where previous studies are limited  to  just a few
magnetic indices, we also study carefully the responses of the
magnetosphere--ionosphere--thermosphere system making use of
a large set of varied measurements described in the following paragraphs.

The data sets used in the present study are obtained by
spacecraft operating during 2002 along the Sun--Earth chain and from
terrestrial stations. In order to properly characterize the events
used in this article, we used the database that we had created for the
period 1996\,--\,2007. It encompasses a great variety of datasets from
various available catalogs and sources.\footnote{
\href{http://www.ias.u-psud.fr/gmi}{\textsf{www.ias.u-psud.fr/gmi}}
(login: \textsf{gmi}, password: \textsf{cme}). This database is a working tool for a
multidisciplinary research group and as such the presentation and
description of its elements, figures, and plots are not
at a level suitable for publication. We initially used the catalog
of CMEs (available at
\href{http://cdaw.gsfc.nasa.gov/CME\_list/}{\textsf{cdaw.gsfc.nasa.gov/CME\_list/}}), which was
established on the basis of \textit{Solar and Heliospheric Observatory}
(SOHO) observations to build our own list of
halo CMEs, from April 1996 to January 2007.}
We concentrate on the specific interests, strengths, and limitations of the various
datasets specifically used in the present article. We have considered
four main regions all along the Sun--Earth chain: Sun, solar wind,
magnetopause--magnetosphere, ionosphere--thermosphere.

In 2002, \textbf{solar} remote-sensing observations were still
relatively limited. We used data from the \textit{Extreme Ultraviolet
Imaging Telescope} \citep[EIT:][]{Boudine1995} providing continuous 2D images of the Sun
in EUV wavelengths and the \textit{Large Angle and Spectrometric Coronagraph} \citep[LASCO:][]{Brueckner1995}, experiment providing images of the solar corona, 
both onboard the \textit{Solar and Heliospheric Observatory} (\citealp[SOHO:][]{Fleck1995}). To characterize
the CMEs associated with SSCs using radio observations, we 
focused mainly on ground-based radio data (dynamic spectra, or fixed-frequency observations) obtained from different terrestrial
observatories and space data from the \textit{Wind}/WAVES experiment
\citep{Bougeret1995}. The radio imaging observations were provided,
in a limited frequency range (150\,--\,432 MHz) and with a limited temporal
coverage, by the \textit{Nan\c{c}ay Radio Heliograph} \citep[NRH:][]{Kerdraon1997}.

Monitoring of the \textbf{solar wind} is in general achieved
by the \textit{Advanced Composition Explorer} \citep[ACE:][]{ace1998} magnetic field experiment (MAG) and plasma particle experiment (\textit{Solar Wind Electron Proton Alpha Monitor}, SWEPAM).
These specific experiments
allow us to monitor the interplanetary magnetic field (IMF) characteristics, in particular its component orthogonal to the ecliptic plane ($B_{\rm z}$), the magnetic field magnitude, and the solar-wind plasma
velocity and density which are known to be major parameters
controlling the coupling with the magnetosphere. The
plasma velocity is also an important parameter to examine
the propagation of the perturbation caused by the CMEs in the solar
wind. Finally, to study more specifically the geoeffectiveness of the
different solar events impinging on the Earth, we used the OMNIWeb service
(\urlurl{omniweb.gsfc.nasa.gov/}), which reconstructs the solar-wind
properties at the Earth's bow-shock nose from ACE data and from other
spacecraft.

We chose to characterize the effect of solar activity on the
\textbf{magnetosphere} by i) detecting the magnetopause position
to evaluate its variation due to the arrival of solar events and ii)
detecting Earth's radio emissions caused by accelerated electrons during
intervals of magnetospheric activity triggered by these solar events. In
2002, the available \textit{in situ} measurements for these purposes
were mainly limited to those performed by the ESA--NASA \textit{Cluster}
mission \citep{Escoubet2001} and by the ISAS--NASA \textit{Geotail} \citep{Nishida1994} spacecraft. With apogees roughly
opposite, \textit{Cluster} and \textit{Geotail} provide complementary
datasets in different magnetospheric regions. 

We characterize the electrodynamic activity in the coupled
magne\-to\-sphere-iono\-sphe\-re system by means of different
geomagnetic indices computed from combinations of local measurements
continuously available in time, at stations of different sub-networks
of the worldwide network of geomagnetic observatories \citep[see,
\textit{e.g.},][]{Svalgaard1977, Mayaud1980, Cid12}: PCN (polar cap),
AU, AL, AE (auroral zone), am (sub-auroral zone), ASY-H, Dst, and SYM-H
(low latitudes) (see the International Service of Geomagnetic Indices, ISGI: \urlurl{isgi.unistra.fr/}). The reader is also referred to
review articles, \textit{e.g.}, \cite{MenvielleBerthelier1991},
\cite{McCreadie2009},  \cite{Menvielle2011} and references therein).

Low-orbiting satellites devoted to ionosphere--ther\-mosphere
regions are limited to observations along their tracks. Moreover,
ground-\-ba\-sed observations of the \textbf{ionosphere} are restricted
to a latitude and/or longitude sector and at a local time varying with the Earth's
rotation. To ensure good coverage of the ionospheric response, we
use the \textit{Super Dual Auroral Radar Network} \citep[SuperDARN:][]{Greenwald1995} of high-frequency
(HF) coherent radars dedicated to the monitoring of auroral ionospheric
convection in both hemispheres.  They provide
valuable information on the temporal evolution of the polar and auroral
convection patterns in the ionosphere and on the related variations of
the potential across the polar cap.

The density variations of the \textbf{thermosphere} are monitored by means
of thermospheric-disturbance coefficients, deduced from measurements of
accelerometers onboard the \textit{Challenging Minisatellite Payload} spacecraft \citep[CHAMP:][]{Reigber2002}. They
are proportional to the mean density of the thermosphere along the CHAMP
orbit between latitudes $-50$ and $+50^\circ$, normalized
in order to eliminate variations due to altitude, local time,
and solar EUV flux. They are  estimated by
comparison to thermospheric models reproducing quiet conditions (see
\cite{Menvielle2007} and \cite{Lathuillere2008} for a complete description
of these coefficients).

In Section \ref{section:relatedSunEarthEvents_methodology}, the
methodology to link CMEs and SSCs is described. 
The whole chain of events (CME, if any, L1 structure, magnetospheric and ionospheric activity, \ldots) related to a given SSC or which we guess will lead to a given SSC is  named as an SSC-led event (an event which we suppose lead to the SSC).
We first classify the SSC-led events as a function of the characteristics of the perturbation
at L1.  Then we try to relate each of the L1 events to
a possible solar source. This is done by the use of different criteria,
the starting point being a propagation-time analysis of CMEs towards Earth
with different tools, complemented by carefully combined observations
of radio wavelengths and imagery.

In Section \ref{section:EventCharacterization}, for each dataset,
we perform  statistical classifications of the selected events, as
they are seen at the different positions along the Sun--Earth path. The
propagation of the events is discussed in Section \ref{section:Discussion}
as well as their geoeffectiveness, as a function in particular of their
properties at L1, from a statistical point of view. Section
\ref{section:SummaryResults} describes the results and we conclude in
Section \ref{section:ConcludingRemarks} with some final remarks.

This work was only possible because the authors come from different communities, which are generally working independently.

\section{Identification of Related Sun--Earth Events: Methodology and Results} 
\label{section:relatedSunEarthEvents_methodology}

To link a shock observed upstream of the Earth at L1, which is seen as a SSC at the Earth, to a given solar source is not straightforward. Thus we describe in this section how we proceed to make and validate our choices, after having characterized the solar wind structures at L1.

\subsection{SSC Events and their Origin at L1}
\label{section:ssc_and_L1}

We start the study with the list of SSCs that occurred in 2002. Table \ref{table:SSC_list}  details the properties of these 32 SSCs. 

\begin{table}[htb]
	\begin{center}
		\caption{The 32 SSCs observed during 2002, and three additional sudden secondary events (SSEs, see definition in the text). SSC and SSE numbers are given in chronological order. 
		SSCs are characterized using their duration, amplitude, and quality of detection at five low-latitude magnetic observatories. 
		The classification value is the mean of the codes (from 0 to 3) given by each observatory, 2 or 3 corresponding to a very clear event. Date and time of the SSEs are provided in Section \ref{section:ssc_and_L1}. Last column gives the event tag at L1.		}
		
		\begin{tabular}{ l  l l  r  r r  r  r    }
			SSC    &      Date        & Time & Duration        & Amplitude     & Classification & SSE &  L1 \\  
			No.      &    [2002]   & [UT] & [minutes]        & [nT]         &                & No. & Cat.  \\ \hline  
			SSC01 & 31 Jan  &21:25 & 6.0         & 13.3         & 1.8      && MC \\   
			SSC02 & 17 Feb  &02:53 & 9.8         & 21.2         & 1.8        && SIR \\   
			SSC03 & 28 Feb  &04:50 & 7.7         & 37.4         & 2.0        && MC \\   
			SSC04 & 18 Mar  &13:21 & 3.6         & 60.8         & 2.0        & &MC \\   
			SSC05 & 20 Mar  &13:27 & 3.3         & 15.4         & 1.8      &  & ICME \\   
			SSC06 & 23 Mar  &11:34 & 4.9         & 21.5         & 2.0        && MC \\   
			SSC07 & 29 Mar  &22:36 & 8.0         & 30.7         & 2.0        && CIR \\   
			SSC08 & 14 Apr  &12:34 & 6.0        & 10.1         & 0.6       & & Shock \\   
			SSC09 & 17 Apr  &11:05 & 3.6         & 52.7         & 2.2        &SSE09 & MC \\   
			SSC10 & 19 Apr  &08:34 & 5.0        & 25.1         & 2.4        &SSE10 & MC \\   
			SSC11 & 23 Apr  &04:47 & 4.0         & 42.6         & 2.6        && Misc. \\   
			SSC12 & 10 May  &11:22 & 5.8         & 29.4         & 2.0        && Misc. \\   
			SSC13 & 11 May &10:13 & 4.6         & 25.9         & 2.2     &   & ICME \\   
			SSC14 & 18 May  &20:08 & 5.0        & 40.7         & 2.0        && MC \\  
			SSC15 & 20 May  &03:39 & 4.2         & 13.6         & 1.0        && ICME \\   
			SSC16 & 21 May  &22:02 & 6.6         & 17.6         & 1.0       & & Shock \\   
			SSC17 & 23 May  &10:49 & 12.8         & 77.6         & 2.0        && MC \\   
			SSC18 & 30 May  &02:04 & 5.2         & 9.0            & 0.6       & & Shock \\   
			SSC19 & 08 Jun  &11:39 & 7.8         & 17.4         & 0.6       & & CIR \\   
			SSC20 & 17 Jul  &16:02 & 5.0        & 40.1         & 2.0        && ICME \\   
			SSC21 & 19 Jul  &10:08 & 7.6         & 18.5         & 1.6       & & ICME \\  
			SSC22 & 29 Jul  &13:21 & 5.8         & 25.1         & 2.0        && Shock \\   
			SSC23 & 01 Aug & 05:09 & 5.6         & 16.8         & 2.0       & & MC \\   
			SSC24 & 01 Aug  &23:09 & 3.6         & 27.7         & 2.0        && MC \\   
			SSC25 & 18 Aug  &18:45 & 4.0         & 36.6         & 2.0       & & MC \\   
			SSC26 & 26 Aug  &11:30 & 7.0        & 21.1         & 1.0       & & Misc. \\   
			SSC27 & 07 Sep  &16:36 & 4.0        & 23.4         & 2.2        && ICME \\   
			SSC28 & 30 Sep  &08:15 & 6.8         & 15.9         & 1.2        &SSE28 & MC \\   
			SSC29 & 09 Nov  &17:51 & 5.4         & 8.2         & 0.8        & &CIR \\   
			SSC30 & 09 Nov  &18:49 & 5.6         & 16.6         & 2.0       & & id.\\   
			SSC31 & 11 Nov &12:29 & 2.8         & 16.4         & 0.6        & &  CIR  \\   
			SSC32 & 26 Nov & 21:50 & 4.2         & 25.3         & 2.0       & & Misc.\\ \hline  
		\end{tabular}
		\label{table:SSC_list}
	\end{center}
\end{table}

\begin{figure}
	\centering
	\includegraphics [angle=0, scale=0.65]{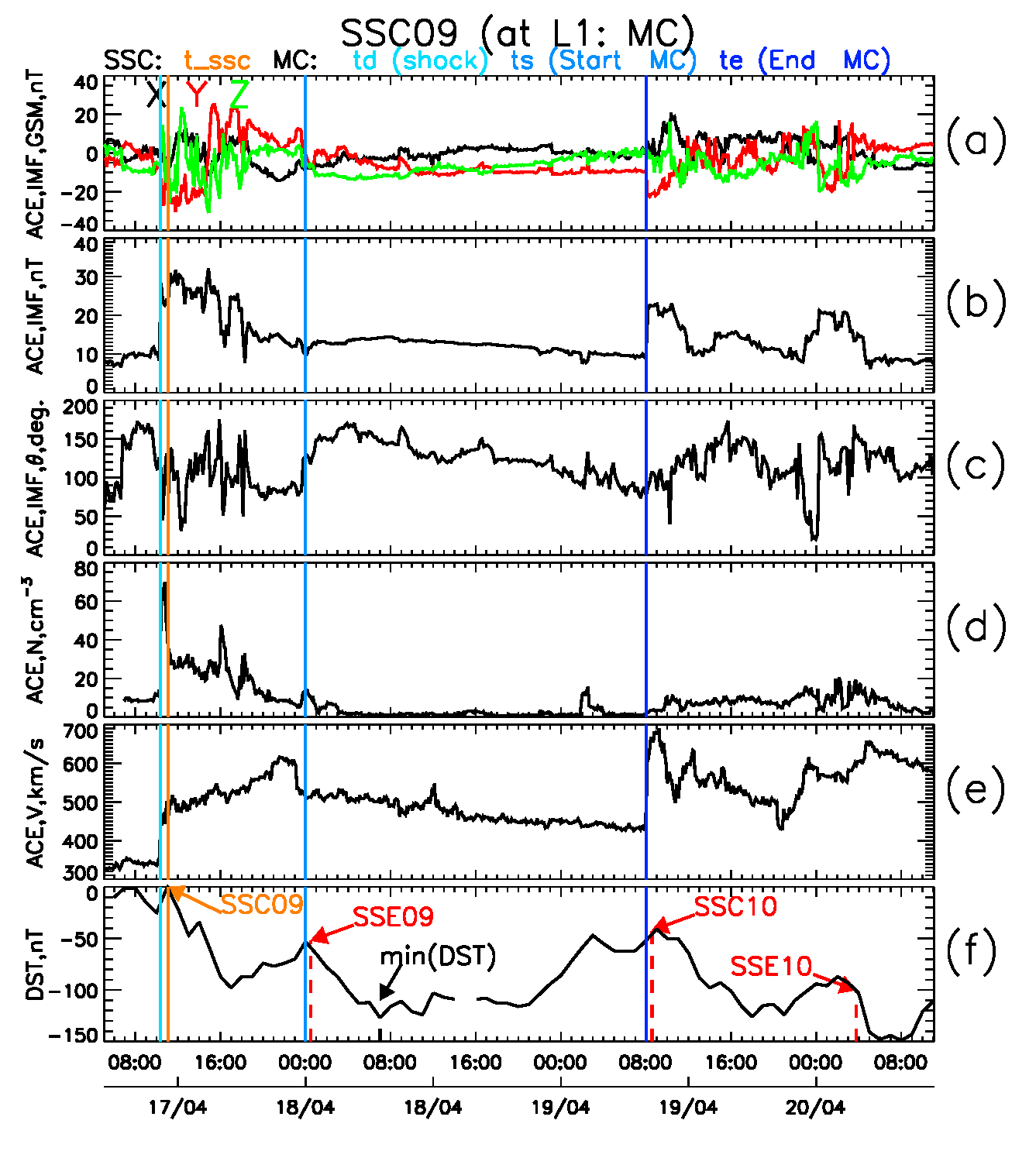}
	\caption{    
		Observations at L1 (ACE data) of an MC together
		with the Dst index associated with the SSC09 of 17 April 2002. (a) The
		three components of the IMF in the GSM coordinate system. (b) The IMF intensity. (c) The IMF inclination with respect to the $Z_{\rm GSM}$ axis. (d) and (e) The solar wind density and velocity, respectively. (f) The simultaneously observed variations of the Dst index along with the indication by dashed lines of  the other SSC observed during the
		time window, the SSE events (see text) and the min(Dst) associated with the SSC09 event.
	}
	\label{figure:Figure_l1_ssc09}
\end{figure}

\label{section:L1_characterization}
For each of the 32 SSCs, we classify the events by the
characteristics of the solar-wind structure that follows the shock sheath
at L1, as shown in the last column of Table \ref{table:SSC_list}. After
having compared our database plots associated with each SSC-led event
(see Figure \ref{figure:Figure_l1_ssc09}), we define five categories:
ICME--MC labeled ``MC", ICME (non-MC) labeled ``ICME", ``CIR/SIR",
Shock labeled ``Shock", and miscellaneous labeled ``Misc.".  ICME,
MC, CIR are defined previously, SIR stands for stream interaction region. Shock is self explanatory, and
Misc. refers to a perturbation that cannot be classified with a high
confidence level in any of the previous categories. In other words a
Misc. event is more similar to an ICME than to a Shock event.

As this categorization is not always straightforward, we choose to rely on lists of classified events given by previous authors in order to confirm our choices. Specific solar events are identified in the solar wind for the year 2002 from: i) ICME catalogs (\citealp[updated by]{Cane2003} \citealp{Richardson2010}; \citealp{Jian2006b}); ii) MC catalogs \citealp{Huttunen2005,Lepping2006,Zhang07,Li2011} and subsets of the above ICME lists); iii) CIR and SIR catalog \citep{Jian2006a}; iv) Interplanetary (IP) shock catalog and associated solar sources \citep{Gopalswamy2010c}.

When an event is listed as an MC at least once, we tag it as MC; an MC listed at least twice is considered as a ``well defined" MC in what follows.
When an event is never listed as an MC but listed as an ICME in two of the relevant catalogs, we use the tag ICME.

The tag CIR/SIR is applied to all CIR/SIR events found by \cite{Jian2006a}.
The other events are split into Misc. and Shock categories following our evaluation. The results of this categorization are given in the last column of Table \ref{table:SSC_list}. More details can be found in  Table \ref{table:Observations_at_L1} in Appendix \ref{section:App_SolWindObs}.

The 32 SSCs are due to 31 events sorted as follows: 12 MCs, 6 ICMEs, 5 CIR/SIR events (4 CIRs and 1 SIR), 4 Misc., and 4 Shock events.
SSC30 is recorded less than one hour after SSC29 and is linked with the same CIR structure at L1. We do not tag the L1 category for SSC30 and  consequently we will only consider 31 events.

For three MCs (linked with SSC09, SSC10, and SSC28),
the detailed analysis of the ground magnetograms revealed the presence
of a second sharp increase of the north magnetic component a few hours
after the SSC. This increase occurs after the end of the decreasing
phase that immediately follows the SSC and is followed by a reinforced
decreasing phase (see an exemple in Figure \ref{figure:Figure_l1_ssc09} bottom panel that we describe in the following paragraph). 
We call them sudden secondary events (SSEs): SSE09 (18 April 00:30 UT),
SSE10 (20 April 03:45 UT), and SSE28 (30 September19:30 UT). Those SSEs are related to sharp pressure increases at L1 at the time of the cloud arrival, the SSC being due to the shock preceding the sheath.

Figure \ref{figure:Figure_l1_ssc09} shows an example of ACE observations
at L1 in relation to an SSC-led event, namely SSC09 associated
with a magnetic cloud (MC) and an SSE (SSE09). The temporal
interval starts 6 hours before the SSC and ends 78 hours later. From
top to bottom, we present the three geocentric solar magnetic (GSM) components of the IMF (Figure \ref{figure:Figure_l1_ssc09}a), the
IMF intensity (Figure \ref{figure:Figure_l1_ssc09}b), the IMF deviation angle ($\theta_{\rm IMF}$)
from the $Z_{\rm GSM}$ axis (Figure \ref{figure:Figure_l1_ssc09}c), the solar-wind proton density
(Figure \ref{figure:Figure_l1_ssc09}d), and the solar-wind bulk flow velocity (Figure \ref{figure:Figure_l1_ssc09}e), and the Dst index to quantify the storm intensity (Figure \ref{figure:Figure_l1_ssc09}f). The
continuous-orange line marks the SSC time ($t_{\rm SSC}$). The blue line
just before it marks the arrival time of the solar-wind (SW) discontinuity
at L1 ($t_{\rm d}$ is the sheath start time). The magnetic-cloud observation time
([$t_{\rm s};t_{\rm e}]$) is delimited by the two vertical-blue lines to the right of the
orange line. We use the same timing definition for all
L1 events (see Table \ref{table:Observations_at_L1} in the Appendix
\ref{section:App_SolWindObs}). The discontinuity is observed by
an increase in the IMF intensity, in the SW density, and in the SW
velocity. This pulse of kinetic and magnetic pressures drives SSC09. The low density, the
decreasing velocity and IMF intensity, and the IMF rotation observed
between $t_{\rm s}$ and $t_{\rm e}$ are characteristics of magnetic clouds. One can
notice SSE09, indicated by a dashed-red line on the bottom panel,
which coincides with the arrival of the MC. The end of the MC
is seen simultaneously with the SW discontinuity at the origin of
SSC10. This is a nice example of an MC, or to be
more exact the shock ahead of the following MC sheath, catching up
with another one. 
Along the same lines, \citet{Lepping2006} reported observations of possible merging of two MCs in the same ejecta, followed by a shock.
It is worth noting that the Dst index has two local
minima following SSC09: one linked to (or associated with) the sheath
($\approx -100$ nT) and the second one to the cloud itself ($\approx -130$
nT), in the present case after the SSE. In what follows, when relevant,
\textit{i.e.} for MC and ICME, we take the lowest of these two values
to define the Dst minimum of a storm.

\subsection{Solar Sources of the L1 Structures Associated with the SSC}
\label{subsection:association}
Our aim is to identify a possible solar source with each of the 31 L1 structures related to the 32 SSC-led events.
As CMEs are the most probable causes of terrestrial disturbances,
we use a list compiling all front-side halo CMEs
(CMEH) observed in 2002, obtained from already existing works based on LASCO data as
well as some partial halo CMEs (angular width larger than
$120^\circ$, CMEP), and non-halo CMEs (angular width smaller
than $120^\circ$, CMEN). Our list contains 60 CMEs with an identified
source detected on the solar disk (among them, 28 halo CMEs). It has been shown
that it could, however, happen that front-side CMEs without an identified solar source
can be mislabeled as backside CMEs \citep{Webb2012}, which 
can be the case in our dataset.

Combining this dataset with EIT images, we estimate the location of
all  active regions and/or filaments related to the CMEs onset in the
Earth-facing solar hemisphere, and deduce propagation directions of the CMEs in
the plane perpendicular to the Earth--Sun line. Since the
earthward component of a CME propagation direction cannot be computed, it is not
possible to determine unambiguously whether the associated ICMEs will encounter
the Earth or not. However, the so-called halo CMEs, \textit{i.e.}
with an angular extent of $360^\circ$, are assumed to move earthward if
their source is located on the visible face of the Sun (front-side CMEs)
\citep{Zhang07,Gopalswamy10a, Gopalswamy10b}, and not if their source
is located on the non-visible face of the Sun (backside CMEs). Finally,
front-side, non-halo CMEs can also be Earth-directed and geoeffective.

To identify the solar source of the SSCs for 2002, we first build 18 time-related groups  of SSCs and CMEs that are taken in a temporal window of five days before an SSC, regardless of L1 observations. Each group contains one or more SSCs and one or more CMEs. 
The refined association relies on several parameters detailed in the following sections. They are: velocity considerations (Section \ref{ballistic}), drag-based model estimations (Section \ref{drag}), chirality observations, wherever appropriate, for MCs (Section \ref{subsection:chirality}), and radio observations (Section \ref{section:L1identification_radio}).

The results are summarized in Table \ref{table:associations_L1}, which
lists the 31 L1 structures associated with the 32 SSC-led events and
their possible solar source(s). Solar sources can be ``Leading" (a Leading
CME is a CME whose characteristics observed at the Sun fit best
with the observations at L1 or ``Contrib." (a contributing CME could
somehow interact with the Leading CME during the propagation).
\citet{Lepping2006} noted that ejecta observed in the interplanetary medium could contain more than one ICME. \\
We also identify interactions based on radio signatures. ``Alter." CMEs (to distinguish between the contributing and alternative CMEs in Table \ref{table:associations_L1} we will use italics for the alternative ones) are alternative Leading, observed in a different solar quadrant than the Leading CME.
Note that chirality only applies to MC and that velocity considerations are the primary criterion for choosing the Leading CME.

No CME candidate could be found for three SSCs (SSC02, SSC07,
and SSC19) that are, in any case, associated with a CIR/SIR
at L1 based on existing catalogs. The characteristics of the 44
CMEs (21 halo CMEs) called above either Leading, Contrib., or Alter. are given in
Table \ref{table:CME_retenues_proprietes}. This table lists the date and time of the beginning of the solar event at the origin of the CME seen in SOHO/EIT images at 30.4 nm or 19.5 nm, the source coordinates and nature of the source seen in SOHO/EIT as well
(active region, filament, flare, coronal hole),  the final
height ($h$) in units of solar radii in the SOHO/LASCO field of view (FOV) at which
the CME is observed, the corresponding velocity ($V_\odot$) and acceleration ($a$) and the exit time of the LASCO FOV, and finally the flare class when available.

\begin{landscape}
	\begin{table}[htb]
		\begin{center}
			\caption{The 32 SSCs detected in 2002 and their identified sources at L1 and at the Sun. In the fourth column Contrib. means contributing CME, and Alter. alternative CMEs (these Atler. are indicated in italics). In the fifth column G10 stands for \cite{Gopalswamy2010c}, a Y in this column means that we agree with G10, and Else that the solar source in G10 is not in our list of CMEs. For an explanation of the velocities, see Section \ref{ballistic}.   $\gamma$ is the drag parameter. From the LASCO data we show: the height ($h$) in solar radii and the CME acceleration ($a$) in the tenth and eleventh columns respectively. The solar source position (S or N for southern or northern hemisphere respectively) and MC chirality at L1 (RH or LH for right or Left handed respectively). 
			The two last columns correspond to the radio signature of the Leading and Contrib. or Alter. CMEs. The radio signatures are A (microwave burst), B (long duration radio continuum), interaction, type II, and type IV. Signatures are marked in that order with a ``y" when present and with a ``-" when not observed (see Table \ref{table:radio}). Values underlined or followed by a question mark show questionable observations for the Leading CME. Values underlined show unexpected observations in the columns $V_{\rm bal}$, Source {\it vs.} Chirality and Radio. 
			}
			\tiny
			\begin{tabular}{r r r c r r r r r r r r c c   c }
				SSC & L1&  \multicolumn{3}{c}{CME} & \multicolumn{3}{c}{Velocities [km\,s$^{-1}$]}   & $\gamma \times 10^{-7}$ & $h$ & $a$ & Source {\it vs.}			&\multicolumn{2}{c}{Radio}\\
				No. & Cat. & Leading & Contrib./{Alter.}  &  G10 & $V_{\odot}$ & $V_{\rm bal}$ & $V_{\rm L1}$ & [km$^{-1}$] & [R$_{\odot}]$ & [m\,s$^{-2}]$ &  Chirality & Leading & Contrib./{Alter.}\\
				\hline
				SSC01 & MC & N02 & {\it H01} & Y & 738 & \emph{350} & 360 & 1.5 & 12 & 35 & S-RH &-y-\,-\,- & \textit{-yyy-} \\
				SSC02 & SIR &  &  & Else &  &  &  &  &  &  & n/a &      &  \\
				SSC03 & MC & N03 &  & Y & 258 & \emph{420} & 390 & $<$0.1 & 6 & 5 & S-RH &-yyy- &  \\
				SSC04 & MC & H04 & P05 & Y & 784 & 487 & 420 & 0.34 & 30 & -17 & S-RH &yy-y- & -\,-\,-\,-\,- \\
				SSC05 & ICME & H06 & P07 P08 & Y & 971 & 468 & 440 & 1.7 & 20 & -3 & n/a &yy-y- & -\,-\,-\,-\,- \\
				SSC06 & MC & H10 & H11 P09 & Else & 1685 & 795 & 440 & 0.30 & 18 & -23 & S-RH &yy-yy & -\,-yy- \\
				SSC07 & CIR &  &  & ./. &  &  &  &  &  &  & n/a & &  \\
				SSC08 & Shock & \emph{N12?} &  & ./. & 497 & \emph{618} & 420 & $<$0.1 & 20 & -3 & n/a & y-\,-y- &  \\
				SSC09 & MC & H13 &  & Y & 742 & 580 & 510 & 0.18 & 26 & 2 & S-RH &yy-yy &  \\
				SSC10 & MC & H15 & N14 & Y & 1103 & \emph{527} & 600 & 1.3 & 28 & -20 & \emph{S-LH} &yy-yy & -\,-\,-\,-\,- \\
				SSC11 & Misc. & H16 &  & Y & 2388 & 742 & 600 & 0.79 & 25 & -1 & n/a &yy-yy &  \\
				SSC12 & Misc. & N17 & {\it H18} & H18 & 1266 & 456 & 400 & 1.1 & 28 & 6 & n/a &-\,-\,-\,-\,- & \textit{\emph{yy-\,-y}}  \\
				SSC13 & ICME & H19 &  & Y & 697 & 561 & 440 & 0.17 & 5 & 79 & n/a  &yy-\,-y &  \\
				SSC14 & MC & H20 &  & Y & 506 & \emph{531} & 450 & $<$0.1 & 28 & -7 & \emph{S-LH} &yy-yy &  \\
				SSC15 & ICME & N21 &  \it{N22} & Y & 532 & 507 & 470 & 0.27 & 20 & 6 & n/a &-\,-\,-\,-\,- & \textit{\emph{y-yy-}}      \\
				SSC16 & Shock & N23 &  & Y & 614 & 492 & 400 & 0.31 & 6 & 45 & n/a &y-\,-y- &  \\
				SSC17 & MC & H25 & P24 & Y & 1504 & 961 & 800 & 0.18 & 28 & -10 & \emph{S-LH} &-yyyy & y-y-\,- \\
				SSC18 & Shock & P26 &  & Y & 1122 & 652 & 510 & 0.53 & 18 & 4 & n/a &-y-y- &  \\
				SSC19 & CIR &  &  & ./. &  &  &  &  &  &  & n/a &  &  \\
				SSC20 & ICME & H27 & P28 & Y & 973 & 598 & 450 & 0.51 & 28 & -26 & n/a &y-\,-y- & -\,-\,-\,-\,- \\
				SSC21 & ICME & H31 & H29 P30 & H29 & 1788 & 1164 & 900 & 0.15 & 28 & ./. & n/a &-\,-\,-\,-\,- & \emph{yy-yy}  \\
				SSC22 & Shock & H32 &  & Y & 816 & 627 & 500 & 0.53 & 30 & 0 & n/a &y-\,-y- &  \\
				SSC23 & MC & \emph{N33} &  & N34 & 409 & \emph{550} & 430 & $<$0.1 & 25 & 3.8 & S-RH & y-\,-y- &  \\
				SSC24 & MC & N35 & N34 & Else & 998 & 533 & 510 & 1.2 & 15 & 32 & S-RH &-\,-\,-\,-\,- & \emph{yyyyy} \\
				SSC25 & MC & H36 &  & Y & 1239 & \emph{504} & 510 & 1.4 & 23 & -67 & \emph{S-LH} &yy-yy &  \\
				SSC26 & Misc. & H38 & {\it P37} & Y & 2066 & 651 & 400 & 2.2 & 26 & 44 & n/a &yy-yy & -\,-\,-\,-\,- \\
				SSC27 & ICME & H39 &  & Y & 1855 & 656 & 470 & 0.51 & 17 & 43 & n/a &yyyyy &  \\
				SSC28 & MC & N40 & {\it N41} & Else & 1300 & 415 & 370 & 0.93 & 21 & -61 & S-RH  & y-\,-\,-\,- & {\it \emph{yy-yy}} \\
				SSC29\,--\,30 & CIR & \emph{N42?} &  & Y & 485 & \emph{487} & 360 & 0.11 & 16 & -6 & n/a &-\,-\,-\,-\,- &  \\
				SSC31 & CIR & H43 &  & ./. & 1977 & 817 & 560 & 0.51 & 26 & 35 & n/a &yy-yy &  \\
				SSC32 & Misc. & H44 &  & Y & 1179 & 813 & 580 & 0.41 & 20 & 21 & n/a &-\,-\,-y- &  \\
			\end{tabular}
			\label{table:associations_L1}
		\end{center}
	\end{table}
\end{landscape}

\begin{table}[htb]
	\begin{center}
		\caption{
			The 44 CMEs associated with the 32 SSCs and
			mentioned in Table \ref{table:associations_L1}: 21 halo
			CMEs (CMEH), 9 partial-halo CMEs (CMEP), and 14 non-halos
			CMEs (CMEN), detected in 2002 with an identified source
			at the Sun. Starting from the second column on, we list the 
			date and time of the beginning of the solar event at the origin of the CME as seen in SOHO/EIT images at 30.4 nm or 19.5 nm, source coordinates and nature of the source seen in SOHO/EIT (AR = active region, Fi = filament, FL = flare, CH = coronal hole),  the final height ($h$) in units of solar radii in the SOHO/LASCO FOV at which the CME is observed, the corresponding velocity $V_{\odot}$ [km\,s$^{-1}$], the acceleration ($a$) [m\,s$^{-2}$], the exit time from LASCO FOV, and the flare class (GOES).
		}
		
		
		\tiny 
		
		\begin{tabular}{ l rrrlr rr rr r}
			CME		&	\multicolumn{2}{c}{Source (EIT)}& \multicolumn{2}{c}{Source (EIT)} 	&\multicolumn{4}{c}{CME (LASCO)}	& 	{Flare}	\\
			No.		&	\multicolumn{2}{c}{ Date }	& \multicolumn{2}{c}{Coord. \& source type}	&$h$&  $V_{\odot}$ &   $a$	& Exit time	 & Class \\				
			\hline
			CMEH01 & 27 Jan & 12:24 & (375, 850)	 & AR	            	 & 26  	&  1000  	&  -19.2    &   16:30 &  C  \\
			
			CMEN02 & 28 Jan & 9:35 & (-200, -500) 	& AR,  Fi            	 & 12		&   738 	&  35.0      &  13:00 &  C   \\
			
			CMEN03 & 24 Feb &  14:45    & (650, -400)	 & AR, Fi        	 &  6   	& 258  	& 5.2        & 17:50 &   C \\
			
			CMEH04 & 15 Mar & 21:48  & (113, -48) 	&  AR,  Fi  	 & 30  	& 784  	& -17.4    &  4:30 (+1) &  M \\
			&&&&CH&&&& \\
			CMEP05 & 17 Mar & 10:28  & (-273,-233)	 &  AR, Fi     		 & 30  	& 931  	& -6.0 & 15:45   	 &    M  \\
			
			CMEH06 & 18 Mar & 1:48 &  (410, -240)	 & AR, Fi     		 &  20		 & 971  	& -2.9 	&  6:30 & M   \\
			
			CMEP07 & 19 Mar &  9:24 & (770, -70) 	 & AR, Fi      	  &  28  	 &  711  	& -0.9 	& 16:10 & M  \\
			
			CMEP08 & 19 Mar & 11:12  & (770, -70)	 & AR, Fi    		 & 12  	&  1030   &  46.4 	 &   13:45 & M  \\
			
			CMEP09 & 20 Mar &  23:24    & (870, - 270) & AR,  Fi         		&  30 	& 1075 	& -0.2 	&   4:50 (+1) & C   \\
			
			CMEH10  & 22 Mar  & 10:36 &  (980, - 160) & AR, Fi        		 &  18  	&    1685 	& - 22.5  & 12:40	 & C          \\
			
			CMEH11  & 22 Mar & 11:36 & (980, -160) 	& AR,  Fi     		  &  36 	 &    1027	 & 14.6 	& 18:30  & M \\
			
			CMEN12   & 11 Apr & 16:24& (420, -200) 	& AR, Fi        	& 20  	&   497 	 &   -3.4    & 23:30  &C \\
			
			CMEH13   & 15 Apr & 3:12   &  (252, -159)  	&   AR  		& 26 		&   742 	& 2.1  	&  9:45  & M  \\
			
			CMEN14    & 16 Apr & 11:00   & (910, -216) 	& AR, Fi  		& 15 		& 421  	&-9.1  	&15:45 & C \\
			
			CMEH15    & 17 Apr &  7:50& (550, -130)	 & AR, Fi      		 & 28 		& 1103  	& -19.7 	& 12:20 & M  \\
			
			CMEH16     & 21 Apr &00:48    & (916,-229)	 & AR			& 25 		 & 2,388 	& -1.4  	& 3:20 & C   \\
			
			CMEN17      & 06 May & 23:47    & (800, 400) & AR                		& 28  		 &   1266  & 5.6 	 &  4:10 (+1)& C   	 \\
			
			CMEH18      & 07 May & 3:36 & (-214, -109) 	& AR, Fi  		& 6  		& 926 	& 158.1 	& 4:50 & C  \\
			
			CMEH19        & 08 May & 13:13  & (130, -150) 	& AR, Fi 		& 5 		& 697	 & 78.9   	&  14:30 & C \\
			
			CMEH20         & 15 May  & 23:47  & (-197, -316) & AR			& 28 		& 506 	& -6.6 	& 8:45 (+1) & C \\
			
			CMEN21        & 17 May & 00:47 & (-100, -350)   & Fi                		 & 20  		& 532     & 5.5 	& 8:30  & C       	 \\
			
			CMEN22         & 17 May   &7:48    &(-900, 200) &AR    			&15		&616   &  -7.5    	& 11:40		& M  \\
			
			CMEN23          & 18 May & 11:50 & (-388, -455) & AR				& 6		& 614 	& 45.6	&14:00 & C		\\
			
			CMEP24          &  21 May & 23:24 & (881, -319) 	& AR, Fi 		&  28 	& 1341	& 14.2 & 3:45 (+1) 	& C \\
			
			CMEH25           & 22 May &  03:12  & (881, -319)	 & AR, Fi  		& 28   	& 1504 	& -10.4	 & 6:50 &  C \\
			
			CMEP26            & 27 May & 12:23  & (-250, 400)    & Fi           		 & 18   	 &1122  	& 3.8 	&15:50   & C         	\\
			
			CMEH27            & 15 Jul & 19:59   & (15,239) 	& AR, Fi    		& 28 		& 973 	& -25.6 & 0:20 (+1) 	& X \\
			
			CMEP28 & 15 Jul & 21:00  & (15,239) 		& AR, Fi      		& 23  	& 1264 	 & -7.3 &0:20 (+1)	 & M \\
			
			CMEH29 & 18 Jul & 7:59    & (500, 250) & AR, FL 						& 22		 & 919	 & -30.1 & 11:20	& X  \\
			
			CMEP30 & 18 Jul & 11:30   & (-730, 200)	 & AR			& 23 		&  680 	& -14.0 	& 16:20 & C  \\
			
			CMEH31 & 18 Jul  & 18:26  & (-730, 200)	 & AR 			& 28 	& 1788	 & - 	& 21:20	& C  \\
			
			CMEH32 & 26 Jul & 21:12& (-400, -370)	 & AR, Fi 		& 30 		& 816   	& -0.1 & 4:20 (+1) 	& M \\
			
			CMEN33 &   29 Jul  & 02:30  &(150,-350) & AR   			&   25           &409 &   3.8  	& 15:00		&   M   \\
			
			CMEN34 & 29 Jul  & 10:59 & (150,-350) 	& AR			& 11 		& 301   	& - 3.8  &16:00   & M          \\
			
			CMEN35 &  30 Jul &  00:30 & (700, -700) & Fi 				&  15		 & 998   	& 32   &4:20       & C            \\
			
			CMEH36  & 16 Aug & 11:24  & (-250, -200)	 & AR, Fi  		& 23 	& 1239 	& -67.1 	& 15:20   &   M \\
			
			CMEP37 & 23 Aug & 05:47 & (-243, -279) 	& AR                	 &  8 		 & 622  	& 36.6   & 8:20	 &  M           \\
			
			CMEH38 & 24 Aug  &  01:13 & (950, -100)   & AR			&   26   	& 2,066  	&  43.7   &3:20     &  X 	 \\
			
			CMEH39 & 05 Sep & 16:30 & (-400, 23) 	& AR 			& 17 		& 1855 	& 43.0 	& 18:20 & M   \\
			
			CMEN40 & 27 Sep  &  01:36 & (900, -200)  & AR                        & 21  	& 1300  	& -61 &3:40        & M          \\	
			
			CMEN41  &  27 Sep  &13:13 & (-650, 150) &AR  			  & 20       &510   &     -10  &18:45 			& M  \\
			
			CMEN42 & 06 Nov & 05:24 & (-49, -280) &AR				&16 		&485 	& -6.3	&  11:00  & C 		\\
			
			CMEH43 & 09 Nov & 12:54  & (624, -205) 	& AR, Fi 		& 26 		& 1977 	& 35.3 	& 15:40   & M \\
			
			CMEH44 & 24 Nov & 19:13 & (-750, 280) 	& Fi          			& 20 		& 1179  	& 20.5 &	23:20 & C   \\						
		\end{tabular}
		\label{table:CME_retenues_proprietes}
	\end{center}
	
\end{table}

\subsubsection{Ballistic Propagation between the Sun and L1}  
\label{ballistic}
In this study, we have essentially three measurement points
along the Sun--Earth line: the Sun (observed by SOHO), L1(where
the solar wind and the IMF are monitored by ACE), and the Earth. To
put those observations into the right context and make the appropriate
associations, we need to estimate propagation times between these three
points.

The SW propagation between L1 and the Earth lasts about 30 to 60 minutes depending on the velocity at L1, inducing a small uncertainty as compared to the propagation time between the Sun and L1, from approximately two to five days.  The latter is much more difficult to determine for several reasons. First, it may be difficult to isolate a CME in coronagraph data and the velocity inferred from LASCO is subject to errors due to projection effects. Then, the ICME velocity evolution in the interplanetary medium is not yet well understood.  Lastly, when several CMEs are ejected within a short time interval, they may interact with each other.
In a first attempt, for each possible association, we compare three velocities (\textit{cf.} Table \ref{table:associations_L1}): i) the CME velocity as measured by LASCO  at the final height $h$ measured in R$_{\odot}$ in LASCO FOV at time $t_{\odot}$.
ii) { $V_{\rm bal}$, the ballistic velocity, deduced from the time delay between $t_{\odot}$ and the arrival time at L1 of the perturbation {\it i.e.} start time of the ICME/MC ($t_{\rm s}$) or time of the SW discontinuity for Misc., Shock, and CIR/SIR ($t_{\rm d}$). 
iii) $V_{\rm L1}$, the ICME velocity at L1 ($V_{\rm s}$ or $V_{\rm d}$) at time $t_{\rm s}$ or $t_{\rm d}$, depending on the type of event}.
	
	As we used catalogs for the timing, there is no $t_{\rm s}$ nor $V_{\rm s}$ available for Misc., Shock, and CIR/SIR events.  For these events we use $t_{\rm d}$ and $V_{\rm d}$ values, which might not be appropriate for propagation considerations if ICMEs can be part of these events, specially for Misc. We took this into consideration during our analysis, especially when no relevant CME candidate could be found.
	
	If the ICMEs or shocks are decelerated from the corona to L1, the following inequality has to be satisfied, in order to possibly have an association between a given CME and a given SSC:
\begin{eqnarray}
V_{\odot} > &V_{\rm bal} &> V_{\rm L1}
\label{equ:vL1vBalvSun}
\end{eqnarray}
The balistic velocity criterion (Equation \ref{equ:vL1vBalvSun}) is fulfilled for 20/28 associations.
Because of uncertainties in $V_{\odot}$ (velocity projection
onto the Sun--Earth direction),  $V_{\rm bal}$, and  $V_{\rm L1}$
($t_{\rm s}$ often differ from one study to another), this inequality does
not always hold true. Even if it does, this may not be sufficient for
the identification, and we are often left with many possibilities,
especially when there are successive CMEs. That is why we used
complementary information to identify the CME corresponding to an ICME
or shock observed at L1.

\subsubsection{Drag-Based Model Propagation between the Sun and L1}
\label{drag}
To confirm the estimation of the CME arrival time at L1, and thus our CME--SSC associations, we have used the drag-based model \citep[DBM: ][]{Vrsnak2013}. The DBM is based on the fact that a CME does not propagate with a constant velocity but is, most of the time, slowed down by the ambient solar wind. Based on the equation of motion
and considering the drag coefficient to be constant, we calculate that a CME will propagate with a velocity [$v(t)$] that evolves with time:
\begin{eqnarray}
v(t)&=&\frac{( V_{\odot}- V_{\rm{up}})}{(1 \pm \gamma (V_{\odot}-V_{\rm{up}})t)}+V_{\rm{up}}
\label{equ:v(t)}
\end{eqnarray}
where $V_{\odot}$ is already defined, $V_{\rm up}$ is the velocity of the ambient solar wind as measured at L1 and $\gamma$ is the drag parameter. The sign of $\gamma$ corresponds to the two different regimes: $+$ for a decelerated CME, when $V_{\rm up}<V_{\odot}$ and	 $-$ for an accelerated CME, when $V_{\rm up}>V_{\odot}$.

Here we have used the DBM in an inverse mode, \textit{i.e.} not to predict the arrival time at L1 but, knowing the arrival time at L1, to infer what value of $\gamma$
matches the observed propagation time. We recall that typical  $\gamma$ values  are expected to lie between $0.1 \times 10^{-7}$ and $100 \times 10^{-7}$ km$^{-1}$ \citep{Vrsnak2013}.

The results are given in
Table \ref{table:associations_L1} for each SSC--Leading CME association. When there is a choice between different CMEs, it happens that the best association is for a rather low value of $\gamma$. Table \ref{table:validity_DBM} summarizes the results for all of the  associations, except for CIR/SIR for which it is not relevant.
Twenty-two associations have a DBM coefficient within the expected range.

The other four associations have a DBM coefficient lower than expected, and $V_{\rm bal} > V_{\odot}$; either the ICME may have sped up after it left the LASCO field of view or the radial velocity toward the Earth was larger than the estimated $V_{\odot}$  radial component.

\begin{table}[htbp]
	\caption{Validity of DBM coefficient $\gamma$ \textit{vs.} event characterization at L1 (leading CMEs only).} 
		\label{table:validity_DBM}
		\begin{tabular}{ r r r r r  r}
			\hline
			$\gamma$ [km$^{-1}$]                   & MC     & ICME      &   Misc.    & Shock & \textbf{Total}\\ \hline
			$ 10^{-8}<\gamma{}<10^{-5}$        &  9         & 6        & 4        &  3   &  22         \\
			$\gamma<10^{-8}$      & 3          &   0        &   0    & 1 &  4\\
		\end{tabular}
	\end{table}

Note that for the 15 ICMEs and MCs showing valid DBM coefficients, the mean
value is $0.7 \times 10^{-7}$ km$^{-1}$ and the median value is $0.5
\times 10^{-7}$  km$^{-1}$, in the lower part of the expected range of values
not far from the usually proposed standard value of $0.2 \times 10^{-7}$
km$^{-1}$. The validity of the $\gamma$ value being one of our criteria for the association, we tested it for the different possible sources of a given event at L1, as can be seen in the detailed example of association analysis given in Section \ref{section:L1_association_discus_ssc06} (see Table \ref{table:detail_SSC06}).  In particular, the DBM coefficients obtained for the so-called Contrib. and Alter. CMEs are out of the valid range respectively  for 7 out of 11 events and for 3 out of 5 events. The use of the DBM model for predicting the arrival time of
ICMEs is discussed in more detail in Section \ref{section:Discussion}.

\subsubsection{Chirality of the Magnetic Field}
\label{subsection:chirality}
MCs are ICMEs exhibiting a large-scale helical internal magnetic field corresponding to a flux-rope structure. The magnetic field orientation of the flux rope is called chirality and can give useful information on the location of the corresponding source at the solar surface. Indeed, the source of an MC in the northern solar hemisphere would give rise to a flux rope observed with a negative chirality (or left handed) and a positive one (right handed) if the source is in the southern hemisphere \citep[see, \textit{e.g.},][]{Ruzmaikin2003,Luoni2005}.
This agreement, or disagreement, between the hemisphere of the source region (Table \ref{table:CME_retenues_proprietes}) and the chirality observed at L1, is indicated in Table \ref{table:associations_L1}
for the 12 MCs events.
L1 and solar observations match each other in eight cases, \textit{i.e.} for 2/3 of the events. 
This proportion is similar to that noted by, \textit{e.g.}, \citet{Leamon2004}. This criterion is not considered for the other types of L1 events (no flux rope, no chirality).

\subsubsection{Analysis of Radio Emissions}
\label{section:L1identification_radio}
\begin{figure}
	\centering
	\includegraphics [angle=0, scale=0.35]{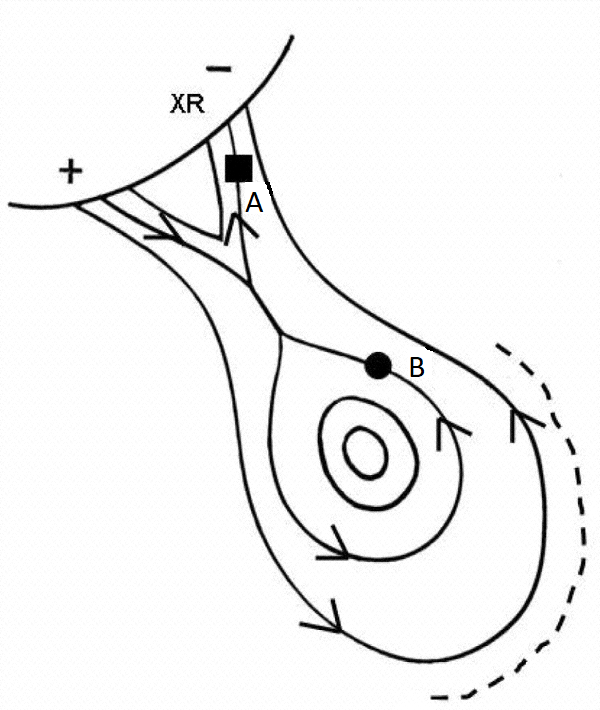}
	\caption{Twisted flux rope driving magnetic reconnection behind it,  during an eruption \citep{Pick2005}. The particles accelerated in the reconnection region propagate along the reconnected field lines, giving the observed main radio type IV sources (A and B), which correspond to the quasi-stationary sources and moving sources, respectively (see Table \ref{table:associations_L1}).}
	\label{figure:Modele-radio}
\end{figure}

We briefly recall our current understanding of the relationship between CMEs and type IV bursts.
The standard picture of the origin of the CMEs is the \citet{Lin2004} model: an initially closed and stressed magnetic configuration, which could be interpreted as a flux rope, overlying a photospheric polarity inversion line that becomes suddenly unstable and erupts. Magnetic lines are then stretched by the eruption and a current sheet (CS) is formed between the inversion lines and the erupting flux rope \citep{Schmieder2015}. Magnetic reconnection occurs along this CS, first at low altitudes then at progressively higher ones explaining the formation of post-eruption loops below the CS.
This model is supported by radio observations of type IV bursts. Reconnection is the source of accelerated electrons: downwards along the reconnected coronal arches, forming the stationary type IV burst component and upwards along the flux-rope border, forming the moving type IV burst (called radio CME by \citet{Bastian2001}), see Figure \ref{figure:Modele-radio} \citep{Pick2005, Demoulin2012b, Bain2014}. The same particles will produce the observed hard X-rays in the low corona \citep{Dauphin2005}.

Analysis of the radio emissions has been undertaken using multi-wavelength radio and white-light CME observations. In the absence of radio imaging observations, we search for the five following signatures:
\begin{itemize}
	\item The existence of a microwave burst near the onset of the event, lasting typically for more than five minutes and reaching a frequency of at least $\approx$ 2 GHz. This emission characterizes the onset of the flare followed by the onset of magnetic reconnection occurring behind the erupting flux rope, hereafter denoted A (see Figure \ref{figure:Modele-radio}). One may occasionally observe in the radio spectra that this A source drifts towards lower frequencies, indicating  that it rises in the corona.
	\item The existence of a long-duration radio continuum detected in a frequency range typically from decimetric to decametric wavelengths, hereafter denoted B (see Figure \ref{figure:Modele-radio}).
	\item The existence in \textit{Wind}/WAVES data of emission across a broad frequency range, which may be considered as the signature of a type IV burst, often following a type II burst.
	\item The existence of a type II burst indicative of a shock propagation (hereafter denoted II), often preceding a type IV burst detected either from ground-based spectral observations and/or from \textit{Wind}/WAVES data.
	It is today widely recognized that the shocks producing radio type II bursts can result from: CME interactions with the ambient medium \citep[see e.g.][]{Demoulin2012a, Pick2016}, CME--streamer interaction \citep[see e.g.][]{Cho2008, Feng2012} or CME--CME interactions \citep[see e.g.][]{Liu2014, Martinez2012}. 
	
	\item The radio signature of an interaction between two CMEs, which could occasionally modify the trajectory of the first one, or even change from one to the other (denoted E). A frequent signature of a CME--CME interaction is the onset of a radio type II burst, approximately at the time that the two CMEs contact, which reveals the beginning of a shock. More rarely observed, is the formation at rather high altitude of a radio continuum following a type II burst which has been also identified in the interplanetary medium as the consequence of the encounter of two CMEs \citep{Gopalswamy2001}.
\end{itemize}

The radio signatures found for the Leading and for the Contrib. or Alter. CMEs are
summarized in the two last columns of Table \ref{table:associations_L1}.
(see also Table \ref{table:radio} for the CME number of the radio
signatures). We found no more than one Alter. or one Contrib. CME with a radio signature  as a possible source for a given SSC. The presence of a type IV burst, and to a lesser extent of a type II burst,
makes the association more plausible. A detailed analysis of radio
signatures is given in Section \ref{section:radio_stat}. One can also
consult \urlurls{sites.lesia.obspm.fr/gmi-radio-cme/} for more details.

\begin{figure}[htb]
	\centering
	\includegraphics [width=10cm,angle=0, scale=0.88]{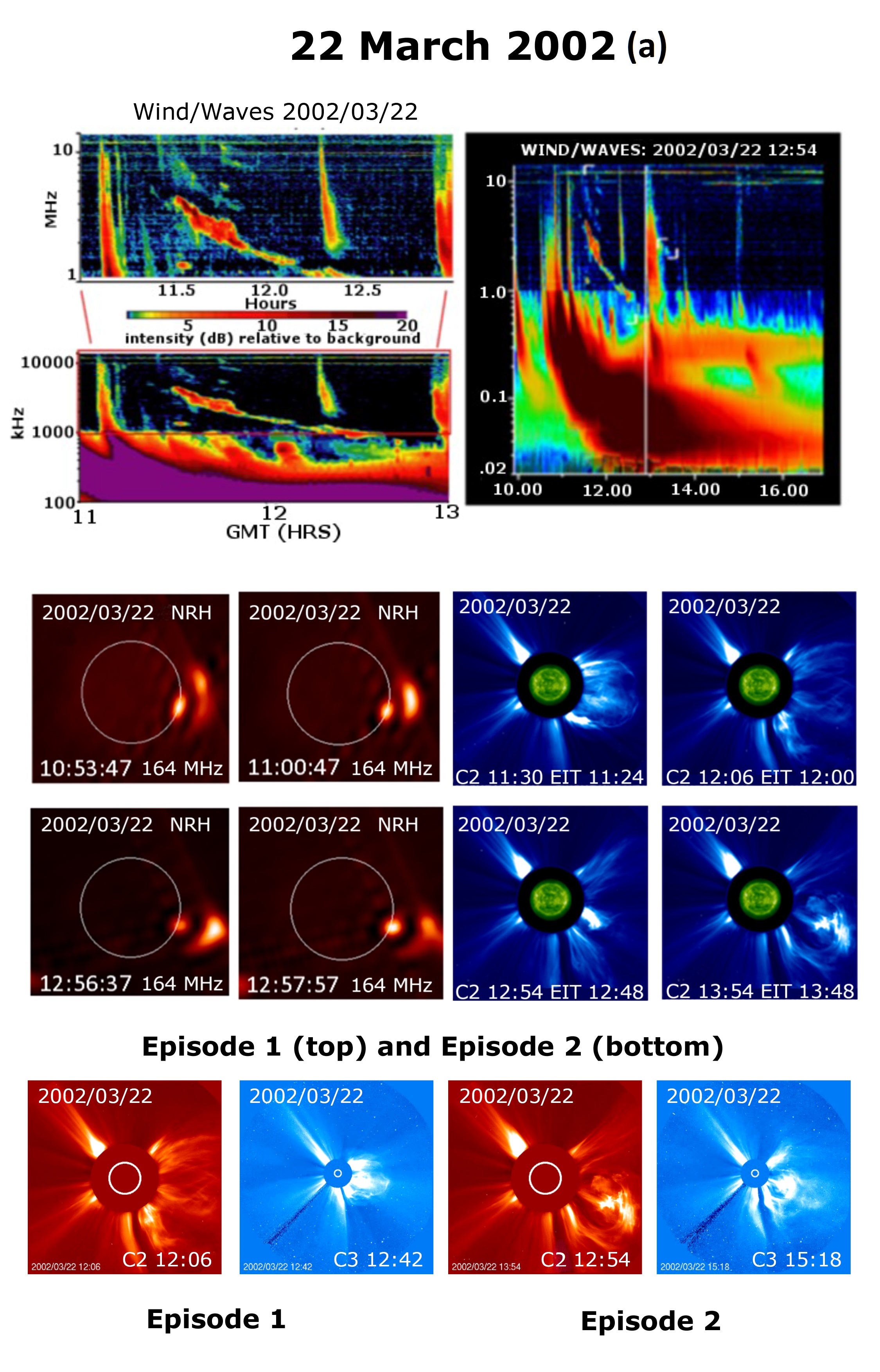}\hfill
	\caption{
		Radio and white-light CME observations on 22 March 2002
		illustrating the two episodes (see text), Part 1. (top panels): Dynamic
		spectrum from \textit{Wind}/WAVES. At the right we show the overall evolution where two successive episodes can be distinguished. The vertical-white line marks the
		onset of the second episode with the onset of a type II
		burst of very short duration (bracketed by two white-light signs). On the left
		side we show an expanding view of the first episode. (central panels)
		NRH images at 164 MHz of the radio source and the SOHO/LASCO-C2
		coronagraph CME images observed during episodes 1 and 2. (bottom panels) A comparison between episodes 1 and 2 of SOHO/LASCO-C2 and
		-C3 coronagraph CME images. Part 2. is in Figure \ref{figMarch22Septb}. }
	\label{figMarch22Septa}
\end{figure}

\begin{figure}[htb]
	\centering
	\includegraphics [width=10cm,angle=0, scale=0.88]{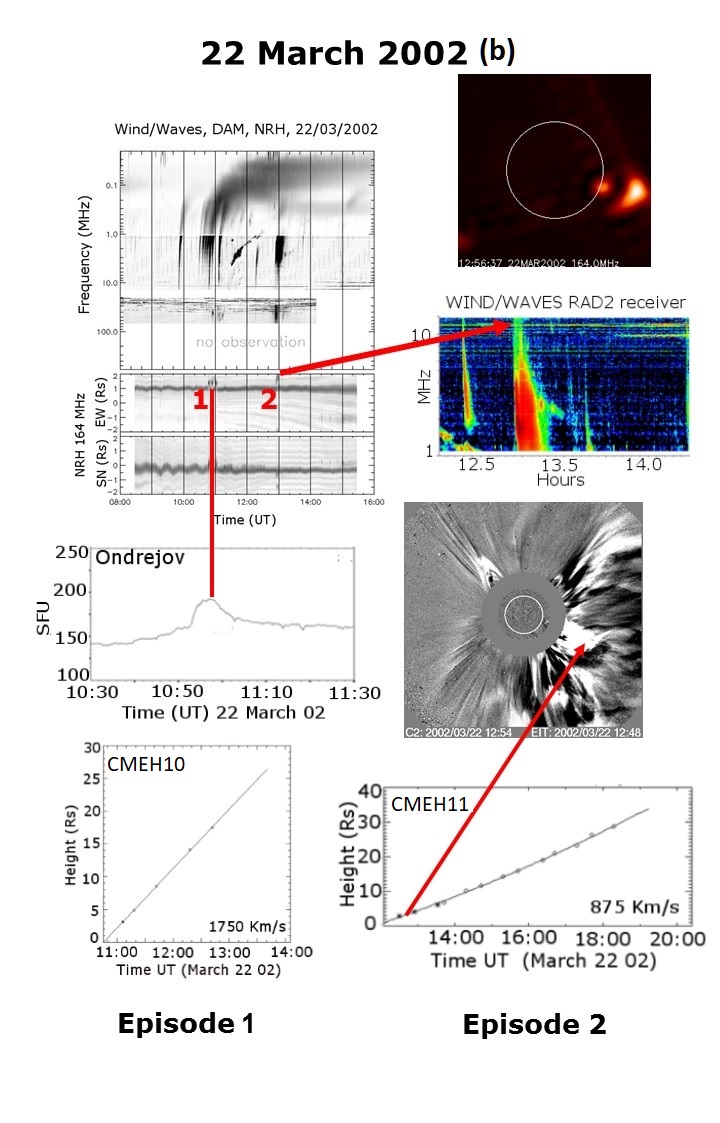}
	\caption{Radio and white-light CME observations on 22 March 2002
		illustrating the two episodes (see text), Part 2.
		(top-left panel) Composite dynamic
		spectrum from \textit{Wind}/WAVES (100 kHz\,--\,13.8 MHz) and from the Nan\c{c}ay Decameter Array (DAM)
		spectrograph (20\,--\,70 MHz) and brightness distribution projected on
		the heliocentric east--west and north--south directions \textit{vs.}
		time from NRH observations at 164 MHz. (central left panel) Microwave
		emission measured at 3 GHz. (top right panel) The radio source, (right panel below the top panel) the \textit{Wind}/WAVES spectrum, (central right panel) a SOHO/LASCO-C2 image during the
		episode 2, and  (bottom panels) CME height \textit{vs.} time of the two CMEs observed during episodes 1 and 2. Part 1. is in Figure \ref{figMarch22Septa}.}
	\label{figMarch22Septb}
\end{figure}

\subsubsection {Detailed Analysis of the Possible Solar Sources for SSC06}
\label{section:L1_association_discus_ssc06}
We discuss here in some details for one example (SSC06) how we try to determine the association between an event at L1 and its potential source(s) (four CMEs) at the Sun. This association is one of the seven cases of disagreement with \citet{Gopalswamy2010c} (see Table  \ref{table:associations_L1}).
The CME selected by \citet{Gopalswamy2010c} (detected by EIT on 20 March 2002 at 16:12 UT) is not included in Table \ref{table:CME_retenues_proprietes} as its association with SSC06, based on our criteria, is not convincing as detailed hereafter. In this section we label this CME as CMEGOP and as Else in Table \ref{table:associations_L1}.
Its source is AR 9871. From LASCO observations, CMEGOP moves at 511 km\,s$^{-1}$, is visible up  to 14 solar radii, and shows a deceleration of -15.2 m\,s$^{-2}$.

\begin{table}[htb]
	\begin{center}
		\small
		\caption{Association between the SSC06 event and the possible solar sources. V means that velocities are consistent
			(see Equation \ref{equ:v(t)}), D indicates coherence with the DBM coefficient $\gamma^{*}=\gamma\times10^{7}$,
			C with Chirality, and R with radio signatures. G10 refers to
			\citet{Gopalswamy2010c} and B17 to the present study. See Section \ref{section:L1identification_radio} for a description of the radio signatures.      }
		\begin{tabular}{l r r r c r c c c c c l}
			\hline
			CME         & \multicolumn{3}{c}{Velocities [km\,s$^{-1}$] }  & \multicolumn{2}{c}{Models} & \multicolumn{2}{c}{Chirality} & Radio      &Assoc.  \\
			No.        & $V_{\odot}$        & $V_{\rm bal}$& $V_{\rm L1}$    & Balistic    & $\gamma^{*}$    & Sun    &  L1        & signatures      &                   \\\hline    
			CMEGOP    &511         &442        &440            &?        & 1.30        &S        &rh        &              &D,C,G10     \\
			P09    & 1075        &446        &440        &?        & 3.14        &S        &rh        &         &V,  C                  \\    
			H10    & 1685        &795        &440        &ok        & 0.30          &S        &rh        &A,B,II    &V,C,D,B17         \\
			H11    & 1027        &823        &440        &ok        & $<$0.10        &S        &rh        &E,II         &V,C,R                    \\
			\hline
		\end{tabular}
		\label{table:detail_SSC06}
	\end{center}
\end{table}

\begin{figure}
	\centering
	\includegraphics [angle=90, scale=1]{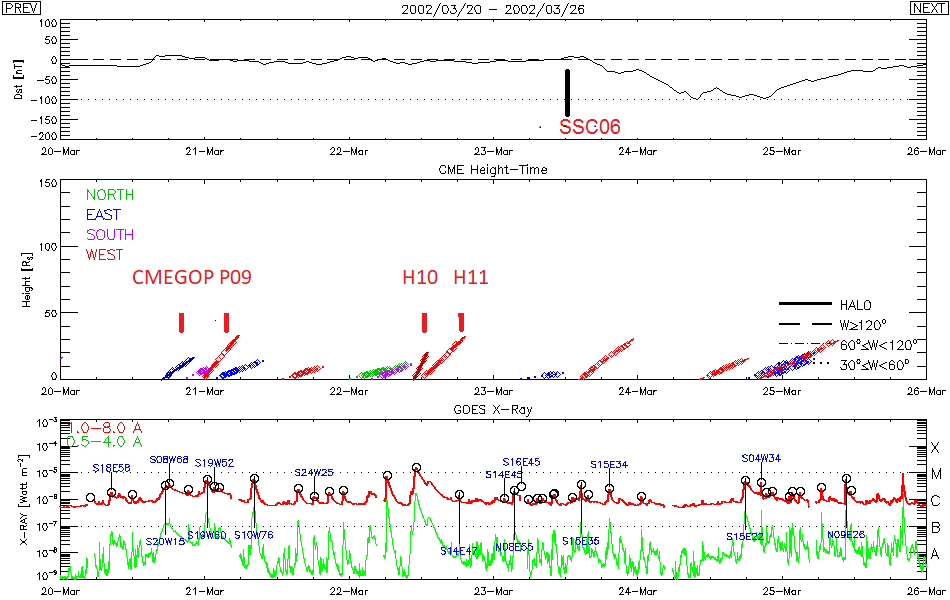}
	
	\caption{ Example of the link between the CMEs and the SSC for the SSC06 event. 
		Top panel: Temporal evolution of the Dst from 20 March to 26 March 2002. Central panel: Height-time plot for CMEs observed from 20 to 26 March 2002. We have indicated in this panel the events CMEGOP, CMEP09, CMEH10, and CMEH11. 	Bottom panel: GOES X-ray flux in W\,m$^{-2}$ on top of which the source coordinates on the Sun are given.
		Figure from the SOHO/LASCO CME catalog web page: \url{https://cdaw.gsfc.nasa.gov/CME_list/daily_plots/dsthtx/2002_03/dsthtx_20020320.html}.}
	\label{figure:dsthtx}
\end{figure}

First, the test on velocities using Equation (\ref{equ:vL1vBalvSun}) permits us to attribute only four possible CME sources to SSC06, which are listed in Table \ref{table:detail_SSC06}
and presented in Figure \ref{figure:dsthtx}. In that figure we use combined plots of the temporal evolution of Dst (first panel), the height--time plots for the CMEs (second panel), and the X-ray flux in W\,m$^{-2}$ (third panel) for the period 20 March\,--\,26 March 2002.

For CMEGOP, the three velocities are so close to one another that it seems unrealistic that CMEGOP could be the source of SSC06, taking into account the strong
deceleration observed at the Sun (-16 m\,s$^{-2}$). CMEGOP is consequently not included in Table \ref{table:CME_retenues_proprietes}. We are left with
three CMEs. For CMEP09, $V_{\rm bal}$ and $V_{\rm L1}$ are very close and
$\gamma$ is rather high. Then CMEH10 and CMEH11  are consistent for the three
velocities, but $\gamma$ for CMEH11  is too low. From the propagation
point of view, CMEH10 is preferred. Considering the chirality, all sources are in the southern
solar hemisphere and at L1 the MC is  right-handed, therefore, all solar sources are consistent and, thus, for this particular event the chirality does not allow us to discriminate.

Taking into account the combined radio
and coronagraph observations, the choice between CMEH11 and CMEH10 is not obvious: one notes the existence
of two episodes, each of them marked by a group of type III bursts
followed by a distinct type II burst with a much shorter duration for
the second episode (Figure \ref{figMarch22Septa}). The onset of
the first episode is also marked with a microwave burst (2\,--\,4.5 GHz
frequency range), with a long post-eruption decrease.  These two episodes
are associated with two distinct CMEs (CMEH10 and CMEH11). CMEH11  appears
in the same region and below CMEH10 at the time when the first type II
burst abruptly disappears. For each CME, the joint radio imaging and
CME analysis reveals a close correspondence between the site of the radio
source, observed at 164 MHz, and the overlying CME. The positions of the
two radio sources are close but different. A difference of approximately
$20^{\circ}$ in the inclination  of the  two CMEs is the only distinction
between these two successive episodes.
As revealed by the observations, these two CMEs appear to
interact during their propagation (see the red arrow in the bottom panels of Figure
\ref{figMarch22Septb}). Indeed, this encounter coincides with the sudden disappearance of the type II burst associated with CMEH10, whereas CMEH10 itself vanishes soon afterward. Meanwhile, CMEH11 continues to be accelerated and to rise in the corona. Besides, the CME trajectories do not appear to be affected by the encounter as CMEH10 propagates along an almost radial direction, while the trajectory of CMEH11 is more inclined (by roughly 30$^{\circ}$).

\begin{figure}
	\centering
	\includegraphics [angle=0, scale=0.45]{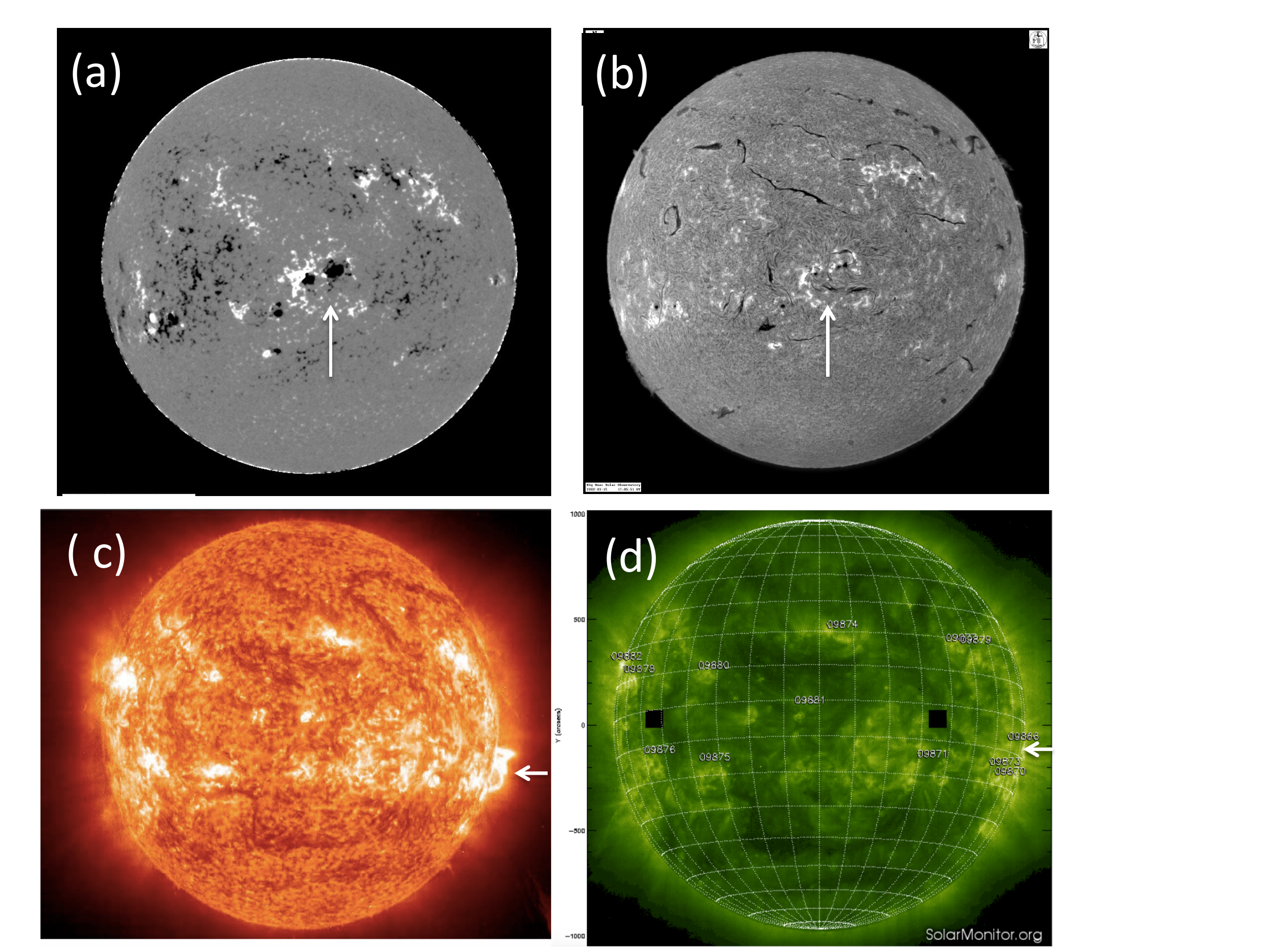}
	\caption{(a)--(b) Full Sun on 15 March 2002 showing the complex region at the central meridian formed by AR 9866 and AR 9870: The left panel is the magnetic field observed by the Global Oscillation Network Group and the right panel shows filaments, plages and sunspots in H$\alpha$ observed by Big Bear Solar Observatory (both images from \textsf{ftp://ftp.bbso.njit.edu}). (c) Full Sun on 22 March 22 2002 at 13:19 UT  in  30.4 nm observed by EIT onboard SOHO (from \url{helioviewer.ias.u-psud.fr/}). Numerous filament channels can be seen as dark corridors among the bright active regions. In the right bottom corner we can see the end of CMEH11, (d)  Full Sun on 22 March at 01:13 UT in 19.5 nm observed by EIT on-board SOHO (extracted from \textsf{www.solarmonitor.org}). The white arrows indicate the complex of regions responsible of CMEH10 and CMEH11 in each panel.
	}
	\label{figure:images_soleil}
\end{figure}

Taking into account both the duration of the two CMEs and the propagationconsiderations, it is unclear if CMEH10 should be favored more than CMEH11. Both CMEH10 and CMEH11 source regions are very likely at the origin of the
geomagnetic SSC06 event, therefore, we do not agree with
\citet{Gopalswamy2010c} who chose CMEGOP. 
\citet{Cid12} discussed the huge disturbance starting on 23 March at 10:53 UT with a minimum Dst = --100nT on 24 March that ended on 25 March. These authors concluded that the limb halo CME of 22 March ``was not related to the disturbance" but that ``the solar source related to the geoeffectiveness was close to the west limb [\ldots]. This large disturbance related to a solar location far from the central solar meridian could be due to the fact that the ejection arose from the interaction between two active regions[\ldots]".
The detailed radio analysis supports this interpretation by showing the existence of two CMEs, one fast CME (1685 km s$^{-1}$) seen at 10:36 UT in EIT images, originating probably from AR 9866 (including probably an erupting filament) and one slower (1027 km s$^{-1}$) CME detected at 11:36 UT in EIT images and originating from the southern active region AR 9870 (see Figure \ref{figure:images_soleil}). 
We conclude that, due to propagation considerations, CMEH10 is the Leading CME associated to SSC06, that CMEH10 and CMEH11  somehow merged, and that CMEP09 (also originating from AR 9866) could also interact with these two CMEs (\textit{cf.} Table \ref{table:associations_L1}). The interaction between the two limb active regions is visible in EIT image as north--south loops along the limb. In this portion of the Sun, there exist three active regions with filaments between them. From the solar point of view, it is the multiple interaction of CMEs that leads to the geoeffectiveness.

\subsubsection {Detailed Analysis of a Few Associations}
\label{section:L1_association_discus_all}
The above discussion demonstrates the difficulty of identifying the solar source(s) of an event seen at L1. We briefly review six more cases of disagreement with \citet{Gopalswamy2010c} (SSC02, SSC12, SSC21, SSC23, SSC24, and SSC28).

The SSC02-led event is associated with a CIR/SIR structure at L1,
and there is no CME that can be taken as its source during the five days
preceding SSC02. The partial halo CME chosen by \citet{Gopalswamy2010c}
was observed since 13 February 20:30 UT on the south-eastern (SE) limb, which gives an arrival
time at L1 earlier than that of SSC02. Finally \citet{Jian2006b} do not
list any ICMEs in that CIR/SIR perturbation.

Concerning SSC12, \citet{Gopalswamy2010c} identified CMEH18, which is is associated with strong radio waves, but it is detected only up to six solar radii. CMEN17, detected up to 28 solar radii, is in our study the most appropriate candidate as the Leading CME, according to chirality and velocity considerations. 
In the same line, the CME (28 September 2002 11:06 UT) associated with SSC28 by \citet{Gopalswamy2010c} has a too low velocity at the Sun. Note that \citet{Gopalswamy2010c} mention ``other possible candidate CMEs exist". CMEN40 is a convincing candidate (successful velocity, chirality  and DBM tests).

The type II and type IV bursts observed in particular by the Hiraiso Radio Spectrograph (HiRAS) in Japan suggest that CMEH29, which was chosen by \citet{Gopalswamy2010c}, is a strong candidate  for an association with SSC21. However $V_{\rm bal}$ is in that case as high as $V_{\odot}$ and $V_{\rm L1}$  (900 km\,s$^{-1}$). The strong deceleration observed at the Sun makes the SSC21--CMEH29 association more difficult. Conversely, CMEH31 is satisfactory from propagation considerations but with less significant radio signatures. We cannot exclude the possibility that CMEH29 and CMEH31 somehow interacted in the interplanetary medium.

SSC23 and SSC24 are separated in time by less than 20 hours and are associated with two different MCs at L1. Associations are thus uncertain. The best candidate for SSC23 is CMEN33, despite mismatching $V_{\rm bal}$. CMEN33 interacted with CMEN34 as revealed by joint SOHO and radio observations.
CMEN35 is the only candidate for SSC24 with a visible solar source. The CME chosen by \citet{Gopalswamy2010c} occurs at almost the same time, expands towards the north probably from a backside source at high latitude. This CME, which interacted with another CME, displays stronger radio signatures than CMEN35. The radio emissions associated with CMEN34 probably result from interaction with these two CMEs with no visible source at the solar surface (see \urlurls{sites.lesia.obspm.fr/gmi-radio-cme} for more details). 

Several other associations need to be discussed.
The SSC08--CMEN12   association would imply a too high $V_{\rm bal}$, and the ICME signature at L1 is not convincing: it is listed in only one catalog \citep{Jian2006b}. A convincing alternative would be that this SSC is not linked to any ICME as suggested by the ``Shock" signature at L1. Similarly, SSC29--30 might be only related to a CIR, and not to CMEN42. The situation is slightly different for the SSC31--CMEH43 association, as the SW discontinuity is observed at the ending time of the CIR. It might be that CMEH43 has caught up with the CIR between the Sun and L1.

Finally, the ballistic velocity criterion (Equation  \ref{equ:vL1vBalvSun}),
defined in Section \ref{ballistic}, is sometimes marginally
or even not fulfilled (as for SSC01, 03, 08, 10,
14, 23, 25, and
29\,--\,30). One should remember that the expansion-velocity value, deduced from the radial velocity measured by SOHO, is
approximate with an expected underestimation, which can be of
the order of 20\,\% or even more.
The SSC01-led event radio observations are commented in Section \ref{interactionCME}.

\subsection {Summary}

We have been able to identify a solar source for all MCs and ICMEs.
All SSCs associated with an MC at L1 have been associated with a CME, as can be seen on
Table \ref{table:associations_L1}. In some cases it is plausible that
two or more CMEs interact or merge before arriving at Earth.
The results concerning the refined relationship between SSCs  and CMEs
can be summarized as follows (see Tables \ref{table:associations_L1},
\ref{table:CME_retenues_proprietes}, and \ref{table:tab_SSC_sources}):

\begin{itemize}
	\item 14/28 (50\,\%) associations agree with all of the criteria relevant to their category, four criteria for MC structures at L1 and three ones for the other categories, as explained in Section \ref{subsection:association}.
	\item From the 28 SSC-led events with at least one source identified at the Sun, let us first consider those linked to MCs, the only category for which the four criteria apply (ballistic, DBM, chirality, and radio): only three of the 12 MC events fulfill these four criteria, four other fulfill three criteria. Ten MCs fulfill the radio criterion. We do not discuss further the two CIRs that may coincide with a CME. For the 14 remaining associations (ICMEs, Misc., Shocks), ten of them fulfill the three relevant criteria (4/6 ICMEs, 3/4 Misc., and 3/4 Shocks). This indicates the difficulty of finding the solar source for an ICME (MC or not) observed at L1.
	
	\item Three SSCs are not related to any CME of our list, but are associated at L1 to an CIR/SIR (Table \ref{table:associations_L1}).
	\item 28 SSC-led events are related to 44 CMEs, of the total of 60 considered CMEs.
	\item 21 halo CMEs are involved (of a total of 28 halo CMEs in 2002). The non-association of the other seven halo CMEs may be due to the fact that their sources are near the east limb.
	\item 17 Leading CMEs  are halo CMEs and 11 non-halo CMEs.
	\item The Leading CME is a halo for seven out of 12 MCs and for five out of six ICMEs.
	\item  From the 28 SSC-led events, only 15 events are related to a single CME and the other 13 events to a succession of several CMEs and to several solar sources.
	
	\item A plausible solar source is found for the four Misc. events and for three Shock events.  At L1 the satellite can travel only through  a part of an ICME, or of its leg.
	
	\item No really convincing solar source is found for the shock related to SSC08 except from the radio point of view (presence of a type II emission).
	\item SSC06 could be due to the interaction of multiple CMEs, for which the source region is at the west limb.
\end{itemize}

\begin{table}
	\caption{Type of solar sources and L1 events for the 31 SSC-led events.}
	\label{table:tab_SSC_sources}
	\centering
	\begin{tabular}{c  |c c c| c c | c| c}
		\hline
		& \multicolumn{3}{c|}{Single solar source} &\multicolumn{2}{c|}{ Multiple solar sources} & No solar & Total \\
		L1       & CMEH & CMEN & CMEP        & CMEH & no CMEH         &     source           &\\
		\hline
		12 MC             & 3 & 2& 0&                5         &        2        &    0            & 12 \\
		6 ICME             & 2 & 0& 0&                3        &        1        &    0            & 6\\
		4 Misc.            & 2 & 0 & 0 &                 2              &          0          &    0            & 4 \\
		4 Shock                        & 1 & 1+1? & 1 &            0        &        0        &     0            & 4 \\
		5 CIR/SIR                    & 1 & 1? &  0 &                0          &        0        &      3            &5\\
		Total                             & 9 & 3+2? & 1  &            10             &         3       &    3            &31   \\
		\hline
	\end{tabular}
\end{table}

\begin{table}
	\begin{center}
		\caption{Summary of radio signatures for the CMEs listed in Table \ref{table:associations_L1}. CIR/SIR-related events are not included, except for the SSC31 event. 
		CME events with no radio signatures are not included.
		A stands for the emission characterizing the onset of the flare followed by the onset of magnetic reconnection occurring behind the erupting flux rope. B stands for the existence of a long-duration radio continuum detected in a frequency range typically from decimetric to decametric wavelengths.
		E stands for CME encounter;
		W for \textit{Wind}/WAVES observations, and y for ground-based observation. ${^*}$ means drift in frequency, $^{\dag}$ means sudden stop, $^{+}$ means very long and complicated event considered as a single complex ejecta, ${^\circ}$ means weak, and ${^-}$ means particular signature. 
		Non leading CMEs are shown in italics, and those with strong radio emissions are underlined.
		For the separation of radio signatures in groups see text in Section \ref{section:radio_stat}.
		}
 
	\label{table:radio}
	\begin{tabular}{  r r| l  l| c c c c c  }
		SSC & min(Dst) &  Date & CME & \multicolumn{5}{c}{Radio signatures}\\
		No. & nT & 2002 & No. & A & B & E& II & IV\\
		
		\hline
		\multicolumn{9}{c}{\textbf{Group I}}\\
		SSC01 & -86 & 27 Jan         &  {\it \emph{H01}} & - & W & y & yW & -  \\
		&       &   28 Jan & N02 & - & y & - & - & - \\
		SSC03 & -71 & 24 Feb & N03 & - & y & y & y & - \\
		SSC04 & -37 & 15 Mar & H04 &  y${^\circ}$ &y& -
		& W &- \\
		
		SSC05 & -13 & 18 Mar & H06 & y${^\circ}$ & y & - & W & - \\
		SSC06 & -100 & 22 Mar & H10 & y & y & - & yW & y \\
		&  &  & {\it H11} & - & - & y & W & - \\
		SSC09 & -127 & 15 Apr & H13 & y	& y & - & W & y \\
		SSC10 & -149 & 17 Apr & H15 & y${^*}$ & y & - & yW & yW \\
		SSC11 & -57 & 21 Apr & H16 & y & y & - & yW & y \\
		SSC12 & -14 & 
		07 May & {\it \emph{H18}}  & y & y & - & - & y \\
		SSC13 & -110 & 08 May & H19 & y & y & - & - & y \\
		SSC14 & -58 & 16 May & H20 & y & y & - & yW & yW \\
		SSC17 & -109  & 21 May &  \it{P24} & y & - & y & - & -\\
		& & 22 May & H25 & - & y & y & yW & W$^{+}$ \\		 
		SSC21 & -36 & 18 Jul & {\it \emph{H29}} & y & y & - & yW & y \\
	SSC24 & -102 & 29 Jul & {\it \emph{N34}} & y & y & y & W & y \\
		SSC25 & -106 & 16 Aug & H36 & y & y & - & yW & yW \\
		SSC26 & -45 & 24 Aug & H38 & y & y & - & yW & y\\
		SSC27 & -181 & 05 Sep & H39 & y & y & y & W & y \\
		SSC28 & -176 & 27 Sep & N40 & y & - & - & - & -\\
		&  &  & {\it \emph{N41}}  & y${^*}$  & y & - & yW & y  \\
		SSC31 & -32 & 09 Nov & H43 & y  & y & - & yW & y \\
		\multicolumn{9}{c}{\textbf{Group II}}\\
		SSC08 & -23 & 11 Apr & N12 & {y} & - & - & {y} & - \\
		SSC16 & -12 & 18 May & N23 & y & - & - & y & - \\
		SSC18 & -13 & 27 May & P26 & - & y$^-$ & - & W & - \\
		SSC22 & 0 & 26 Jul & H32 & y & - & - & yW & - \\
		\multicolumn{9}{c}{\textbf{Group III}}\\
		SSC15 & -36 & 17 May &	{\it \emph{N22}} & y${^*}$ & - & y$^{\dag}$ & y & - \\
		SSC20 & -17 & 15 Jul & H27 & y & - & - & W & - \\
		SSC23 & -51 & 29 Jul & N33 & y & - & - & y & - \\
		SSC32 & -64 & 24 Nov & H44 & - & - & - & yW & - \\
		\hline
	\end{tabular}
\end{center}
\end{table}

\section{Event Characterization along its Sun--Earth Path}   
\label{section:EventCharacterization}

\subsection{Statistical Considerations on the Listed CMEs}
As mentioned in the previous section, we identified one or more possible solar sources for each SSC-led event.
The probability for an SSC to occur is some 75\,\%  if the CME is a halo (21/28).
According to the CDAW list, more than 500 non-halo, front-side CME were observed in 2002 (1.5 {\it per} day on average). Only 23 CMEs (see Table \ref{table:CME_retenues_proprietes}) could be the sources of an SSC, \textit{i.e.} 4\,\% (23/500).

For the 44 CMEs associated with an SSC (as Leading or Contrib.) in Table \ref{table:associations_L1},
\begin{itemize}
\item{18 have their source only in an AR (in one case, EIT detected the accompanying flare), four have their source only in a filament, and 22 have their source in an AR and a filament (including one case near a coronal hole (CH)). Thus 91\,\% (40/44) of the CMEs have their source in an AR (with or without a filament), 60\,\% (26/44) in a filament (in or out of an AR).}

\item {73\,\% (32/44) come from the southern hemisphere of the Sun, and 27\,\% (12/44) from the northern hemisphere of the Sun.}

\item{39\,\% (17/44) come from the eastern side of the Sun, 61\,\% (27/44) from the western side of the Sun, which is not surprising.}

\item{13.5\,\% (6/44) have a velocity less than 500 km\,s$^{-1}$ (non-halo CME), 43\,\% (19/44) a velocity ranging from 500 km\,s$^{-1}$ to 999 km\,s$^{-1}$, 38.5\,\% (17/44) a velocity between 1000 km\,s$^{-1}$ and 2000 km\,s$^{-1}$ (10/17 are halo CME), and 4\,\% (2/44) a velocity greater than 2000 km\,s$^{-1}$ (2 halo CME),}

\item{22 CMEs are associated with GOES C-class events, 19 with M-class events, and three with X-class events. All but one are linked to an AR. In 2002, around 2000 C-class flares, around 200 M-class flares, and 12 X-class flares were seen by GOES.}

\end{itemize}

\subsection{Statistical Analysis of the Radio  Signatures}
\label{section:radio_stat}
Our analysis, including the corresponding radio diagnostics, was performed for all of the 31 SSC-led events. The radio observations are summarized in Table \ref{table:radio}, the CIR events are not included.  Note that the SSC02 event (SIR), which is associated with a type II burst, is not reported in this table, while SSC31 event (a CIR at L1), associated with a halo CME (see Section \ref{section:L1_association_discus_all}), is included in the table. 

One distinguishes in Table \ref{table:radio} three groups: i) Group I that gathers all the radio events that display, at least, a low frequency long-duration component (see column B in Table \ref{table:radio}). 
ii) Group II that gathers the four radio events, which are associated to the four
Shock structures observed at L1 (and labeled Shock, see definition
in Section \ref{section:L1_characterization}). Their radio emissions appear to be quite different from those of Group I.
iii) Group III that gathers the remaining events. At first glance, the first three events look similar to those of Group II. However, this is not exactly the case as
discussed hereafter. The last event is only a type II burst.

\subsubsection{Different Radio Signatures}

	Group I contains the largest number (19) of events.  Seventeen events meet the two main characteristics of type IV burst events listed
	in Section \ref{section:L1identification_radio}, \textit{i.e.} the presence of two components, namely a high frequency component (A) and a lower-frequency (B) one. However, for two radio events, on 15 March (for SSC04) and on 18 March (for SSC05), the component A is unusually weak. In the case of the radio event on 27 January (for SSC01), no component A is observed. Seventeen among these 19 events are associated with a type II burst,
	most often observed by \textit{Wind}/WAVES. A large majority, 16 events, are associated with halo CMEs. We also note that the development of several events result from the interaction of two or maybe even more CMEs.\\
	Note that the radio analysis, for the four events SSC12, SSC21, SSC24, and SSC28,  leads to a choice of the CME origin (respectively CMEH18, CMEH29, CMEN34, and CMEN41 reported in column ``Contrib./Alter.'' of Table \ref{table:associations_L1})  which is different from  the identification essentially based on velocity consideration and drag-based model estimations (respectively  CMEN17, CMEH31, CMEN35 and CMEN40).
	The radio analysis for SSC01-led event leads to a choice between two CMEs, CMEH01, and CMEN02, which is discussed in Section \ref{interactionCME}.

	The four events of Group II in Table \ref{table:radio} correspond to the four so called Shock events detected at L1, and  their properties appear to be quite different from those of Group I. For three of them, their main characteristic is the presence of  a microwave component of short duration, a typical characteristic of the impulsive phase of a flare;  they are associated with a type II burst. The radio signature of the third one, on 27 May, is different (see \urlurls{sites.lesia.obspm.fr/gmi-radio-cme/}):  the NRH observed two moving radio sources detected at 432 MHz and 164 MHz. 
	The SOHO coronagraph images revealed that these four events are associated  with well-identified, narrow, eruptive structures presumably interacting with the CME before propagating again independently of it (see, in particular,  the striking example of SSC18).

	Group III gathers the remaining events.
	The development of 17 May radio event is, at its onset, typical of a type IV burst, \textit{i.e.} a microwave component which drifts towards lower frequencies. It then suddenly stops, presumably after its encounter with a coronal obstacle, as could be inferred from radio observations.
	This event is seen as an ICME at L1. The radio signatures of the 15 and the 29 July events look rather similar to the Group II events.

	However, no distinct  narrow eruptive structure was detected by SOHO, and at L1 one observes an ICME and an MC, respectively. The 24 November event consists in a type II only, and is seen as a Misc. event at L1 (see definition in Section \ref{section:L1_characterization}).

\subsubsection{The Importance of CME--CME Interactions}
\label{interactionCME}
As shown in Table \ref{table:radio}, seven CMEs in six events of Group I (\textit{i.e.} 32\,\%) interact with one or more CMEs during their propagation through the field of view of LASCO aboard SOHO. Some interactions might even be more complex than what we show in our analysis. Because of its  low velocity at the Sun we did not include in our CME list a third CME related to SSC17-led event with a distinct radio signature (A component, E component and type II burst). G. Lawrence (see \href{http://cdaw.gsfc.nasa.gov/CME\_list/}{\textsf{cdaw.gsfc.nasa.gov/CME\_list/}}) already noted that this CME could interact with CMEP24 and CMEH25 and that ``a single complex ejecta at 1 AU is the most likely outcome of these three events".\\
We also found evidence of interactions between CMEH01 and another CME (09:54 UT,  position angle 289).
The last one displays type II and component E radio emissions and is recorded about six days before $t_{\rm s}$, the MC arrival time at L1. That CME was thus not included in our CME list. 
The radio analysis shows that the SSC01-led event might be associated either with CMEH01 or CMEN02.
Consequently we do not consider a disagreement between radio observations and the choice of CMEN02 as Leading CME for the SSC01-led event.
The full presentation of this puzzling event 
(during the 24 hours of ICME sheath duration at L1, we could notice possible ICME signatures) is out of the scope of this study.\\
The analysis of these two events points out the difficulty to estimate the transit time between the Sun and L1.

\subsubsection{Summary}
To conclude, the analysis of radio signatures shows that:
\begin{itemize}
    \item A good agreement in the identification of the CMEs responsible of the SSCs, based essentially on the propagation time and velocity criteria, on the one hand, and, on the other hand, on the radio emissions accompanying their development and propagation, is obtained for 22 events (81\,\%  of the 27 events with a radio signature in Table \ref{table:radio}).	
	\item No or weak radio signatures were found for SSC07 (min(Dst)=-38 nT, CIR at L1), SSC19 (-16 nT, CIR), and SSC29-30 (-28 nT, CIR). 
	\item  {21} \textit{Wind}/WAVES type II bursts are associated with {20} SSC-led events (one event including two type II bursts). Eighteen type II bursts were detected by ground-based instruments, and 13 of them were also detected by \textit{Wind}/WAVES.
	We conclude that 25 SSC-led events (81\% of the total of 31 SSC-led events) are associated with type II bursts.  
	\item 	Among the {19} events of Group I, the type IV burst category, 17 have a minimum Dst value lower than -30 nT. This group gathers all the intense storms (9/9) and most of the moderate ones (5/6) of our study.
	\item {15} type IV bursts are identified (components A plus B, and  \textit{Wind}/WAVES cases).  
	48\,\% of the 31 SSC-led events are associated with type IV emission.	
	\item  Only four type IV bursts were detected by \textit{Wind}/WAVES in 2002. All of them are related to SSC-led events (and thus included in the previously mentioned 14 cases), and more specifically to MCs at L1. 
	\item Among the {15} events associated  with type IV bursts, {14} have a minimum Dst value lower than -30 nT.
	\item Eight of the {15} type IV bursts are associated at L1 with MCs and three with ICMEs. 67\,\% of the 12 MCs associated with an SSC are also associated with type IV bursts.	
	\item Seven CMEs interact  with one or more CME  during their propagation through the LASCO/SOHO field of view. There are two different CME  interactions in association with SSC17. These interactions may lead to a change in the propagation direction (CMEH10 and CMEH11, for example) or may result in the onset of a \textit{Wind}/WAVES shock.

	\end{itemize}

\subsection{Solar Wind} 

We compare the total number of L1 events during 2002 to the number
of events associated with an SSC. Based on the previously presented
bibliography, we distinguish four categories (ICME, MC, CIR/SIR, and
IP shock) the numbers of unique events observed in 2002
(total number) and associated with an SSC (+SSC). We also count the well-observed
MCs and ICMEs. A well-observed ICME (MC, respectively) is an
event reported in two (three or more, respectively) catalogs. The 
results presented below are summarized in Table \ref{table:mc_efficacite}.

\begin{table}[htb]
	\begin{center}
		\caption{MCs, ICMEs, CIRs/SIRs, and IP shocks associated with an SSC in 2002.
			The efficiency gives the percentage of (well-observed, right) events associated with an SSC.
			For the ICME (non-MC) events see definition in the text.
			A well-observed event is an event reported in several publications.   }
		\begin{tabular}{rrrrrrr  }
			\hline
			& \multicolumn{3}{c}{ Events in 2002} & \multicolumn{3}{c}{Well-observed events}   \\
			Structure at L1 & Total & + SSC & Efficiency  & Total & +SSC & Efficiency \\
			\hline
			MC & 17 & 12 & 71\,\% & 11 & 11 & 100\,\% \\
			ICME (non-MC) & 25 & 12 & 48\,\% & 10 & 6 & 60\,\% \\
			CIR/SIR (non-ICME) & 41 & 5 & 12\,\% & - & - & - \\ 
			IP shock (incl. ICME) & 35 & 28 & 80\,\% & - & - & - \\ \hline
		\end{tabular}
		\label{table:mc_efficacite}
	\end{center}
\end{table}

There are at most 17 MCs observed at L1 in 2002.  Eleven of them are
listed in three or more of the seven published studies (see Table \ref{table:Observations_at_L1} in Appendix \ref{section:App_SolWindObs}). We
refer to these 11 MCs as well-observed magnetic clouds.  As mentioned
in the previous section, 12 MCs are associated with an SSC. It is worth
noting that all of the 11 well-identified MCs, \textit{i.e.}
MCs that are well structured, are associated with an SSC.  Among the
five MCs not associated with an SSC, four are reported in only one
study, one in two studies. We also note that more than one-third (12/32) of the SSCs are caused by the encounter of the shock driven by the MC with the magnetosphere. Three MCs cause an SSC--SSE pair each, an SSC by the leading shock, an SSE by the MC itself (SSC--SSE pairs are those numbered as 09, 10, 28).

The two ICMEs studies \citep{Jian2006b,Richardson2010} include 34 unique events, including a total of 25 non-MCs ICMEs.  12 out of those 25 are associated with an SSC (6 ICMEs, 4 Misc., and 2 Shocks) and 10 are listed in both studies. The proportion of non-MCs ICMEs associated with an SSC is larger in the well-observed subset. The fraction of events associated with an SSC is lower for the non-MCs ICMEs category than for the MCs.

\cite{Jian2006a} identified 41 SIR events in 2002, including 20 CIRs. In our study only one SIR event and four CIRs are associated with one or several SSCs (SSC29--30 result from the arrival of the same CIR). The CIRs/SIRs efficiency in terms of SSCs is rather low as compared to the ICMEs and MCs.
In 2002, four events are SIR+ICME hybrids \citep{Jian2006a}; two of them cause an SSC.

\cite{Gopalswamy2010c} listed 35 IP shocks in 2002: 80\,\% caused an SSC.
The three SSCs that are not induced by an IP shock
from this list are linked to CIRs (SSC07, SSC19, and
SSC31). Conversely, the structure at L1 that follows
the seven IP shocks that are not followed by an SSC are identified as six
``sheath only" and one ejecta. The SSC--SSE pairs and the multiple
SSC29--30 (linked to a unique CIR), each group being
associated with a single structure at L1, are never associated with
multiple IP shocks. 88\,\% of the 2002 SSCs are related to
an IP shock.

In summary, IP shocks are a better proxy for SSC prediction than CIRs/SIRs, as noted by \citet{Taylor1994}, and well-observed MCs are the most efficient SSC drivers in 2002.

\subsection{Magnetopause--Magnetosphere} 
\label{Magnetopause-Magnetosphere}
\subsubsection{In situ Observations}
Here, our use of spacecraft observations is twofold. We look first at the magnetopause position after an SSC and second at Earth's radio emissions. We recall that in 2002 only the \textit{Cluster} and \textit{Geotail} spacecraft were operational to record the observation. 

To identify the magnetopause we used
the routine Cluster Science Data System
plots\footnote{\urlurl{www.cluster.rl.ac.uk/csdsweb-cgi/csdsweb\_pick}},
complemented by the \textit{Geotail} Plasma Wave Instrument (PWI) dynamic
spectra\footnote{\urlurl{space.rish.kyoto-u.ac.jp/gtlpwi/gtldata.html}}
\citep{Matsumoto1994},
from a few Hz up to 800 kHz, and orbit plots. For the identification
of terrestrial radio emission we looked at the \textit{Cluster}/Whisper wave data
dynamic spectra \citep{Decreau1997} in the frequency range 2\,--\,80
kHz together with the \textit{Geotail}/PWI dynamic spectra. 

\begin{figure}
	\centering
	\includegraphics [angle=90, scale=0.7]{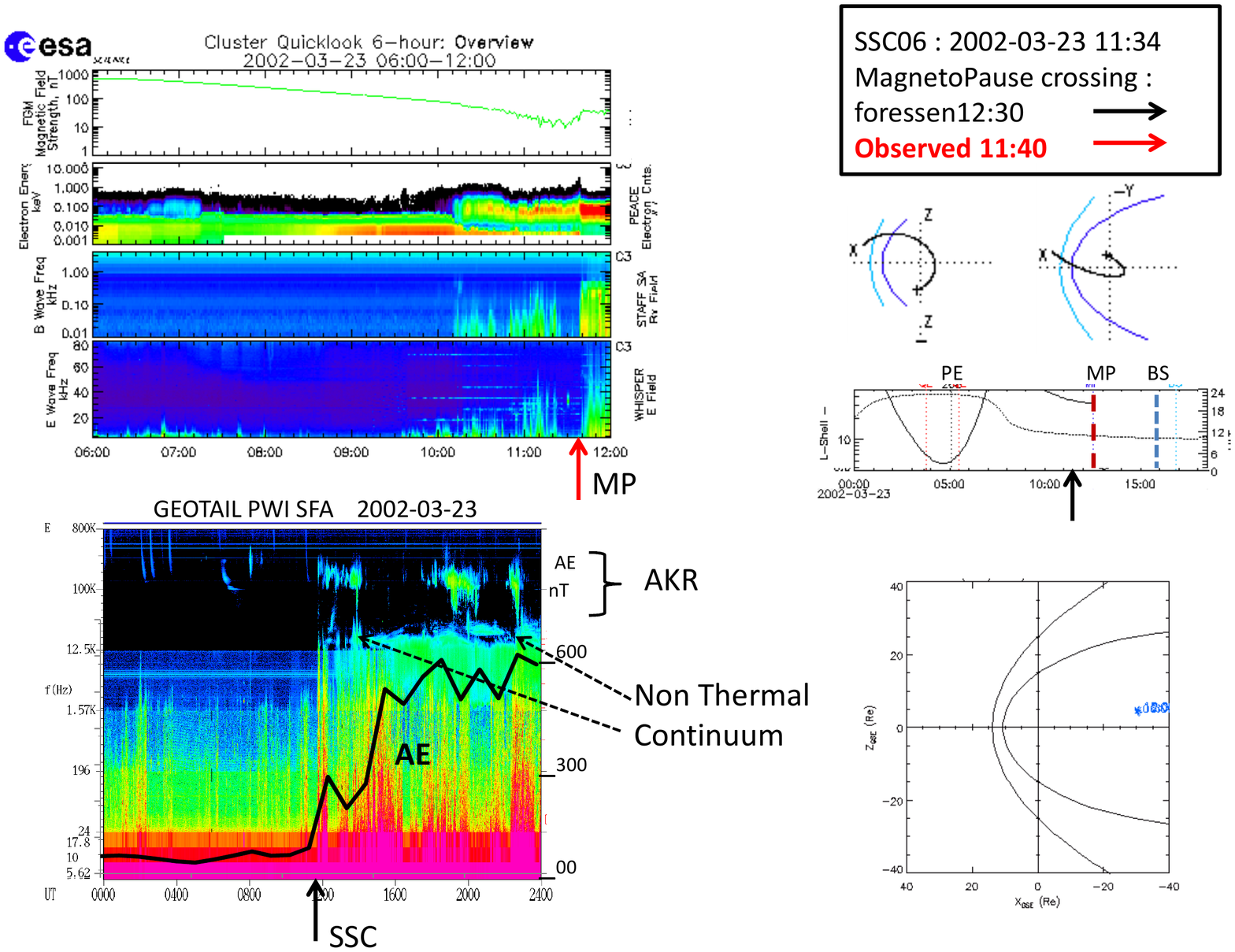}
	\caption{\textit{Cluster} and \textit{Geotail} data at the time of SSC06 on 23 March 2002.
		At the time of the SSC (11:40 UT) \textit{Cluster} crossed the magnetopause (red arrow), going from the high-altitude cusp into the magnetosheath \citep{grison2005}, as shown by the different data plotted (from top to bottom: $B$ field, electron fluxes, electric and magnetic wave data). \textit{Geotail} was far in the Earth's magnetotail. Orbit information on the right shows that the perigee was at 05:00 UT and magnetopause crossing expected at 12:30 UT. After the SSC, the wave activity increases, showing both NTC and AKR. The AE index is superimposed on the plasma wave dynamic spectra.}
	\label{figure:magnetosphere_SSC06}
\end{figure}

An example of data obtained after an SSC by \textit{Cluster} and
\textit{Geotail} is given in Figure \ref{figure:magnetosphere_SSC06}.
In what concerns the magnetopause position, observations are available
for only 15 events, which are distributed over all local times around noon,
from 05:00 to 21:00 local time. It is beyond the scope of the present study
to try to make any statistical study and comparison with prediction models
for magnetopause position as a function of solar-wind parameters, which
has been done recently by \citet{Case2013} for eight years of \textit{Cluster} data.

Table \ref{table:Cluster_obs} gives the local time (LT) of the magnetopause crossing, the delay between the SSC, and the time of the crossing and its distance from Earth. The observed magnetopause position is compared to the predicted one, which is established with the \citet{Sibeck1991} model for a moderate activity, \textit{i.e.} for a 2.1 nPa solar-wind pressure. All distances are given normalized to the Earth's radius [R$_{\rm E}$]. The difference between the observation and the prediction is given, as well as the percentage of variation, normalized to the predicted value. The table is organized by increasing time delays (sixth column), which is quite variable, from three minutes up to nine hours. 

\begin{table}[htb]
	\begin{center}
		\caption{Position of the magnetopause after the SSC when observed by either \textit{Cluster} (CLU) or \textit{Geotail} (GEO) spacecraft, organized by increasing time delay between the SSC and observation times. The SSC amplitude is also given (see text).}
		\label{table:Cluster_obs}
		\begin{tabular}{l r r | l r r r r r r  }
			\multicolumn{3}{c|}{SSC}  &  \multicolumn{7}{c}{Magnetopause crossing}    \\
			Event ID    & Date & $\Delta B$    & Space- & LT  & $\Delta t$ & obs. & mod. & $\Delta R$ &  \%   \\
			& & [nT] & craft & [h] & [mn] &[R$_{\rm E}$]&[R$_{\rm E}$] &[R$_{\rm E}$]&\\ \hline
			SSC16 &21 May& 17.6 &CLU &7 	&3 		&9.5 	&11.4 	&-1.9 	&17 	\\
			SSC14 &18 May& 40.7 &CLU &7 	&38 	&10.3 	&14 	&-3.7 	&36 	\\
			SSC06& 23 Mar& 21.5& CLU &12 	&56 	&10.8 	&13.1 	&-2.3 	&18 	\\
			SSC02 &17 Feb& 21.2 &GEO &10 	&60 	&10.2 	&10.8 	&-0.6 	&6 		\\
			SSC22 &29 Jul& 25.1 &GEO &8 	&60 	&10.3 	&11		& -0.7 	&6 		\\
			SSC29 &09 Nov& 8.2& CLU &20 	&60 	&17.3 	&19.3 	&-2 		&14		\\
			SSC04 &18 Mar& 60.8 &CLU &12 	&94 	&7.5 	&10.9 	&-3.4 	&31		\\
			SSC01 &31 Jan& 13.3 &CLU &13 	&120 	&10.2 	&11 	&-0.8 	&7 		\\
			SSC32 &26 Nov& 25.3 &GEO &4.5 	&150 	&15.5 	&17 	&-1.5 	&9 		\\
			SSC11 &23 Apr& 42.6& CLU &12 	&213 	&8.7 	&10.9 	&-2.2 	&20 	\\
			SSC20 &17 Jul& 40.1 &GEO &21 	&300 	&21.1 	&23.8 	&-2.7 	&11 	\\
			SSC13 &11 May& 25.9 &CLU &8 	&450 	&13 	&13.4 	&-0.4 	&3 		\\
			SSC03 &28 Feb& 37.4& GEO &20 	&460 	&8.9 	&14.3 	&-5.4 	&38		\\
			SSC09 &17 Apr& 52.7 &GEO &21 	&540 	&26.7 	&23 	&3.7 	&16 	\\
			SSC18 &30 May& 9.0 &CLU &5 	&540 	&17 	&15 	&2 		&13 	\\
			\hline
		\end{tabular}
	\end{center}
\end{table}

Considering the first seven events of the list for which the time
delay is smaller than 95 minutes, one notices an average decrease of
the magnetopause distance of 2 R$_{\rm E}$ and some correlation ($\approx$ 0.6,
not shown) between the SSC amplitude (third column) and the amplitude
of the displacement toward Earth, as compared to the
moderate-activity model. The biggest differences (4 and 3 R$_{\rm E}$)
are observed for SSC14 and SSC04, having an amplitude of
40 and 60 nT respectively, and kinetic pressure in the solar wind,
which varied by a factor of four and nine respectively at the time of the
shock. For larger time delays, it is difficult to reach any conclusion,
since there is no obvious correlation, which is understandable as various
factors can contribute in two to nine hours after the start of the storms.
The two cases (SSC09 and SSC18) for which the magnetopause is
observed farther away from Earth correspond to crossings that occurred
respectively around 21:00 and 05:00 LT, on the flanks of the magnetosphere,
and nine hours after the SSC.

One can also observe {terrestrial} radio emissions that either appear
after an SSC, or are stronger than before it, after 29 different SSC
events, \textit{i.e.} for all events where these radio emissions could
be observed by one or both spacecraft except for one case. For another
case, the orbits did not allow such observation, which is indicated as
``not available" (n/a) in Table \ref{table:Synthese_geoffect}. Those
radio emissions are auroral kilometric radiation (AKR) for 23 events
(75\,\%) and non-thermal continuum (NTC) for 26 events (90\,\%); both
being observed in 20 cases as can be seen in the central columns of Table
\ref{table:Synthese_geoffect}. 

An example is given on the bottom-left of Figure \ref{figure:magnetosphere_SSC06}. At the bottom of the \textit{Geotail} wave dynamic spectrum the AE index values (defined below), one-hour averaged data, are superimposed. One can see that AKR activity is in phase with the temporal evolution of the AE index.

Those observations are consistent with the fact that after an SSC
there are injections of energetic particles from the tail toward Earth,
the higher-energy electrons being responsible for AKR generation and
the medium ones for non-thermal continuum (NTC) generation \citep[see,
\textit{e.g.}, ][]{Louarn1994,Decreau2004}. AKR is believed to be emitted
by accelerated electrons of some 5\,--\,10 keV on auroral field lines
\citep[see a review by][]{Louarn2006}, whereas NTC comes from conversion
to electromagnetic waves of electrostatic emissions generated in the
plasmapause gradient in the equatorial region thanks to lower-energy
electrons accelerated in the tail and convected towards Earth \citep[see,
\textit{e.g.}, ][]{Gough1982, Kasaba1998}.

\subsubsection{Ground-Based Observations: Geomagnetic Response}

The global response of the magnetosphere--ionosphere system is monitored and characterized by means of geomagnetic indices. 
\begin{itemize}
\item{\textbf{PCN index:} The northern polar-cap magnetic index is representative of the magnitude of the northern trans-polar convection electric field, which drives the transpolar part of the ionospheric two-cell current system \citep{McCreadie2009}. As a result, increasing PCN values can be interpreted as increasing day-side merging solar-wind electric field \citep[see, \textit{e.g.}, ][]{Hanuise2006}.}
\item{\textbf{AU, AL, and AE indices:} The auroral-activity indices reflect the magnetic activity produced by the auroral electrojets that are mostly related to the magnetosphere--ionosphere coupling through the field-aligned currents: AU monitors the intensity of the electrojet flowing eastward in the magnetic local time (MLT) afternoon sector, AL monitors that of the electrojet flowing westward in the MLT morning sector \citep{DavisSugiura1966}, AE=AU-AL, AL being negative, is a global indicator that is currently used in substorm activity studies.}
\item{\textbf{am-index:} The planetary geomagnetic am-index is a three-hour planetary index derived from K-indices measured at a network of sub-auroral latitude geomagnetic observatories evenly distributed in longitude in both hemispheres. An extensive regression analysis of am and solar-wind data enabled \cite{Svalgaard1977} to show that any am-index is somehow a measure of the energy transfer to the magnetosphere during the corresponding three-hour interval.}
\item{\textbf{Dst and SYM-H indices:} The Dst index monitors the axi-symmetric part of the magnetospheric currents, including mainly the ring current, but also the magnetopause Chapman--Ferraro current. The SYM-H index is mostly similar as the hourly Dst index, but derived from a different set of stations, and with the advantage of being a one-minute index.}
\item{\textbf{ASY-H index:} The ASY-H index aims at monitoring the asymmetric part of the low-latitude geomagnetic field, in particular during geomagnetic storms. The asymmetric disturbance field has usually been attributed to a partial ring current. However, it may also be interpreted in terms of the effect of a net field-aligned current system flowing into the ionosphere near noon and flowing out near midnight.}
\end{itemize}

We use the min(Dst) quantity  to characterize the geomagnetic-storm intensity. For 2002 data, we are only concerned with the following categories of storms:
weak (min(Dst) $>$ -50 nT),
moderate ($-100 <  $min(Dst) $\leq -50$ nT),
intense ($-200 <$  min(Dst) $\leq -100$ nT), where we call very intense those with  min(Dst) $< -150$ nT.
	
	We find that only 15 out of the 31 different SSC-led events that occurred in 2002 are followed by at least a moderate geomagnetic storm; among them nine are tagged as intense, including two very intense ones, and six as moderate (see Table \ref{table:Synthese_geoffect}). We do not consider weak events as storms in what follows.

	Approaching the problem in the opposite sense and searching in the Dst index for geomagnetic storms in 2002 (AMDA\footnote{\urlurl{amda.cdpp.eu/}} data mining), we end up with 30 events, 12 intense and 18 moderate storms, 15 (nine intense and six moderate) of the 30 storms following one of the identified SSCs. 
	In other words,
	75\,\% of the intense storms (9/12) that occurred in 2002 are associated with an SSC,
    and 33\,\% of the moderate ones (6/18) are associated with an SSC.
	
 In addition, we consider  the integral $\Sigma_{i}(e)$ for each SSC-led event $e$, and each index $i$, where $i$ denotes PCN, AU, AL, ASY-H, or  am index. The integration starts at the time of the SSC, and ends when the Dst recovers its initial value.

\begin{figure}
        \centering
        \includegraphics [ 
        scale=0.6]{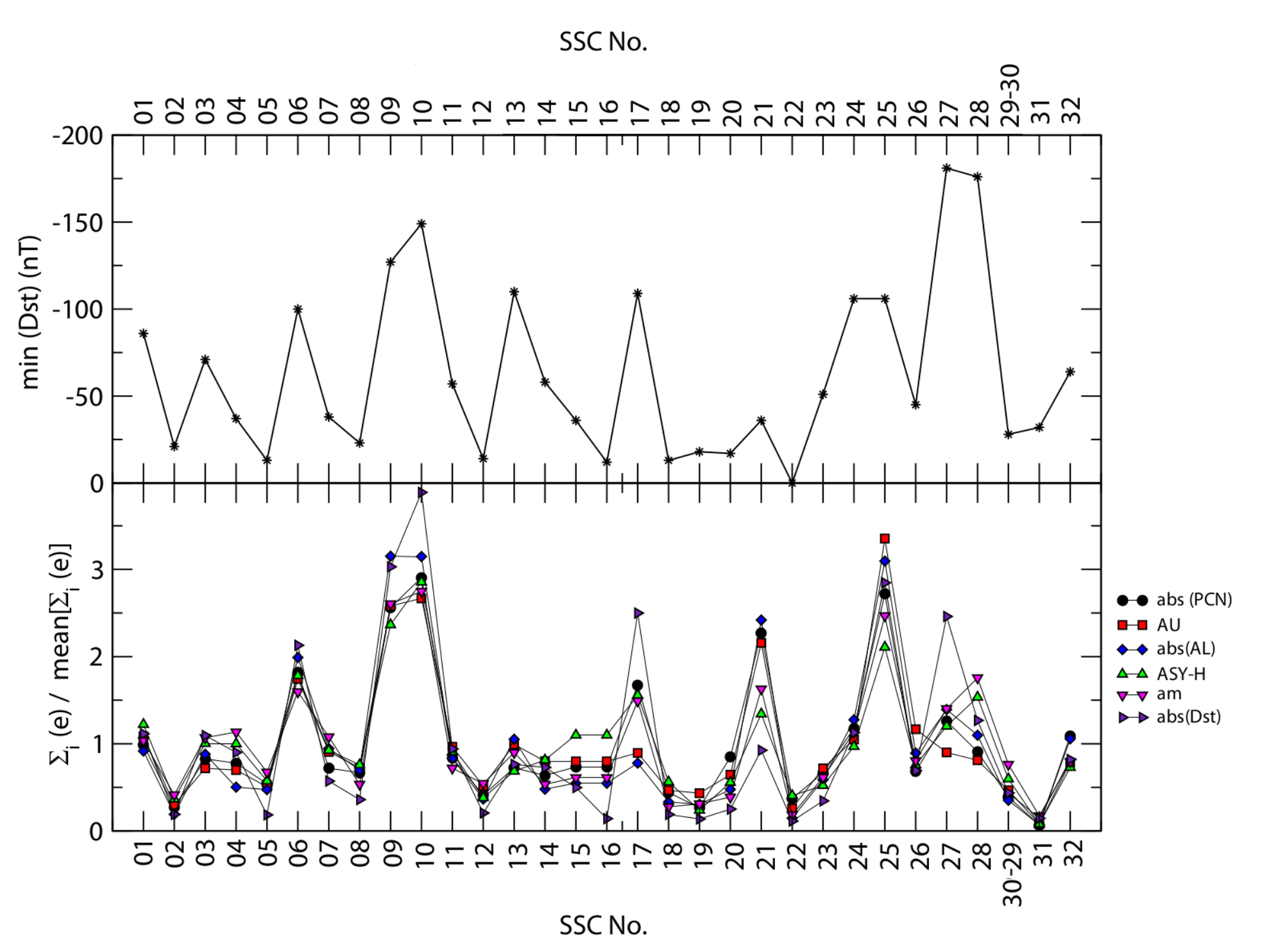}
        \caption{Variation of the min(Dst) and $\Sigma_{i}(e)/\textrm{mean}[\Sigma_{i}(e)]$ (where $i$ means abs(PCN), AU, abs(AL), ASY-H, am, or abs(Dst) quantities as a function of the SSC-led event number.}
        \label{figure:Indice_variation-sum}
        \end{figure}

Figure \ref{figure:Indice_variation-sum} shows that the distribution and relative values of the normalized $\Sigma_{i}(e)$ quantities vary from one event to another, but do not follow the min(Dst) variations exactly
(displayed on the top panel of Figure \ref{figure:Indice_variation-sum}).
min(Dst) is a good indicator of the overall geoeffectiveness of a magnetic storm,
although it is not sufficient to fully characterize the whole system.
A study of the differences between the behavior of the various indices is beyond the scope of this article.

\subsection{Ionosphere}  

To study the ionospheric convection associated with the 31 SSC-led events,
we used data from the SuperDARN radars. We have plotted the global
convection maps \citep{Ruohoniemi1996} for the 48-hour period following
each SSC. We have limited our analysis to northern-hemisphere radars
to keep the SuperDARN dataset comparable to the northern polar cap and
auroral indices used in this study. As expected, the convection associated
with SSC-led events reflects mainly the IMF variations, especially those
of the IMF $B_{\rm z}$ component. It is particularly true on the day-side:
lobe convection cell during northward IMF and intensified poleward
plasma convection during southward IMF. The main results of this study
are summarized in Table \ref{table:SSC_iono_thermo}.

We have particularly studied the time of each SSC arrival, to identify
specific convection signatures. 24 of all SSCs were concomitant with
positive or zero IMF $B_{\rm z}$ and 18 out of these 24 events presented
a night-side vortex, with two to ten minutes duration, 500\,--\,1500 km
in diameter, located around 23:00\,--\,01:00 magnetic local time (MLT)
and $65^\circ$\,--\,$80^\circ$ magnetic latitude (MLAT). For three other
SSCs concomitant with northward IMF, it was not possible to verify the existence of such a vortex due to very sparse data on the night side. 
The remaining three events show the usual Harang discontinuity signature without visible perturbations
caused by the SSCs. Finally, only one night-side vortex was observed for
an SSC concomitant with a southward IMF. These results are in agreement
with \citet{Tian2010} where plasma-flow vortices were observed in the
near magnetotail during a prolonged and intensified compression of the
magnetosphere and during northward IMF.

\begin{figure}
	\centering
	\includegraphics [angle=0, scale=0.75]{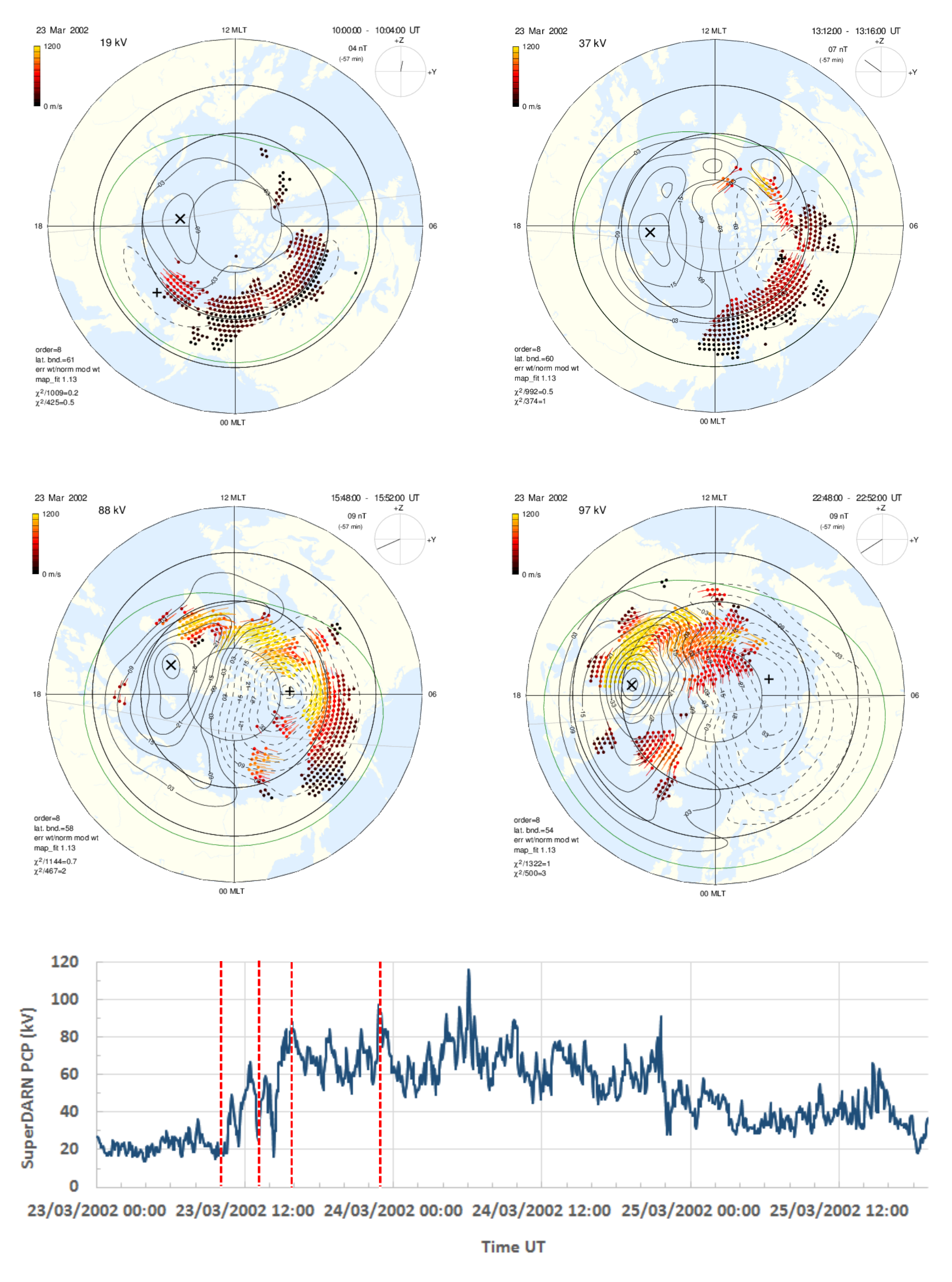}
\caption{Global convection maps deduced from all available radars in the northern hemisphere
for different times on 23 March 2002, before, during, and after SSC06. 
Top left: Map 1 from 10:00 to 10:04 UT.
Top right: Map 2, from 13:12 to 13:16 UT.
Middle left: Map 3 from 15:48 to 15:52 UT.
Middle right: Map 4 from 22:48 to 22:52 UT. 
Lower panel: SuperDARN PCP. The vertical red dashed lines show the times of Map 1 to Map 4.}
\label{figure:SuperDARN_23032002_vfinal}
\end{figure}

As for the other data sets, we focus on the MC event following SSC06 to
illustrate the different signatures identified with SuperDARN. Global
convection maps deduced from all available radars in the northern
hemisphere are shown in Figure \ref{figure:SuperDARN_23032002_vfinal}
for different times, before and after SSC06, together with the polar-cap
potential (PCP) deduced from the successive maps during the event (bottom
panel). Before the SSC, the SuperDARN data were very sparse
(Figure \ref{figure:SuperDARN_23032002_vfinal}, upper left map) and
only located on the night-side close to the Harang discontinuity, where
velocities remain relatively low. This poor coverage may be explained
by the dominant northward IMF prior to the SSC. The PCP is also very
low around 20 kV. At the SSC time, global convection remains quite similar,
except for the appearance of the aforementioned small vortex around 3
MLT and 78 MLAT, which lasts a few minutes, then disappears (not shown
here due to the four-minute average maps, which smooth this signature). During
four hours after the SSC, the PCP increases rather
significantly due to magnetosphere compression but remains very variable,
following the IMF $B_{\rm z}$ variations encountered in the MC sheath. Thus,
the PCP drops three times when IMF $B_{\rm z}$ turns northward and the lobe cell
has even time to develop in the dawn-side (Figure \ref{figure:SuperDARN_23032002_vfinal} upper right map). 
For the following 30
hours, the IMF $B_{\rm z}$ turns southward and the PCP increases around 70 kV
in average, reaching several times 80 to 90 kV. SuperDARN maps show two
well-defined convection cells (Figure \ref{figure:SuperDARN_23032002_vfinal} central panel maps), very dynamic in intensity
and direction in the day-side and often showing an asymmetric pattern in
agreement with dominant duskward IMF (Figure \ref{figure:SuperDARN_23032002_vfinal} central panel left map). During this period, periodic
night-side intensifications are also observed revealing substorm activity
(Figure \ref{figure:SuperDARN_23032002_vfinal} central 
panel right map). After this period, IMF $B_{\rm z}$ turns northward again and the PCP
and the SuperDARN data coverage decrease until the end of the MC. The number
of SuperDARN data points used to create convection maps is relatively high
for this event (see Table \ref{table:SSC_iono_thermo}), and the PCP deduced
is thus reliable during this event. We can see that a large increase
(up to 90\,--\,100 kV) is observed when the IMF turns southward, and not
at the SSC time, confirming the close relationship between ionospheric
convection intensification and IMF southward component.

Considering now the 31 events, columns 4, 5, and 6 in
Table \ref{table:SSC_iono_thermo} show the maximum polar-cap potential
(max(PCP)) reached during the 48 hours following each SSC, the time
of this maximum, and a classification depending on the strength of the
convection response. Strong (S) convection events are defined when
max(PCP) $\geq95$ kV, moderate (M) events are defined when $75\leq$
max(PCP)$<95$ kV, and weak (W) events when max(PCP)$< 75$ kV. It is
important to note that if the associated number of real SuperDARN data
points used to obtain global potential maps is below 300\,--\,350,
these maps are essentially driven by the statistical model used to fit
the data and the deduced PCP value is less reliable (identified by the symbol * in column 6 of Table \ref{table:SSC_iono_thermo}).

Among the 31 different events, 39\,\% (12 out of 31) fall in the
S-category, 51.5\,\% fall in the M-category (16 out of 31), and 9.5\,\%
fall in the W-category (3 out of 31). Although max(PCP) does not
reflect the average value over the entire event, statistically,
PCP is around 60 kV during negative IMF $B_{\rm z}$ and around $20$
kV during positive IMF-$B_{\rm z}$ component \citep{Ruohoniemi2005}. It seems
then that more sustained convection is observed following SSC-led
events, as can been seen for the SSC06 event (Figure
\ref{figure:SuperDARN_23032002_vfinal}, bottom panel). For all of these events, max(PCP)
was almost always associated with the strongest southward IMF period,
reflecting again the strong and almost immediate coupling between the IMF
(direction and magnitude) and the resulting ionospheric convection. This
explains why max(PCP) is reached at variable times after the SSC since
it depends on the IMF variations during each event, although the delay
was generally smaller for SSCs occurring during southward IMF periods.

\begin{landscape}
\begin{table}
	\caption{Ionospheric--convection and thermosphere-disturbance coefficients. Classification for ionospheric convection: S for strong if (PCP  $ \ge $ 95 kV); M for moderate if ($75 \le $ PCP $< 95$ kV) and W for weak if (PCP $< 75$).
	The symbol * indicates that the number of SuperDARN data points used to generate the global convection maps of this event is low ($<$ 300 points). For IMF $B_{\rm z}$ component close to 0, the IMF dominant component is indicated. Classification for thermosphere coefficients: A if (night $> 2$); B if ($1.5 <$ night $< 2$) and C if (night $< 1.5$). Letters n/a means no data available. $-$ means no significant variation. * indicates that even if no thermospheric coefficients are available, estimations are made using raw densities and the article by \cite{Forbes2005}.}
\tiny
\begin{tabular}{ c | c |c| r |c| c | c | r | r | c   }
  &	& \multicolumn{4}{ c|}{Ionospheric-convection coefficients} & \multicolumn{4}{c}{Thermosphere disturbance coef.}   \\
SSC 	& IMF		& Vortex type	& max(PCP)	& $t_{max(PCP)}$ & Class.	& $t_{max(night)}$	& Night		& Day		& Class.     \\
	No 	& B$_{\rm z}$	& at $t_{\rm SSC}$	& after $t_{\rm SSC}$		& after $t_{\rm SSC}$		&	& after $t_{\rm SSC}$		& coef.		& coef.	&       \\
\hline
SSC01	& South	& night-side		& 105	& 35:20 	& S	& 40:40	& 1.43	& 1.30	& C 	\\
	SSC02	& South	& no			& 75	& 10:30 	& M	& -		& -		& -		& - \\
	SSC03	& North 	& dawn-side (TCV)	& 85	& 22:00 	& M	& 21:37	& 1.35	& 1.24	& C \\
		&     	& night-side		&    	&       	&     &		&		&		&	\\
	SSC04	& South	& no    		& 60	& 00:15	& W	& 16:57	& 1.53	& 1.26	& B \\
	SSC05	& North	& night-side		& 90	& 02:30	& M	& -		& -		& -		& - 	\\
	SSC06	& North	& night-side		& 105	& 11:00	& S	& 22:20	& 1.65	& 1.29	& B \\
	SSC07	& North	& night-side		& 90	& 05:45	& M	& 12:31	& 1.42	& 1.21	& C \\
	SSC08	& South	& no			& 95	& 02:00	& S	& -		& -		& -		& - \\
	SSC09	& North	& night-side	?	& 90	& 07:30	& M*	& 07:55*	& n/a		& n/a		& B* \\
		&        	& (sparse data) 	&   	&     	&    	&   		&		&		& 	\\
	SSC10	& 0(By<0)	& night-side	& 100	& 22:00	& S & 21:20	& 1.75	& 1.29	& B \\
	SSC11	& North	& night-side		& 85	& 11:20 	& M*	& 04:46	& 1.53	& 1.22	& B \\
	SSC12	& North	& night-side		& 90	& 11:00	& M*	& -		& -		& -		& - \\
	SSC13	& South 	& no			& 105	& 04:00	& S*	& 10:16	& 1.67	& 1.28	& B \\
	SSC14	& South	& no			& 100	& 00:25	& S	& 12:52	& 1.75	& 1.30	& B \\
	SSC15	& North	& night-side		& 80	& 06:30	& M	& -		& -		& -		& - 	\\
	SSC16	& North	& night-side? 	& 85	& 07:30	& M*	& -		& -		& -		& - 	\\
		& 		&(sparse data) 	& 	& 		& 	& 		&		&		&	\\
	SSC17	& North	& night-side		& 95	& 01:15	& S*	& 05:32	& 2.15	& 1.39	& A \\
	SSC18	& North	& night-side		& 65	& 07:30	& W*	& -		& -		& -		& -\\
	SSC19	& North	& night-side		& 85	& 05:30	& M	& 19:69	& 1.35	& 1.17	& C \\
	SSC20	& South	& no			& 75	& 00:15	& M*	& -		& -		& -		& - 	\\
	SSC21	& North	& no			& 110	& 12:00	& S	& 21:35	& 1.49	& 1.48	& C \\
	SSC22	& North	& no;sparse data	& 65	& 00:15	& W	& -		& -		& -		& - 	\\
	SSC23	& North	& no;sparse data	& 75	& 00:30	& M*	& 11:20 	& 1.35	& 1.27	& C \\
	SSC24	& North	& night-side 	& 80	& 01:15	& M*	& 07:09	& 1.68	& 1.34	& B \\
	SSC25	& South	& no			& 95	& 11:15	& S*	& 14:42	& 1.71	& 1.37	& B 	\\
	SSC26	& 0 (By<0)	& night-side		& 90	& 07:00	& M	& 07:38	& 1.30	&1.12		& C \\
	SSC27	& 0 (By>0)	& dawn side(TCV)	& 100	& 05:00	& S*	& 11:07	& 2.36	& 1.53	& A \\
	SSC28	& North	& night-side		& 110	& 35:30	& S	& 34:30 & 2.46	& 1.82	& A \\
	SSC29-30	& North	& night-side		& 85	& 11:45	& M	& -		& n/a 	& n/a		& n/a \\
	SSC31	& North	& dawn-side(TCV)	& 90	& 03:45	& M	& -	& -		& -		& - 	\\
	SSC32	& North	& night-side?	& 100	& 06:45	& S*	& -		& n/a 	& n/a		& n/a \\
	\hline
	\end{tabular}
\label{table:SSC_iono_thermo}
\end{table}
\end{landscape}

\subsection{Thermosphere} 
\label{section:Thermosphere}
The density variations of the thermosphere are monitored by means of
thermospheric-disturbance coefficients, deduced from measurements of
accelerometers onboard the CHAMP spacecraft \citep{Reigber2002} along
orbit segments between $-50^\circ$ and $+50^\circ$ latitude that lasted
about 30 minutes.

For each orbit segment, the disturbance coefficient is estimated as the
observed mean density of the thermosphere normalized with respect to the
one computed using a semi-empirical thermospheric model (either the Jacchia-Bowman 2008 (JB2008)
model described in  \citet{bowman2008} or the NRLMSISE-00 model described
by \citet{Picone2002}) with magnetically quiet conditions, in order
to eliminate variations due to altitude, local time, and solar EUV flux. A value close to $1$ indicates that the actual
thermospheric conditions are in agreement with the model and can thus be
considered as those that would prevail in the absence of magnetic activity. A
value of $2$ indicates an overall increase in thermospheric density at low
and medium latitudes of 100\,\% \citep{Lathuillere2008}. For each orbit,
two coefficients are thus obtained for the northbound and southbound parts of the orbit, each one corresponding therefore to a local time differing by 12
hours (called night and day). The coefficients are associated with the UT
time at the Equator, with a sampling time of about 90 minutes corresponding
to the orbital period.

We examined the behavior of the thermosphere during the 24- to 48-hour period following each SSC-led events. 
Typically, the thermospheric disturbance-coefficients are enhanced
during storms with higher values for the night-side coefficient than for
the day side. The results are presented in the four last columns of
Table \ref{table:SSC_iono_thermo}. It gives the night and day maximum disturbance-coefficients following each SSC, as well as the time of
the maximum night-value after the SSC. For SSC29-30
and SSC32 too many missing values in the original data may
make it impossible to calculate the thermospheric coefficients. In this
case n/a (not available) is indicated in the table.
The last column
shows a classification depending on the density increase. Strong events
that correspond to a night thermospheric coefficient above two are labeled
A, moderate events that correspond to a night coefficient comprised
between 1.5 and 2 are labeled B, weak events correspond to a night
coefficient between 1.3 and 1.5 are labeled C. All of the events with a
coefficient below 1.3 are considered as showing no significant increase
in the thermosphere coefficients and are labeled -; no thermosphere
response is associated with these events. We have also not been able to
calculate the thermospheric coefficients in the period following SSC09,
but using raw densities \citep{Forbes2005} allowed us to tag this period
as moderate thermospheric storms, labeled B.

Among the 29 events with available thermospheric data, 19 (65.5\,\%) show a thermospheric-storm response, the other ten events do not show significant increase in the thermosphere coefficients ($-$). Among the 19 thermospheric storms events, the B class (moderate events) is the most populated one with 9 events (47\,\%).
There are 7 weak thermospheric storms labeled C (37\,\%), and only three strong thermospheric storms labeled A (16\,\%). In summary, we can say that about 40\,\% of the SSC-led events with thermospheric data (12 out of 29) are associated with a large or moderate thermospheric storm (labeled A and B), while the thermospheric response to the remaining 60\,\% (17 out of 29) is nonexistent or weak.

\section{Discussion}
\label{section:Discussion}

Before discussing the geoeffectiveness of the solar-wind events associated with the SSCs observed during 2002, we briefly compare our propagation observations with different simple or empirical models.

\subsection{Propagation from Sun to L1 for the 2002 Events giving Rise to SSCs}
\label{prop}
The radial propagation with constant velocity, using time and velocity observed by LASCO, is the simplest method to estimate the arrival time at L1 of any CME (Section \ref{section:relatedSunEarthEvents_methodology}).  In this section we focus on more sophisticated models: the \citet{Huttunen2005} empirical model and its complement by \citet{Schwenn2005}, and the DBM model already mentioned in Section \ref{section:relatedSunEarthEvents_methodology}.

We have used the empirical model proposed by \citet{Huttunen2005} for estimating the arrival time at L1 of the shocks and of the MC leading edge, when relevant, possibly associated with CMEs observed at the Sun: 
\begin{eqnarray}
T_\textrm{tr} &=& 236.7- 25.94 \ \ln(V_\textrm{exp})\hbox{ for the shocks,} \\
T_\textrm{tr} &=& 233.9 -23.55 \ \ln(V_\textrm{exp}) \hbox{ for the MC leading edges.}
\end{eqnarray}
 
Where $T_{tr}$ is the travel time [hours] and $V_\textrm{exp}$  is the expansion speed [km\,s$^{-1}$].

 We also used the relation obtained by \citet{Schwenn2005} for the estimation of the shock arrival time:
\begin{equation}
T_\textrm{tr-s}= 203-20.77 \ \ln(V_\textrm{exp})\hbox{ for the shocks.}
\end{equation}

For those three different  propagation-time calculations, we estimated the expansion velocity of the CME by applying a correction factor to the measured radial CME velocity,
using the one proposed by \citet{Schwenn2005}:
\begin{eqnarray}
V_\textrm{exp} &=& \frac{V_\textrm{rad}}{0.88}
\end{eqnarray}

\begin{figure}[htb]
	\centering
	\includegraphics [angle=0, scale=0.5]{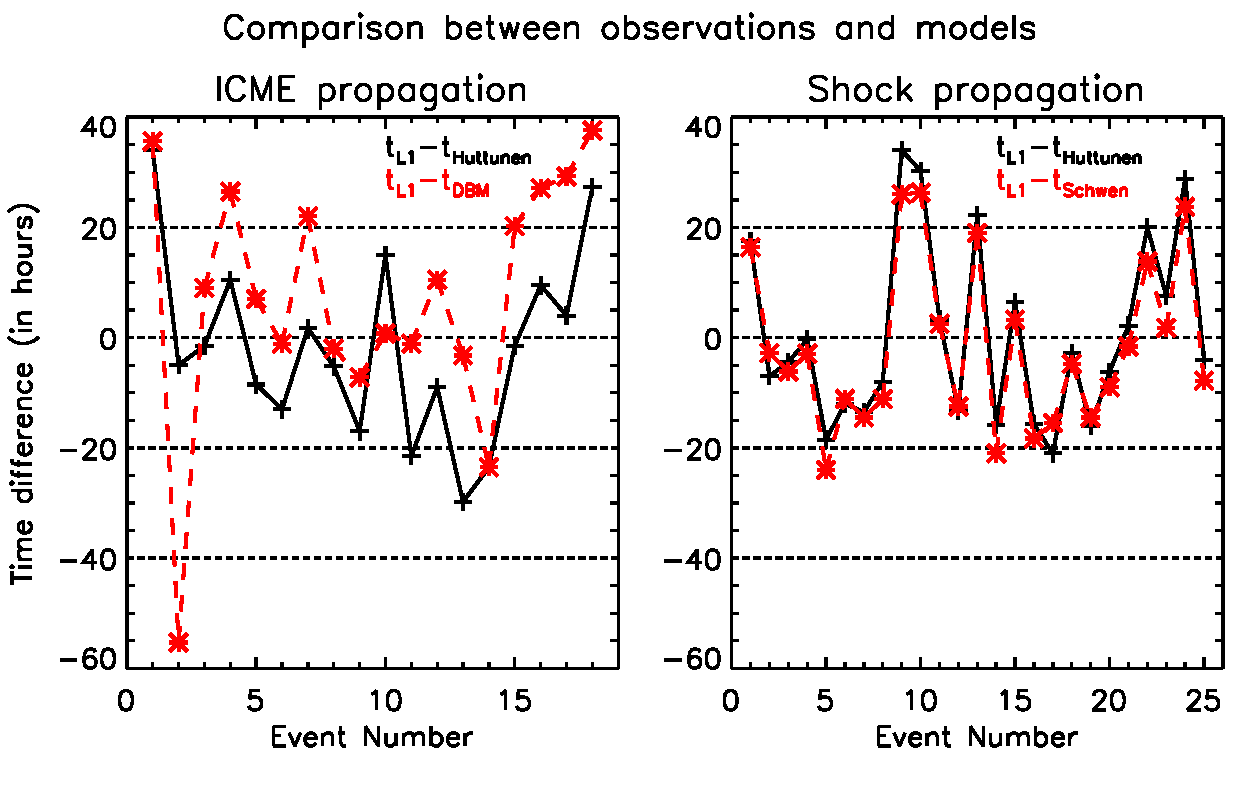}
	\caption{ICME (left) and shock (right) propagation. Accuracy [hours] of the three different models and the empirical relation is presented in Section \ref{prop}. Reference times are obtained from ACE observations.}
	\label{figure:PropaICME_Shocks}
\end{figure}

The DBM model \citep{Vrsnak2013} (see Section \ref{section:relatedSunEarthEvents_methodology}) also allows us to evaluate the arrival time of the MCs and ICMEs and further compare it with not only the actual observations from ACE but also with the \citet{Huttunen2005} equation results. Here we use the DBM as a prediction tool, with a standard drag coefficient, fixed to be $0.2\times10^{-7}$ km$^{-1}$ (see Equation \ref{equ:v(t)}). 
The results of the calculation are reported in Figure \ref{figure:PropaICME_Shocks}. They are given in terms of the difference of travel time between those calculated by the models and the observations by ACE at L1. Positive (negative) differences mean that the event (shock or ICME) arrived at L1 after (before) the model predicted time.

The comparison of the two empirical models of \citet{Huttunen2005} and  \citet{Schwenn2005} with ACE observations for shock-arrival times is given on the right-hand side of
Figure \ref{figure:PropaICME_Shocks}. There are 25 events, the five
CIRs/SIRs being not taken into account. One shock event is also excluded. Not surprisingly,
the results are similar. There are roughly as many
negative (15) as positive (nine) delays,  and one being
less than one hour, which is our temporal precision. 44\,\% of the absolute
delay values are greater than 10 to 14 hours,
which is the estimated error given by those authors
respectively.

For the ICME leading edges (the 12 MCs and the six ICMEs), the
results given by the \citet{Huttunen2005} empirical equation and
by the DBM model are given on the left-hand part of Figure
\ref{figure:PropaICME_Shocks}. For \citet{Huttunen2005} equation
(black curve), there are nine negative delays and seven positive ones
within their 14 hours error bars, and seven delays are out of this time range. We have
looked at the possible effect of the number of CMEs preceding a given
MC or ICME and found no particular trend. On the same left--hand part of
 Figure \ref{figure:PropaICME_Shocks} the DBM results are plotted
with a red dashed curve, for comparison. There are both similarities and
differences between the two models. Limiting ourselves to only MC does
not improve the results. Taking into account the number of the preceding
CMEs  does not change much the result,
but in this case we can say that the increasing number of CMEs before
a given MC slightly reduces the dispersion of time delays for the case
of the DBM.

We considered all ICME
and MC events of our list, and we have not restricted
ourselves to the L1 events clearly associated with only one halo CME,
contrary to \citet{Huttunen2005} who found more precisely a standard
deviation of 11.4 hours for 26 CME--MC leading edge pairs, and 9.7 hours
for 25 CME--shock pairs. Whereas limiting ourselves to well-identified
events with a single halo CME source would have lowered the number of events from 18 to five,
our results show that it would have not decreased systematically the
difference between prediction and observation.

In the present study we used models to get first-order rough estimates. As
noticed by \citet{Forsyth2006}, acceleration and deceleration may not be
important at large distances from the Sun, but other effects have to be
taken into account as, for example, the interaction of the CME with the
solar wind, ongoing reconnection, and CME mass \citep[for more details see
][Chapter 9]{Howard2014}. Taking into consideration more complicated
models, \textit{e.g.}, \cite{Xie2004}, or theoretical works, \textit{e.g.,}
\cite{Demoulin2008}, is beyond the scope of this study.

\subsection{Geoeffectiveness} 
\label{section:Geoeffectiveness}

In the following, we discuss the geoeffectiveness of the events associated with SSCs in 2002 from their characteristics in the interplanetary medium and the level of their responses in the magnetosphere, ionosphere, thermosphere. 

\subsubsection{ Characteristics of Geoeffective Events in the Interplanetary Medium }
\label{section:geoeff_interplanetary}

We start from the event types observed at L1: ICME-MC and non-MC events (58\,\% of all), Misc. events (13\,\% of all), Shocks (13\,\% of all), and CIRs/SIRs (16\,\% of all). Their geoeffectiveness is generally estimated by the the intensity of the storm that they trigger, \textit{i.e.} intense for min(Dst)$\leq -100$ nT and moderate for $-100<$min(Dst) $\leq$ $-50$ nT, as defined previously.

\begin{table}[htb]
	\begin{center}
		\caption{Geoeffective events starting by an SSC as a function of their classification at L1 (2002)}
		\label{cat_L1_geoeffectiv}
		\begin{tabular}{ r c c c c  }
			& Intense storm & Moderate storm & No or Weak Storm & Total \\
			\hline
			MC & 7 & 4 & 1 & 12 \\
			ICME (non-MC) & 2 &	- &	4 & 6 \\
			Miscellaneous &  -  & 2 & 2 &	4 \\
			Shock  &  -  & - & 4 & 4 \\
			CIR/SIR  &  -  & - & 5 & 5 \\
			\hline
			Total  &  9 & 6 & 16 & 31 \\
			\hline			
		\end{tabular}
	\end{center}
\end{table}

The results in Table \ref{cat_L1_geoeffectiv} show that almost all MCs (92\,\%) and only 33\,\% of ICMEs trigger an intense or moderate storm, whereas Shocks and CIRs/SIRs cause, at best, weak storms.
There are 15 SSC-led storms in 2002. The nine intense storms are all linked to ICMEs (seven MCs and two non-MCs). For the moderate storms, four out of the six are caused by ICMEs, which are all MCs, and two by miscellaneous structures. 

It is worth noting that the three more intense storms linked to MCs are those for which we identified an SSE at the beginning of the MC itself (SSC09-SSE09, SSC10-SSE10, and SSC28-SSE28). Those events clearly show two successive Dst minima, interpreted as the successive impacts of the sheath and then of the event core.

In the following, we examine several effects known to impact the geoeffectiveness, such as the orientation of the IMF north--south component, the respective role of the sheath and the central structure of ICME,
and their magnetic field rotation inside MCs.
\\

\paragraph{Role of a Southward IMF Component\\}
The presence of a southward magnetic component ($B_{\rm z}<0$) in
the interplanetary medium is generally recognized as a driver of
geoeffectiveness in the terrestrial environment. This orientation, opposite
to that of the terrestrial magnetic field, provides favorable conditions at the
magnetopause to trigger solar-wind/magnetosphere interaction processes,
such as the reconnection between the interplanetary and planetary magnetic
fields. The efficiency of these processes is generally expected to depend
on the amplitude and/or duration of the southward IMF component. In
order to estimate these effects, we considered for each event a time
interval ($dt$), defined from the arrival time of the SW discontinuity at
the bow-shock nose location to the time of the minimum Dst. We computed
two parameters from the five-minute averaged OMNI dataset:

\begin{itemize}	
	\item { $B_{\rm pz}$: the peak value  is the lowest negative $B_{\rm z}$ value recorded during the interval; this is the parameter tested by \citet{Echer2008} that correlates the best with min(Dst). $B_{\rm pz}$ is set to 0 if no negative value is recorded.} 
	\item {$B_{\rm z<0}^*$, computed by integration of $B_{\rm z}<0$ during the same interval $dt$, divided by the total interval duration (including periods of northward IMF).}
\end{itemize}

The parameter $B_{\rm pz}$ focuses on the effects due to large values of the southward component, while the parameter $B_{\rm z<0}^*$ is related to both its intensity and its duration. It takes larger values for longer periods of
southward magnetic field. This is further explained by comparison to mean$(B_{\rm z}<0)$, the computed integration of negative $B_{\rm z}$ divided by the total time for which $B_{\rm z} <0$.
If we consider two events with the same mean$(B_{\rm z}<0)$ value, the normalization over the whole interval gives a largest negative $B_{\rm z<0}^*$ for the event with the
largest fraction of time of negative $B_{\rm z}$. 

\begin{figure}
	\centering
	
	\includegraphics [angle=0, scale=0.5]{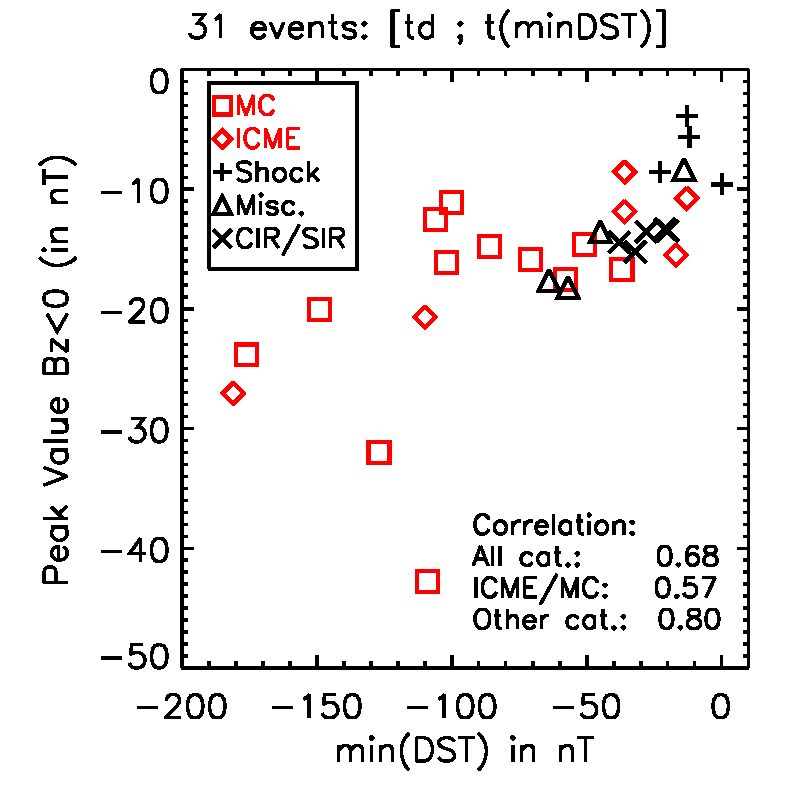}
	\includegraphics [angle=0, scale=0.5]{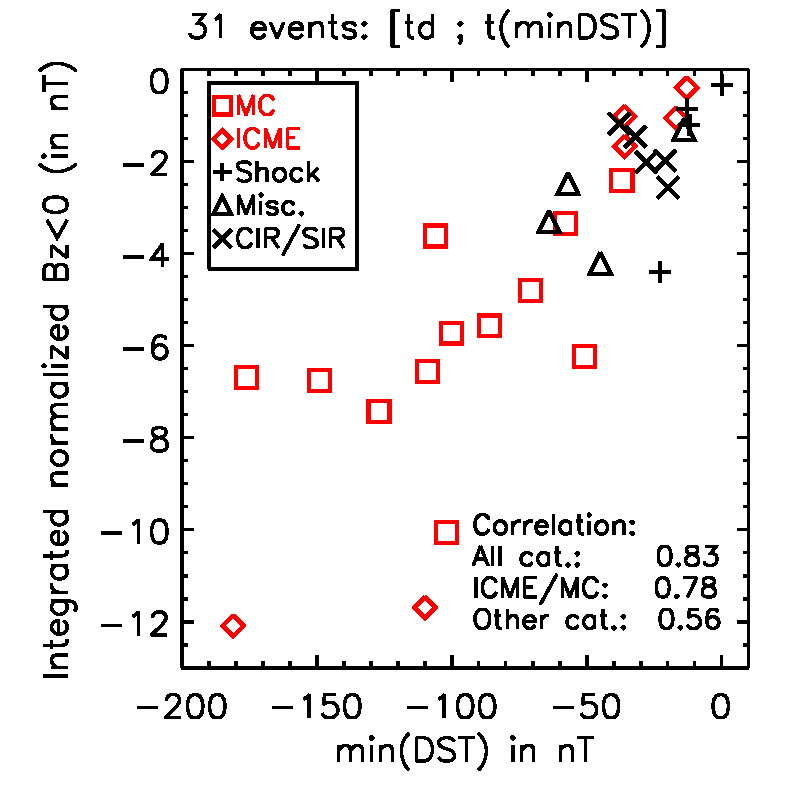}
	\caption{Minimum of Dst as function of $B_{\rm pz}$ (left) and of $B^*_{\rm z<0}$ (right) for all 31 SSC-led events.
	}
	\label{figure:BZ_DST}
\end{figure}

In Figure \ref{figure:BZ_DST}, $B_{\rm pz}$ ($B_{\rm z<0}^*$, respectively) is plotted \textit{vs.} the minimum of Dst on the left panel (right panel, respectively).
All events in each of the five L1 categories have the same symbol. ICME symbols (MC and non-MC) are drawn in red. 
For min(Dst) values larger than $-70$ nT, $B_{\rm pz}$ and min(Dst) are well correlated. The correlation coefficient ($\alpha$) for the non-ICME events is 0.80. A linear correlation can be visually noted. For more intense storms, the scatter is larger and the correlation less obvious.
Accordingly to this, $\alpha=0.57$ for the ICME events (MC or not).

When considering all 31 events, $\alpha$ is larger for $B_{\rm z<0}^*$ (0.83) than for $B_{\rm pz}$ (0.68). We can note that the high concentration of events in the upper-right corner makes the $\alpha$ values larger.  From that perspective, the 0.78 $\alpha$-value obtained for the ICME events, which are not limited to the upper-right corner of the figure, is indicative of a high correlation over the whole data range.
As previously noted, events associated with a min(Dst) value lower than $-100$ nT are only ICMEs and MCs, and no-MC ICMEs have the lowest $B_{\rm pz}$. Above $-100$ nT,
there is a linear correlation between $B_{\rm z<0}^*$ and min(Dst). 
For min(Dst)$<-100$ nT, the linear correlation is again less obvious: one should take also into consideration other drivers as, for example, the state of the magnetosphere or the
SW pressure ($P_{\rm SW}$). However our results demonstrate that $B_{\rm z<0}^*$ is a good indicator of the potential geoeffectiveness of a solar-wind structure. Down to $-100$ nT min(Dst) values, the solar-wind appears as the main contributor to the Dst decrease.
\\

\paragraph{Geoeffectiveness of ICME Core Structures and Sheaths\\}
For the specific ICME (MCs or non-MCs) category, the sheath preceding the structure can play an important role to trigger the storm. In order to estimate the geoeffectiveness of ICME structures and sheaths, we consider all MCs (12) and non-MC ICMEs (6). The results are presented in Table \ref{table:SW_ICME} (Appendix) and organized with increasing values of the global minimum Dst for each ICME category. Among the nine intense storms, three are caused by the sheath, four by the central ICME structure, and two by both regions. The four moderate storms are due to the MC core.

Table \ref{table:SW_ICME} also lists other parameters separately computed over the duration of each sheath and each central structure. The mean values of the parameters, such as $\beta$ (ratio of the plasma pressure to the magnetic pressure) and the Alfv\'en Mach number ($M_{A}$), reach their lowest values in the central structure. The values are relatively scattered; only the sheaths causing intense storms show limited variations for $0.4<\beta<0.75$ and $3.6<M_{A}<6.5$ and a trend of increasing $B_{\rm z<0}^*$ with increasing values of the minimum Dst.

\paragraph{Orientation of the Helical Structure of MCs\\}
Finally, the rotation of the magnetic field in MCs is given in Table \ref{table:SW_ICME}. It reveals that the south--north bipolar events are more numerous (seven events), than the north--south bipolar events (three events) and unipolar south events (two events).
There is no unipolar north MC. We note that the MC that triggered the largest storm, as well as the one that did not trigger any storm are both north--south bipolar events. The right-handed rotation dominates the left-handed one but cannot be related to the storm intensity. It should be linked to the fact that all the sources of the CMEs identified as being at the origin of MCs are located on the southern hemisphere of the Sun and no one in the northern hemisphere.

\paragraph{Geoeffectiveness of Other Events\\}
Among the four SSC-led events driven by miscellaneous structures,
two cause a very moderate geomagnetic storm (min(Dst)$>$ -64
nT). As a reminder, miscellaneous structures correspond either to not-well-defined 
ICMEs (not referenced as such in many articles) or to
more usual solar-wind structures but with some peculiarities, such
as for example strong magnetic field or sustained periods of negative $B_{\rm z}$. N
one of these events show specific characteristics in the
solar wind (see Table \ref{table:SW_others} in the Appendix), which could
help to understand why two of them are geoeffective and the two others
are not. None of the solar-wind parameters have the characteristic values
found for ICME events.

None of the four SSC-led events driven by Shock structures caused a geomagnetic storm, which can be explained by the solar-wind properties: moderate $B_{\rm z<0}^*$ value, very moderate mean $P_{SW}$ and low jump in solar-wind velocity across the shock (see Table \ref{table:SW_others}). 

None of the five SSC-led events driven by CIR/SIR (with one double event) causes a geomagnetic storm. The solar-wind parameters associated with these events are given in Table \ref{table:SW_CIR} in the Appendix. In detail, the CIR/IR response is more often stronger after the stream interface (SI) than before, and displays a lower min(Dst) if the jump in solar-wind velocity and the maximum of solar-wind pressure reached across the SI are stronger.

However, the low number of events for these peculiar categories may  not be statistically significant in terms of a solar-wind driver.

\subsubsection{Responses of Solar Events in the Magnetosphere, Ionosphere, Thermosphere }
\label{section:response_MIT}

The response of the different components of Earth's environment to these solar events can be described from the data libraries presented in Sections \ref{section:Introduction} and \ref{section:EventCharacterization}. 
The results summarized in Table \ref{table:Synthese_geoffect} allow quantifying different processes, as: 
\begin{itemize}
	\item {the magnetosphere compression, as characterized by the variation of its radius, and also by the amplitude of the SSC;}
	\item {the acceleration of energetic electrons in the magnetosphere, as manifested by the triggering of AKR and NTC terrestrial radio emissions;}
	\item {the current circulation, as characterized by geomagnetic indices such as the Dst index monitoring the magnetospheric current intensity and the AE index in close relation with energetic processes such as substorms occurring in the magnetotail;}
	\item {the ionospheric plasma transport, as characterized by the maximum PCP reached during the 24 to 48 hours following the SSC;}
	\item {the thermospheric dynamics, as characterized by the increase in thermospheric density at low- and mid-latitudes.}
\end{itemize}

The high pressure of solar events impinging on the {\bf magnetosphere} contributes to compressing it as shown by Table \ref{table:Cluster_obs}, when this information is available.
By considering only the first seven largest events of this table, the comparison with Table \ref{table:Synthese_geoffect} shows that the strongest compressions (-3.7 R$_{\rm E}$ $<\Delta{}R<-2.3$ R$_{\rm E}$) are caused by MCs. Here again the very large variability of the compression does not result in a simple correlation with the storm intensity.\\

Inside the magnetospheric cavity, the geomagnetic activity may be
manifested in the triggering of substorms in the geomagnetic tail and
the Earthward acceleration of magnetospheric particles. 

This particle acceleration also contributes to generate \textbf{radio waves}. Depending on their propagation, these waves might be observed far from their emission sites. This is the case for NTC and AKR emissions (see Section \ref{Magnetopause-Magnetosphere}). Table \ref{table:Synthese_geoffect} shows that, when observations are available:  

\begin{itemize}
	\item{NTC is observed during all ICMEs, MCs, and miscellaneous structures, whatever the intensity of the storm,} 
	\item{ AKR is observed during all intense ICME and MC storms and during five out of the six moderate storms. The case without AKR is consistent with the low value of its max(AE) index (570 nT). Also, the presence of AKR activity for all four miscellaneous events correlates well with their rather high AE index.}
	\item{For Shocks and CIRs/SIRs, the wave activity is more sparse. In most events, either NTC or AKR emissions are detected but not both, and one event (SSC22) is found without any wave signature (explained by the low AE index (198 nT) and a min(Dst)=0).} 	
\end{itemize}

For most MCs, ICMEs, and Misc. structures, the simultaneous detection of NTC and AKR emissions -- caused at different locations in the magnetosphere by particles of different energies, indicates a widely spread activity of acceleration processes, extended over a large sector of the plasmasheet. Conversely, the detection of one of them, as in the case of most Shocks and CIR/SIR events, suggests more local effects.\\

The enhanced particles acceleration and transport in the magnetosphere also results
in an enhancement of the strongly coupled \textbf{electrojets in the
auroral ionosphere}. Very intense auroral activity with an AE index, such as: max(AE) $>
1000$ nT occurred during the events associated to all nine intense storms and two moderate ones.  
Thus ten (out of 12) of the most intense auroral-activity events
are observed during storms driven by ICMEs (MCs or not) and one during a
storm driven by a miscellaneous structure, the latter being associated
with a relatively weak (geomagnetic) storm (min(Dst) = -45 nT). We note that
the auroral activity indicated by the max(AE) index is relatively high
for the four miscellaneous events, between 886 nT and 1136 nT.

The \textbf{ionospheric transport} depends on the polar-cap potential
imposed by the solar-wind/magnetosphere coupling. The maximum values of
the PCP (deduced from SuperDARN) during the events associated with the nine intense storms and the six moderate storms reach
values larger than or equal to 95 kV (strong convection) for ten events, and
between 75 and 95 kV (moderate convection) for five events. The PCP response
is globally stronger for the strongest geomagnetic storms. Concerning
ICMEs, the only case of weak convection concerns the only non-geoeffective
MC. Finally, 67\,\% of events leading to intense and moderate storms are associated with PCP
values larger than 95 kV and 33\,\% with PCP values between 75 and
95 kV. This is significantly different from the statistics for the
overall 31 SSC-led events (39\,\% for max(PCP)$\geq 95$ kV and 51.5\,\%
for $75\leq$ max(PCP)$<95$ kV). This shows the reinforcement of auroral
ionospheric convection during events associated with intense and moderate storms.\\

For the same 15 events associated with strong and moderate storms, 14 \textbf{thermospheric responses} were available, and significant enhancements of the nocturnal neutral density were identified in all cases (100\,\%) (see Section \ref{section:Thermosphere}). All three strong responses
(labeled A with a density increase of more than a factor of two) are associated with three intense storms driven by ICMEs (including two MCs); eight moderate responses (57\,\%, labeled B) correspond to six intense and two moderate storms driven by ICMEs and one miscellaneous event), and three weak responses (21.5\,\%, labeled C) correspond to three moderate geomagnetic storms driven by ICMEs. To conclude:
\begin{itemize}	
	\item {100\,\% of the events associated with strong and moderate storms induced a thermospheric storm. Conversely, 78\,\% of the thermospheric storms are linked to such events. } 
	\item {Concerning ICMEs (MCs or not), 83\,\% of them induce a thermospheric storm, and conversely, 79\,\% of thermospheric storms are linked to ICMEs. The three strong (100\,\%) thermospheric storms are associated with intense ICMEs (including two MC) driven geomagnetic storms } 

	\item { CIRs/SIRs and Shocks have almost no impact on the thermosphere}
\end{itemize}
Finally, the thermospheric response correlates rather well with storms. All events with very negative min(Dst) ($\leq -100$ nT) exhibit a strong thermospheric density enhancement (labeled A or B), consistently with \citet{Krauss2015} using {\it Gravity Recovery and Climate Experiment} (GRACE) spacecraft data. Conversely, the five non-geoeffective MC and ICMEs (min(Dst) $> -50$ nT ) show globally weak or non-existent responses for auroral activity, convection, and thermospheric activity. More generally, 
except for rare cases labeled C, most events associated with min(Dst)$> -30$ nT do not affect the thermosphere.

\begin{landscape}
\begin{table}
\tiny
\begin{center} 
\caption{\tiny{Characterization of the geoeffectiveness of the events associated with the SSCs, classified first by their category at L1 (MC, ICME, Misc., Shock, and CIR/SIR) and then by increasing values of the min(Dst) value. Shown from left to right: For the SSCs, No, date, time, amplitude value (A in nT), and amplitude rank (by decreasing amplitude). For the geomagnetic activity, min(Dst) value and geomagnetic storm intensity (C, S+ means strong for min(Dst)$\leq$-100 nT, S means moderate for -100$<$min(Dst) $\leq$-50 nT, and nothing otherwise), max(AE) value and AE activity intensity characterization (C, S for strong max(AE)$>$1000 nT, W for weak max(AE)$<$350 nT, nothing otherwise). For the thermosphere neutral density response (A means strong if $>2$, B means moderate if between 1.5 and 2, and C means weak if $<1.5$, - means not noticeable, and n/a not available). For the ionosphere, PCP maximum value (S for strong, {\i.e.} PCP $\geq$ 95 kV, M for moderate, W for weak, {\it i.e.} PCP $< 75 kV$). For the magnetosphere, AKR and NTC (Y for yes, - for no signature, and n/a if not available). Category of event at L1. Solar event, CMEH, CMEN,  or CMEP shown as H, N, and P, respectively. Flare class in GOES. Radio signatures marked as y if present or - if absent in the following order: A, B, interaction, type II, type IV. 
				}	}

	\begin{tabular}{lrrrr|rcrc|c|c|rr|lr|rrc}	

            \multicolumn{5}{c|}{SSC}      & \multicolumn{4}{c|} {Geomag. activity} &Th.& Iono. &\multicolumn{2}{c|}{Magneto.}   & \multicolumn{2}{c|}{L1}   & \multicolumn{3}{c}{Solar event}     \\
            No.  & Date & Time &A &  R &\multicolumn{2}{c} {min(Dst)} & \multicolumn{2}{c} {max(AE)}    &  N.     &PCP     &AKR     &NTC   &Ev. &$B_{\rm z<0}^*$    & CME   &  Flare  & Radio \\
            &         &        & [nT]  &      & [nT] &C & [nT]                                &  C     &         & max  &           &  &Type &  [nT] &              &  Class & leading  \\
            \hline 
			SSC28-SSE28&30 Sep&08:16&16&26&-176&S+&1088&S&A&S&Y&Y&MC&-6.7&N&M& y-\,-\,-\,- \\
			SSC10-SSE10&19 Apr&08:34&25&14&-149&S+&1639&S&B&S&Y&Y&MC&-6.8&H,N&M,C &yy-yy\\
			SSC09-SSE09&17 Apr&11:06&53&3&-127&S+&1356&S&B&M&Y&Y&MC&-7.4&H&M&yy-yy\\
			SSC17&23 May&10:49&78&1&-109&S+&1480&S&A&S&n/a&n/a&MC&-6.6&H,P&C,C&-yyyy\\
			SSC25&18 Aug&18:45&37&8&-106&S+&1095&S&B&S&Y&Y&MC&-3.6&H&M&yy-yy\\
			SSC24&01 Aug&23:10&28&11&-102&S+&1125&S&B&M&Y&Y&MC&-10.1&N,N& ./.; C&-\,-\,-\,-\,-\\
			SSC06&23 Mar&11:35&22&17&-100&S+&1025&S&B&S&Y&Y&MC&-5.7&H,P,H&C,M,./.&yy-yy\\
			SSC01&31 Jan&21:26&13&29&-86&S&570&&C&S&-&Y&MC&-7.8&N,H&./.,C&-y-\,-\,-\\
			SSC03&28 Feb&04:50&37&7&-71&S&1015&S&C&M&Y&Y&MC&-4.8&N&C&-yyy-\\
			SSC14&18 May&20:08&41&5&-58&S&701&&B&S&Y&n/a&MC&-3.3&H&C&yy-yy\\
			SSC23&1 Aug&05:10&17&23&-51&S&989&&C&M&Y&Y&MC&-6.2&N,N&M,M&y-\,-y-\\
			SSC04&18 Mar&13:21&61&2&-37&&692&&B&W&-&Y&MC&-2.4&H,P&C,M&yy-y-\\
			SSC27&7 Sep&16:36&23&16&-181&S+&1413&S&A&S&Y&Y&ICME&-12.1&H&M&yyyyy\\
			SSC13&11 May&10:13&26&12&-110&S+&1287&S&B&S&Y&Y&ICME&-11.7&H&./.&yy-\,-y\\
			SSC15&20 May&03:39&14&28&-36&&661&&-&M&Y&Y&ICME&-1.0&N&C,./.&-\,-\,-\,-\,-\\
			SSC21&19 Jul&10:08&19&20&-36&&972&&C&S&Y&Y&ICME&-1.7&H,H,P&C,C,X&-\,-\,-\,-\,-\\
			SSC20&17 Jul&16:02&40&6&-17&&503&&-&M&Y&Y&ICME&-1.1&H,P&X,M&y-\,-y-\\
			SSC05&20 Mar&13:27&15&27&-13&&713&&-&M&-&Y&ICME&-0.4&H,P,P&M,M&yy-y-\\
			SSC32&26 Nov&21:50&25&13&-64&S&990&&n/a&S&Y&Y&Misc.&-3.3&H&&-\,-\,-y-\\
			SSC11&23 Apr&04:48&43&4&-57&S&1136&S&B&M&Y&Y&Misc.&-2.5&H&C&yy-yy\\
			SSC26&26 Aug&11:30&21&19&-45&&1166&S&C&M&Y&Y&Misc.&-4.2&H,P&X,M&yy-yy\\
			SSC12&10 May&11:22&29&10&-14&&886&&-&M&Y&Y&Misc.&-1.3&N,H&./.,C&-\,-\,-\,-\,-\\
			SSC08&14 Apr&12:34&10&30&-23&&940&&-&S&Y&Y&Shock&-4.4&N?&C&y-\,-y-\\
			SSC18&30 May&02:04&9&31&-13&&335&W&-&W&Y&Y&Shock&-0.9&P&./.&-y-y-\\
			SSC16&21 May&22:02&18&21&-12&&681&&-&M&-&Y&Shock&-1.2&N&C&y-\,-y-\\
			SSC22&29 Jul&13:21&25&15&0&&198&W&-&W&-&-&Shock&-0.3&H&M&y-\,-y-\\
			SSC07&29 Mar&22:36&31&9&-38&&780&&C&M&-&Y&CIR&-1.2&./.&&\\
			SSC31&11 Nov&12:30&16&25&-32&&171&W&-&M&Y&-&CIR&-1.5&H&M&yy-yy\\
			SSC29-30&9 Nov&51:00&17&32&-28&&673&&n/a&M&-&Y&CIR&-2.0&N?&C&\\
			SSC02&17 Feb&02:54&21&18&-21&&331&W&-&M&Y&Y&SIR&-2.0&./.&&\\
			SSC19&6 Jun&11:39&17&22&-16&&581&&C&M&Y&n/a&CIR&-2.6&./.&&\\
		\end{tabular}
\label{table:Synthese_geoffect}
	\end{center}
\end{table}
\end{landscape}

\subsubsection{Synthesis}
\label{section:geoeff_synthesis}

For each of the 31 SSC-led events, we consider the different indices (PCN, AU, AL, ASY-H, am) for which we compute the time integral of their absolute value, from the shock arrival time until the time of the final recovery of Dst.

\begin{figure}
	\centering
	\includegraphics [angle=0, scale=0.4]{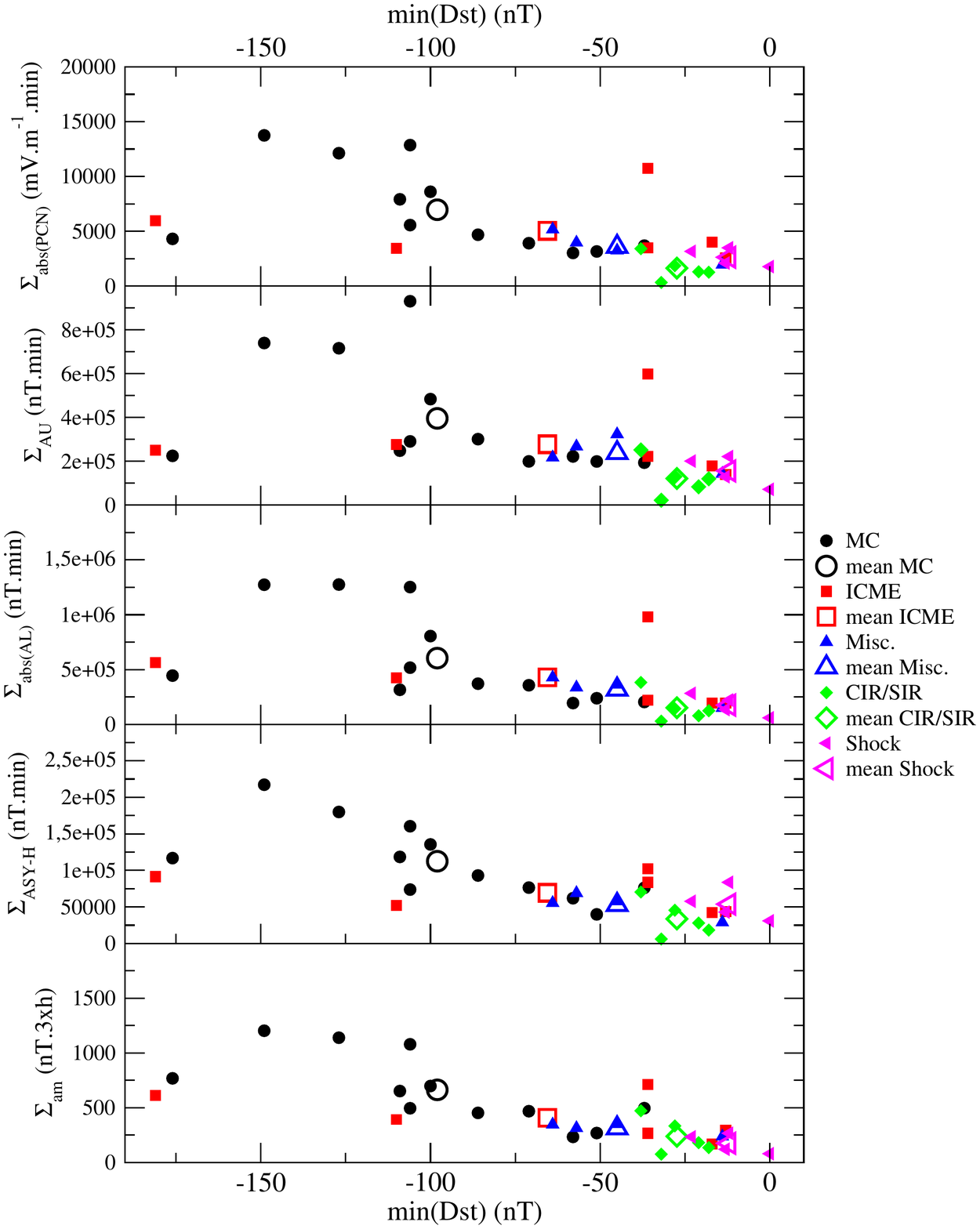}
	\caption{Minimum of Dst as a function of integrals of different indices over the event duration (from top to bottom: PCN, AU, AL, ASY-H, am).
	}
	\label{figure:indices_DST}
\end{figure}

The results, plotted with respect to min(Dst) in Figure \ref{figure:indices_DST}, not surprisingly show the lowest integrated indices for Shocks, followed by CIRs/SIRs,
confirming their very weak effect on all current systems of the magnetosphere and the ionosphere. Miscellaneous structures (\textit{e.g.}
not well defined ICME) reach somewhat higher integrated indices for values of min(Dst) approaching the geoeffectiveness level (-50 nT). For these types of events,
the integrated indices follow some kind of slope increasing with decreasing values of the min(Dst).    

The ICMEs, MCs and non-MCs, are the most geoeffective events as seen by
the largest integrated indices. The relation between their integrated
indices and the min(Dst) continues the linear trend initiated by the other
event categories but with a steeper slope around min(Dst) $\approx$ -100
nT and then a plateau down to min(Dst) $\approx$ -150 nT. This plateau may
be compared with the effect of polar-cap potential saturation observed
at its highest values. This is explained by a reduced efficiency of the
coupling between the solar wind and the Earth's environment during the
strongest events \citep[\textit{e.g. }][]{Siscoe2002}. The two largest
events (one MC, SSC28, and one ICME, SSC27) show
surprisingly smaller integrated indices, especially for high-latitude
indices (PCN, AU, AL). For the auroral indices, this may result from a geometrical effect, because of the equatorward shift of the auroral-electrojet
location the AE magnetic network may be located inside the
polar zone and may no longer capture the auroral activity with the same
efficiency. The same increasing slope with decreasing min(Dst) is also
observed for the mean in events of the same type.

Almost all of the geomagnetic storms following an SSC in 2002 are caused by ICMEs (MCs or non-MCs). Only two moderate storm events are caused by unclear solar-wind signatures. Geoeffective ICME (MCs and non-MCs) events (min(Dst) $\leq -50$ nT) show globally similar responses, characterized by:
\begin{itemize}
	\item {strong $B_{\rm z<0}^*$ in the preceding sheath and/or in the ICME itself}
	\item {lower mean $\beta$ and more moderate mean Mach number than in the usual solar wind in the sheath and the ICME itself}
	\item {stronger SSC amplitude than for the overall SSC-led events caused by a stronger magnetosphere compression}
	\item {stronger geomagnetic activity than for the overall SSC-led events probably induced by the triggering of substorms and particle accelerations in the magnetosphere}
	\item {emission of both NTC and AKR radio-waves indicating a magnetic activity covering a large sector of  the magnetospheric plasmasheet.}
	\item {larger auroral activity and ionospheric convection than for the overall SSC-led events.}
	\item {stronger thermospheric response than for the overall SSC-led events in particular on the nightside and in good correlation with geomagnetic storms.}
\end{itemize}

Finally, for 2002 none of the studied storms are only caused by an isolated Shock or CIR/SIR.
These solar events may affect the magnetosphere but the wave activity essentially shows local effects rather than a global perturbation. They have almost no impact on the thermosphere.
As already observed in the past, Shock storms are not geoeffective, . They are in general characterized by a low jump in solar-wind velocity and low maximum of solar-wind pressure reached across the shock.
If CIRs/SIRs have been proposed as possible drivers of geomagnetic
storms in the past, they often give a weaker response of the
geospace environment and are very infrequently associated to an SSC
\citep{Borovsky2006}. Moreover, CIR-driven storms have been found
to be more geoeffective during the declining phase of the solar
cycle \citep{Echer2008} than during solar maximum. For
high solar activity, the magnetosphere--ionosphere system is almost
continuously buffeted by ICMEs (two to three {\it per} month) and is consequently in
a state where CIR/SIR action could be largely attenuated,
thus inhibiting the storm response. In 2002, only four intense geomagnetic
storms were caused by CIRs/SIRs and none of them were preceded by an SSC
\citep{Echer2008}. The only important result for the five CIR/SIR-related
events of our study is that globally the geospace response is slightly
stronger if the jump in solar-wind velocity and the maximum of solar-wind
pressure reached across the SIR are stronger.

\subsection{Energy Input and Coupling Functions} \label{section:coupling_function}

Estimating the forcing of a given solar-wind disturbance on the Earth's magnetosphere--ionosphere coupled system has always been a challenge. First, it evidently depends on the power that the solar wind transfers to the magnetosphere. Also, it depends significantly on the parameter, or magnetospheric state variable (magnetic index, location of the polar cusp, size of the polar cap, \textit{etc.}), that one considers to estimate its response.

The power that the solar wind provides to the Earth's magnetosphere has been estimated by the so-called coupling functions since the late 1970s.
These are simple functional forms that take into account the solar-wind and/or IMF parameters. Most of them consist of an expression of the Poynting flux ($vB^2$: with $v$ the solar-wind velocity and $B$, the magnitude of the IMF) under various forms and of a clock-angle dependence to account for the state of coupling between the
IMF and the Earth's magnetic field. \\

\begin{figure}
	\centering
		\includegraphics [angle=0, scale=0.6]{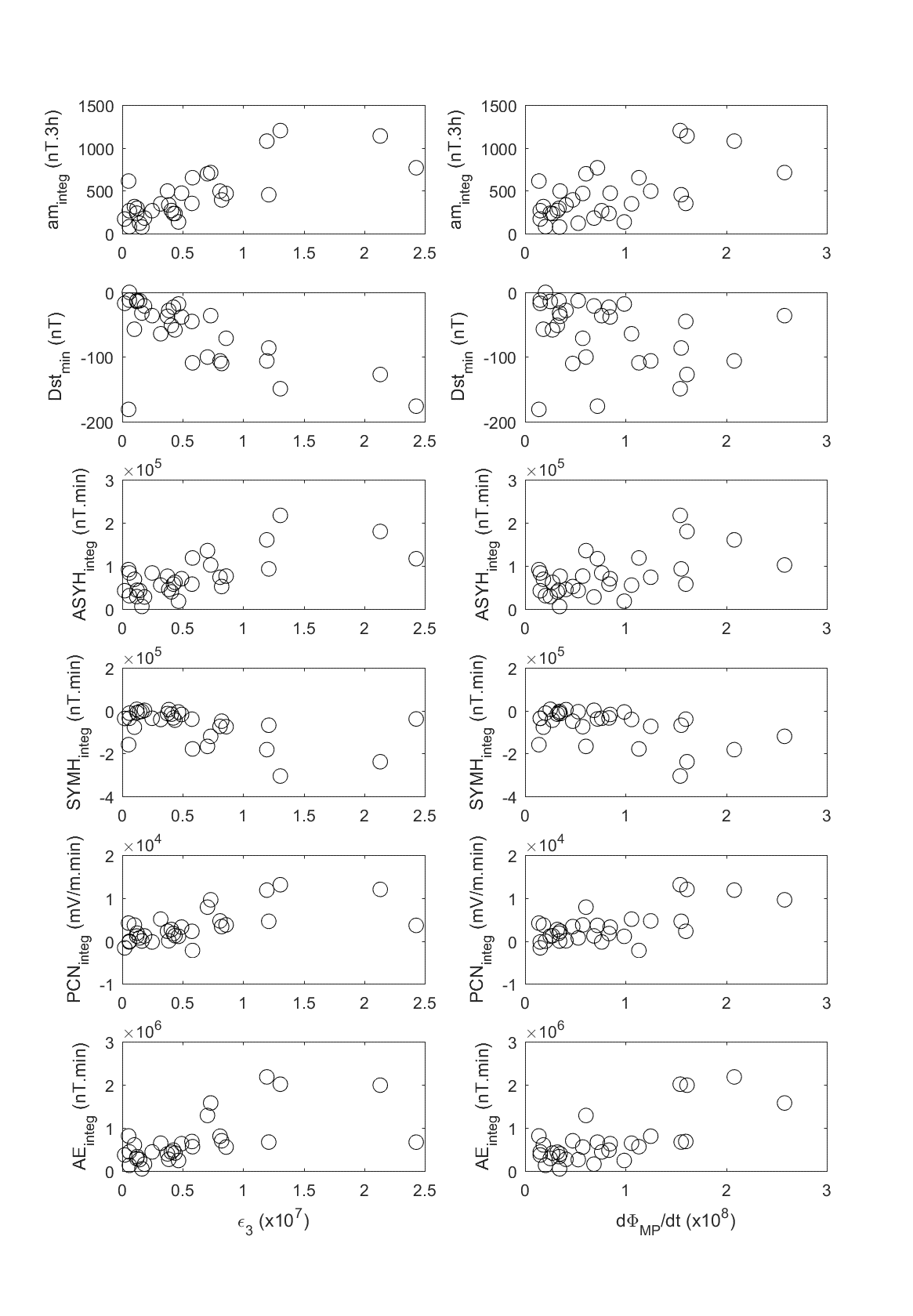}
		\caption{Integrated values of various magnetic indices as functions of integrated values of  the coupling functions $\epsilon_3$ (on the left) and
		$\frac{\textrm{d}\Phi_{MP}}{\textrm{d}t}$ (on the right). From top to bottom: am, Dst, ASY-H, SYM-H, PCN, AE. For the case of the Dst,
		its minimum value is plotted. }
	\label{figure:indices_epsilon_defisurdt}
\end{figure}

In this study, we chose two of those coupling functions. The Akasofu parameter ($\epsilon$) \citep{Perreault1978} is one
of the best known and most used coupling functions. It comes in three variants. It is the latest variant ($\epsilon_3$) that
we consider in this work as it is the variant that gives the best correlations with the magnetospheric state variables
\citep[\textit{e.g.}]{Newell2007}, $\epsilon_3 = vB\sin^4 \left(\frac{\theta_c}{2}\right)$. In their study, \citet{Newell2007} also proposed a new
coupling function [$\frac{\textrm{d}\Phi_{MP}}{\textrm{d}t} = v^{4/3}B_T^{2/3}\sin^{8/3}(\frac{\theta_c}{2})$], which is supposedly better in the sense that it correlates better
with magnetospheric state variables. According to these authors, this coupling function quantifies the rate of reconnection at the
day-side magnetopause and has therefore a better physical meaning. In these relations, $v$ and $B$ are the velocity and the magnetic field amplitudes of the solar wind,
${\theta_c}$ and $B_T$ are the IMF clock angle and transverse component, respectively.

For each of the SSC-led events, we have thus:
i) Calculated the corresponding maximum value and integrated value of $\epsilon_3$ and  $\frac{\textrm{d}\Phi_{MP}}{\textrm{d}t}$ over the whole ICME interval,
ii) Extracted the extrema values of the magnetic indices as well as their integrated values over the whole temporal interval.

We show in Figure \ref{figure:indices_epsilon_defisurdt} our results for the integrated values of a selection of magnetic indices as functions of integrated values of $\epsilon_3$ and  $\frac{\textrm{d}\Phi_{MP}}{\textrm{d}t}$. The expected trends are found, the larger the coupling function, the stronger the magnetospheric response. Those coupling functions seem to be a suitable indicator for forecasting or nowcasting the strength of the magnetospheric response to a given solar-wind discontinuity, on a statistical scale at least.

\begin{table}[ht]
\begin{center}
\caption{Correlation coefficients between integrated values (represented with the symbol $\Sigma$) of magnetic indices and coupling functions (for the Dst, the min(Dst) value is taken). The first column shows the correlation coefficients for all events, the second column for events identified as ICMEs or MCs.}
\label{table:correl_coupling_functions}
\begin{tabular}{lccccccccccc}
\hline
                         & & & All events & MC, ICME&&&&& \\
\hline
 $\Sigma_\textrm{am}$ /  $\Sigma_{\epsilon}$ & &&  0.75  & 0.68 &&&&&\\
 min(Dst) /$\Sigma_{\epsilon}$ & && -0.65 & -0.62&&&&& \\
 $\Sigma_\textrm{ASYH}$ / $\Sigma_{\epsilon}$ &&   &0.68 &  0.63 &&&&&\\
 $\Sigma_\textrm{SYMH}$ / $\Sigma_{\epsilon}$ && & -0.52 & -0.39&&&&& \\
 $\Sigma_\textrm{PCN}$ /$\Sigma_{\epsilon}$ & & & 0.63  & 0.56 &&&&&\\
$\Sigma_\textrm{AE}$ / $\Sigma_{\epsilon}$ & &  &0.63  & 0.53&&&&& \\
\hline
$\Sigma_\textrm{am}$ / $\Sigma_\textrm{Newell}$ & &  &0.62 &  0.66 &&&&&\\
min(Dst) /  $\Sigma_\textrm{Newell}$ & & &-0.20  &-0.09 &&&&&\\
$\Sigma_\textrm{ASYH}$ /  $\Sigma_\textrm{Newell}$ & & & 0.55 &  0.61 &&&&&\\
$\Sigma_\textrm{SYMH}$ / $\Sigma_\textrm{Newell}$ && & -0.52  &-0.54 &&&&&\\
$\Sigma_\textrm{PCN}$ /  $\Sigma_\textrm{Newell}$ & & & 0.67  & 0.68 &&&&&\\
$\Sigma_\textrm{AE}$ /  $\Sigma_\textrm{Newell}$ &  & &0.72  & 0.75&&&&& \\
\hline
\end{tabular}
\end{center}
\end{table}

We then compare in Table \ref{table:correl_coupling_functions} the correlation coefficients between the energy input (coupling functions)
and the magnetospheric response (magnetic indices).
For each pair of magnetic index and coupling function, the first column of Table \ref{table:correl_coupling_functions}  shows the correlation coefficient for all events, and the second only for those events identified as MCs or ICMEs. All of the results show relatively low correlation coefficients, significantly lower than those reported by previous studies \citep[\textit{e.g. }][]{Newell2007}. There are two reasons for this. The first and obvious one is the low number of events in our sample, implying that a few outliers are enough to degrade the correlation.
The second reason, for the integrated values, is that we integrate over
the duration of our events at L1 for solar-wind $\epsilon$-related
quantities, and at Earth for geomagnetic indices. Although the start
time is often well defined, the end time is often more problematic to
establish. As a consequence, for a given event, the integration interval may be
different at L1 and at Earth. 

One can also notice that if only MCs and ICMEs are taken into account (last column of Table \ref{table:correl_coupling_functions}), most of the correlations involving  $\epsilon_3$ are degraded while most of the correlations with $\frac{\textrm{d}\Phi_{MP}}{\textrm{d}t}$ are slightly improved, but the one for min(Dst).
It is intriguing the very weak correlation coefficients between  min(Dst)-index and $\frac{\textrm{d}\Phi_{MP}}{\textrm{d}t}$.
We note that have no clear interpretation for the results shown in Table \ref{table:correl_coupling_functions}. 

\section{Summary}
\label{section:SummaryResults}
In this article, we focus on the year 2002, a period of maximum solar activity. We propose a novel, multidisciplinary, and statistical approach to the complete chain of processes from the Sun to the Earth (Sun, L1, magnetosphere, ionosphere, thermosphere) in order to study the geoeffectiveness of solar events. In contrast to previous statistical or case
studies, the starting point is neither the coronal mass ejection (CME) emission at the Sun, nor the value of the min(Dst) index used to evaluate the intensity of the geomagnetic storm, but the storm sudden commencements (SSCs): near-Earth signatures produced by shocks impinging on the magnetosphere and followed by geomagnetic activity. 
This study then aims first to associate an SSC with a possible source at the Sun, and then to characterize the propagation of the solar event along the entire chain from the Sun to the Earth. It exploits existing catalogs and observations of SSCs,
solar activity, solar wind at L1, magnetosphere/magnetopause, and the coupled ionosphere/thermosphere system.\\
We start with the Observatori de l'Ebre/ISGI list of 32 SSCs detected during 2002.  The 32 SSCs are linked to 31 perturbations observed at L1 due to the fact that two SSCs are associated with the same event.  These two SSCs are considered as a single event in our statistics. We considered four criteria which enable us to associate 28 SSCs with 44 CMEs from the larger list of possible CMEs with a visible source on the Sun.

\subsection{Relationships between SSCs and the CMEs at their Origin} 
The statistical analysis of the 44 CMEs, including 21 halo CMEs possibly responsible for the 28 SSCs led  to the following results:
\begin{itemize}
\item We confirm that the solar sources of these 44 CMEs are active regions (ARs) mainly located at the central part of the disk, some of which (60\,\%) have filaments. The presence of a filament indicates that the magnetic field of the region is strongly sheared which is a good indicator of destabilization and eruption. CMEs are associated with small, medium, or large X-ray flares with no preference. The relationship between extreme solar events  (large sunspot areas, filaments, large free magnetic energy) and extreme geoeffective events observed on the Earth is not obvious \citep{Schmieder17}.  This can be the reason of the difficulty to forecast whether or not a  solar  active region will be  the source of  a geoeffective event.   
\item In 54\,\% of the cases (15/28 SSCs) a single CME is related to one SSC and in 46\,\% of the cases two CMEs at least are related to one SSC. Twenty-one are halo CMEs, \textit{i.e.} 75\,\% of the 28 halo CMEs with a visible source on the Sun in 2002 originated an SSC in the Earth environment, most of the time the faster are the more geoeffective. 
According to the CDAW list, more than 500 front-side non-halo CMEs were recorded in 2002 (1.5 per day on average). Half of them are front-sided and only 23 (5\%) could be associated with an SSC (see Table \ref{table:CME_retenues_proprietes}). \\

	\item Radio observations allowed us to classify the events at the Sun in three categories, the largest group includes events displaying type IV radio emissions. 
	The presence of the type IV burst component that we call B (a long-duration radio continuum detected in a frequency range typically from decimetric to decametric wavelengths), which is physically linked with the development of the CME current sheet, is statistically the most important factor for SSC prediction.
A second group assembles the four events related to Shocks at L1. Taking only the B-components would have led to predict 85\,\% of the SSC-led events with a minimum value of Dst less than -30 nT, in the present study. 	 \textit{Wind}/WAVES observed only four type IV radio emissions in 2002; those correspond to CME--SSC associations that are related to an MC.
Generally, 25/31 events are associated with a type II event, which is indicative of electrons accelerated by a shock.
\end{itemize}

Our analysis also underlines the importance of joint white-light and  multi-wavelength-radio observations, in  particular of the radio imagery, for revealing and explaining the complex interactions between different CMEs or between a CME and the ambient medium. 
These interactions take place most frequently in the corona, \textit{i.e.} in the field of view of LASCO onboard SOHO. They may lead to a modification of the CME structure and/or of their trajectories, so that both have to be followed step by step. 

\subsection{Propagation between the Sun and L1}
The uncertainty in the arrival time prediction at L1 results mainly from the acceleration/deceleration of CMEs in the ambient interplanetary medium.
As a first approach to associate L1 events with their solar source, we start with the ballistic model and a time window to account for propagation uncertainties (Section \ref{ballistic}). 
In addition, we also use the drag-based-model (DBM) and calculate the drag coefficient (Section \ref{drag}). 85\,\% of the Leading CMEs show a DBM coefficient $0.11 \times 10^{-7}< \gamma <  2.2 \times 10^{-7}$ km$^{-1}$,
included in the range predicted by \cite{Vrsnak2013} (between $0.1 \times 10^{-7}$ and $100 \times 10^{-7}$ km$^{-1}$). 
Both mean and median values are not far from the commonly used value in the models of $0.2 \times 10^{-7}$km$^{-1}$, which validates the resulting association between most L1 events and their solar sources.

In Section \ref{prop}, we calculate the propagation delay from the Sun to L1  using different simple propagation models \citep{Huttunen2005,  Vrsnak2013, Schwenn2005} for shock propagation (25 events) and for the ICME and MC propagation (18 events). We compare the results with our observations. There are roughly as many negative as positive delays. Half of them arrive in $\pm$ 14 hours, which is considered as the uncertainty of the models. These statistics are not improved by restricting our event set to halo CMEs or to isolated CMEs. 

The results demonstrate the need for a reliable propagation tool to properly relate an ICME observed at L1 and a CME detected within the five days preceding its arrival at L1.

\subsection{Relationships between SSCs, Signatures at L1 and CMEs}

We characterize the perturbation in the solar wind observed at L1 associated with each SSC. 
The 31 L1 events are sorted as follows: 12 MCs, 6 (non-MC) ICMEs, 4 Misc., 5 CIRs/SIRs, and 4 isolated Shocks (see Section \ref {section:relatedSunEarthEvents_methodology}).

In order to further investigate the link between signatures at L1 and CMEs  that could be associated to SSCs, we consider four criteria: \\
i) propagation considerations based on the ballistic model,
ii) estimations from the drag-based model (DBM),
iii) radio emissions as signatures of acceleration processes linked to solar sources, and 
iv) the compatibility of the flux rope chirality observed at L1 with the location of the solar source. The last criterion only applies to MCs.

Most shocks observed in 2002 at L1, either isolated ones (Shocks) or part of other events, cause an SSC. This is the case for 80\,\% of the 35 IP shocks listed by \cite{Gopalswamy2010c}. Conversely, only 5 out of 41 CIRs/SIRs reported by \cite{Jian2006a} in 2002 were associated with SSCs, and for three of them there are no CME candidates.

For 28 SSC-led events observed at L1, a plausible solar source is identified. Concerning MCs and ICMEs, different catalogs and studies exist. We identified 11 MCs and 10 non-MCs ICMEs listed by respectively three or more and two or more studies. All 11 MCs (100 \%) and six ICMEs (60 \%) caused SSCs. There is no obvious correlation between the solar source properties and the  L1 categories.

Finally, we emphasize that 14 of the 28 associations mentioned above fulfill all the relevant criteria for the considered category (\textit{i.e.}, four criteria for the MCs, three for the other categories). In particular, the criterion based on the ballistic velocity is not satisfied in seven cases. These mismatching cases result not only from the complexity of the ICME
velocity evolution during its propagation (interaction with the ambient solar wind and/or with other ICME) but also from the lack of direct observation of the radial velocity along the Sun--Earth direction.\\

\subsection{Geoeffectiveness}

The geoeffectiveness is first discussed as a function of the minimum Dst values, generally used as an indicator of the storm strength
(intense, $-200$ nT $<$ min(Dst) $ \leq -100$ nT, moderate, $-100$ nT $<$ min(Dst) $ \leq -50$ nT).
The analysis of the SSC-related events in 2002 shows that:
\begin{itemize}
	\item If the CME velocity ($V_{\odot}$) is larger than 1000 km\,s$^{-1}$ there is a greater probability to trigger a moderate or intense storm.  No particular rule is found with the nature of the  CME source (halo or not, single or multiple, flare class), but the most geoeffective events are associated with type IV radio bursts.

	\item The most efficient storm drivers are MCs, followed by ICMEs: 11 out of the 12 (92\,\%) MCs cause storms (seven intense and four moderate).  The two other intense storms that follow an SSC were caused by ICMEs.
	\item The three most geoeffective MCs induce a sudden secondary event (SSE) with a magnetic signature similar to an SSC.
	\item Among the six moderate storms that follow an SSC, four are due to MCs and two to the so-called miscellaneous events. Our statistics differ from those of \citet{Echer13} who reported that the interplanetary structure (CIRs and pure high-speed streams) are responsible for 30\,\% of these storms, ICMEs being the second major cause. Our study shows that when the storm is preceded by an SSC, the main driver of a moderate storm is in general an ICME (including MCs, non-MCs, and Misc.). 
	\item The presence of a southward IMF component and its duration are generally considered as favorable conditions for geomagnetic activity. In order to account for it, we computed a normalized and time-integrated parameter ($B_{\rm z<0}^*$) and we find that this is a good indicator of the potential geoeffectiveness of a solar-wind structure.
	\item For geoeffective ICMEs (11 MCs and two non-MCs) triggering intense and moderate storms, we separate the effects of their sheath and central core. These statistics are limited, but in general, the central core is responsible for the minimum values of the Dst. We note that for events related to intense storms, the sheath plays an important role in about half of the events and that this role becomes dominant for three (33 \%) events since they are the ones with the lowest min(Dst) values. The sheaths causing intense storms show small variations and low values of the $\beta$  parameter  (0.4 $<$  $\beta$ $<$ 0.75) and of the Alfven Mach number (3.6 $< M_A <$ 6.5).

	\item Five SSCs of our study are related to CIRs/SIRs. CIRs/SIRs are possible drivers of weak geomagnetic storms (-38 nT $<$ min(Dst) $<$ -16 nT), not considered as geoffective, as already noted \citep{Borovsky2006}.
	The analysis of these events shows that the geospace response tends to be slightly stronger if the jump in solar-wind velocity and the maximum of solar-wind pressure reached across the stream interface, are stronger. As already observed in the past, Shocks are not geoeffective (-25 nT $<$ min(Dst) $<$ 0 nT).
\end{itemize}

The analysis of the perturbations at L1 and their associated geomagnetic response enabled us to estimate the power that the solar wind provided to the Earth's magnetosphere by means of two coupling functions. We find that their correlation with the different magnetic indices remains relatively weak. The function proposed by \citet{Newell2007} ($\frac{\textrm{d}\Phi_{MP}}{\textrm{d}t}$), might better account for the effect of the magnetic field within the discontinuity that hits the Earth's environment. 
The Akasofu parameter \citep{Perreault1978}, ($\epsilon_3$), correlates with mid-latitude and global indices better than does $\frac{\textrm{d}\Phi_{MP}}{\textrm{d}t}$, whereas the latter correlates best with auroral indices.

Finally, the response of the magnetosphere--ionosphere--thermosphere system is expectedly enhanced with the geomagnetic activity level. Among them, we emphasize the following issues:
\begin{itemize}
	\item {A combined NTC and AKR wave activity develops in the magnetosphere during ICMEs (MCs, non-MCs and Misc.), suggesting the presence of acceleration processes over a large sector of the plasma sheet. Conversely, this effect appears more local for most CIRs/SIRs or Shocks with the enhancement of only one of these emissions.}
	\item All events associated with strong and moderate storms induce a thermospheric storm, mostly identified by a significant enhancement of the night time neutral density. This occurs during most ICMEs. Conversely, CIRs/SIRs and Shocks have almost no impact on the thermosphere.
\end{itemize}

\subsection{Concluding Remarks}
\label{section:ConcludingRemarks}

Major breakthroughs in our understanding of how the Sun creates and controls the heliosphere will be provided by
\textit{Solar Orbiter} \citep{Muller2013}, scheduled to be launched in 2019.  Combining remote-sensing and \textit{in-situ} observations in the inner heliosphere as close as 0.28 AU to the Sun, {\it Solar Orbiter} will provide the data to determine
the properties of CMEs and will establish how they expand and rotate into the inner heliosphere. It will relate the properties of an ICME to those of the CME and provide better insight into the geoeffective potential of the event.
Furthermore, the NASA  \textit{Parker Solar Probe} mission \citep{Fox2016}, expected to be launched in 2018, will measure ICMEs \textit{in situ} down to 10 solar radii. Joint observations from both missions will enable
tracking the evolution of an ICME throughout the inner heliosphere and better constrain the current expansion and propagation models.

\section*{Acknowledgments}
The results presented rely on geomagnetic indices calculated and made available by ISGI Collaborating Institutes from data collected
at magnetic observatories. We thank the national institutes involved, the INTERMAGNET network and ISGI (\urlurl{isgi.unistra.fr}).
The authors would like also to acknowledge the CDPP-Plasma Physics Data Centre (\urlurl{cdpp.eu/}) and MEDOC for SOHO
data (\urlurl{medoc.ias.u-psud.fr}). EIT movies can be found at \urlurl{www.ias.u-psud.fr/eit/movies/} and the list of CME at \urlurl{cdaw.gsfc.nasa.gov/}.
The \textit{Cluster} data used are on the Cluster Science Data System (CSDS) Web site (\urlurl{www.cluster.rl.ac.uk/csdsweb/}). H. Kojima from RISH,
Kyoto University, is thanked for making available {\it Geotail Plasma Wave Instrument} (PWI) dynamic spectra (\urlurl{space.rish.kyoto-u.ac.jp/gtlpwi/}).
We thank the ACE/SWEPAM and the ACE/MAG instrument teams, and the ACE Science Center for providing the ACE data.
SOHO is a project of international collaboration between ESA and NASA. 
The SOHO/LASCO data used here are produced by a consortium of the
Naval Research Laboratory (USA), Max-Planck-Institut f\"ur Aeronomie
(Germany), Laboratoire d'Astronomie Spatiale (France), and the University
of Birmingham (UK). \textit{Wind}/WAVES radio products and plots are
provided by  the National Aeronautics and Space Administration (GSFC). The
{\it Nan\c{c}ay Radioheliograph} (NRH) and {\it Decameter arrays} (DAM) are operated
by the Paris  Observatory; the data  are accessible through the “Radio
Monitoring site” (\urlurl{secchirh.obspm.fr}) and DAM data by request
through the Nan\c{c}ay site.  The {\it Radio Solar Telescope Network} (RSTN)
is operated by the U.S. Air Force Weather Agency. We acknowledge the
access to the radio data archives from several sites managed by solar
observatories: ARTEMIS (Thermopylae, Athens University), Hiraiso (Japan),
Nobeyama (Japan), Ondrejov (Czech Republic), ETH Zurich Radioastronomy
(Switzerland).  We thank G. Mann and J. Rendtel for the Potsdam radio
spectra. Operation in 2002 of the northern SuperDARN radars was supported
by the national funding agencies of the United States, Canada, the United
Kingdom, France, and Japan. M. Pick thanks A. Hamini, R. Romagnan, and
M.P. Issartel for their help in  the data analysis  and A. Lecacheux
for fruitful discussions. N. Cornilleau-Wehrlin thanks P. Canu and O. Le
Contel for fruitful discussions. 
B. Grison acknowledges support of GACR grant No 18-05285S, of the Praemium Academiae Award, and of the Europlanet 2020 research infrastructure. Europlanet 2020 RI has received funding from the European Union’s Horizon 2020 research and innovation program under grant agreement No 654208.
The authors thank the referee for their
many helpful suggestions allowing us
to improve this long article. The authors thank the Programme National
Soleil-Terre. Finally, the authors want to express their warmest thanks
to C. Lathuill\`ere and N. Vilmer for their important collaboration at
the beginning of this work.

\section *{Disclosure of Potential Conflicts of Interest}
The authors declare that they have no conflicts of interest.

\appendix

 \section{Event Timing and Solar-Wind Observations}
\label{section:App_SolWindObs}

We look into solar-wind data a few hours before each SSC in order to find the cause of the compression.
Taking into account the propagation time, we were able to find an increase in the density and/or velocity that causes the SSC.
We refer to these jumps as  discontinuities of solar-wind perturbations. The discontinuity times at L1 are gathered in Table \ref{table:Observations_at_L1} (ACE data),
column $t_{\rm d}$; they usually happen from 30 to 60 minutes prior to the SSC. Then we identify later on if the perturbations following these shocks are ICMEs or other perturbation.
$t_{\rm s}$ and $t_{\rm e}$ when different from $t_{\rm d}$ indicate the start and end times of the perturbation. The $[t_{\rm s};t_{\rm e}]$ interval marks the sheath of the solar-wind perturbation, when observed.

To look at the geoeffectiveness of the solar-wind structures, we used OMNIWEB data to transport data from L1 to the bow-shock nose.  Although the bow-shock crossing can modify SW properties, these data do not contain the exact cause
of the magnetosphere compression, nevertheless one can reasonably believe that an increase of the SW density or velocity upstream of the bow shock is still present downstream, even with a different intensity.\\
In the interplanetary medium, ICMEs are generally first identified by a shock (also called {leading shock}) in velocity, density, and magnetic field, then followed by a highly fluctuating region, the so-called ICME sheath.
MCs appear in the central part of ICMEs with a magnetic structure that is well defined and resembles a flux rope with a slowly rotating magnetic field and low temperatures. 
We thus consider the following dataset at L1 (\textit{cf.} also Figure \ref{figure:Figure_l1_ssc09}): the date, density, temperature, and bulk flow velocity of the solar wind,  the solar-wind pressure, and the IMF intensity and orientation. The orientation is defined by two angles ($\theta_{\rm IMF}$ and $\phi_{\rm IMF}$). $\theta_{\rm IMF}$ is defined by the deviation of the IMF from the $(X,Y)_{\rm GSM}$ plane
	($\theta_{\rm IMF}=90^\circ$ is aligned with $Z_{\rm GSM}$); $\phi_{\rm IMF}$ is the azimuth in the $(X,Y)_{\rm GSM}$ plane ($\phi_{\rm IMF}=0^\circ$ is parallel to the Earth--Sun axis),

By coupling these measurements with several lists describing the properties of the different solar-wind processes, it is possible to identify the cause of the terrestrial magnetosphere response.

At L1 we sort the SSC-related events into five categories: MC, ICME, Misc., Shock, and CIR/SIR (see Section \ref{section:L1_characterization}). Table \ref{table:SW_ICME} gathers ICME and MC properties observed at L1.
We compare parameter values in the sheath to the values in the cloud itself.  
Similar parameters are presented for the Misc. and Shocks in Table \ref{table:SW_others}, and for CIR/SIR events in Table \ref{table:SW_CIR}. In the latter case, we compare properties before and after SI.
Many results presented in Section \ref{section:Geoeffectiveness} are based on these tables.

\begin{table}[htb]
	\tiny
	\begin{center}
		\caption{Observations at L1 for each SSC in 2002.  Events are classified in five categories: ICME with (MC) or without (ICME) magnetic clouds, streaming and co-rotating interaction regions (CIR/SIR), shocks (Shock), and miscellaneous (Misc.) when we cannot decide. $t_{\rm d}$ is the discontinuity time that causes later on the SSC. For each MC and ICME, $t_{\rm s}$ and $t_{\rm e}$ are the start and end times of the ICME. Velocities are given: before the discontinuity ($V_{\rm up}$), at the discontinuity ($V_{\rm d}$), at $t_{\rm s}$ ($V_s$) and at $t_{\rm e}$ ($V_{\rm e}$). $B{^*}_{\rm z<0}$ is a normalized value of the negative part of the IMF z-component (see Section \ref{section:Geoeffectiveness}). We provide reference to catalogs that list the events. The references to these catalogs have been numbered as 1, 2, 3, 4, 5, 6, 7, and 8, and the numbers correspond to \citet{Huttunen2005}, \citet{Jian2006a}, \citet{Jian2006b}, \citet{Lepping2006}, \citet{Li2011}, \citet{Richardson2010}, \citet{Zhang07}, and \citet{Gopalswamy2010c}, respectively.
		}
		\begin{tabular}{l l   c   c  c c c c c c l    }
			\multicolumn{2}{c}{SSC}& {$t_{\rm d}$} & {$t_{\rm s}$}& {$t_{\rm e}$} & $V_{\rm up}$&  $V_{\rm d}$ &  $V_{\rm s}$ &  $V_{\rm e}$  & $B{^*}_{\rm z<0}$ & Litt.     \\
			No.	& Cat. & date [hh:mm] & date [hh] & date [hh] &    \multicolumn{4}{c}{[km\,s]$^{-1}$} & [nT] &     \\\hline
			
			SSC01 & MC & 31 Jan  20:36 & 02 Feb  02 & 02 Feb  16 & 300 & 360 & 360 & 400 & -8.5 & 5,8 \\
			SSC02 & SIR & 17 Feb  02:08 &  &  & 330 & 400 &  &  & -3.9 & 2,8 \\
			SSC03 & MC & 28 Feb  04:00 & 28 Feb  17 & 02 Mar  00 & 300 & 380 & 390 & 390 & -4.8 & 1,3,6,8 \\
			SSC04 & MC & 18 Mar  12:35 & 19 Mar  05 & 20 Mar  16 & 300 & 460 & 420 & 450 & -2.4 & 1,3,4,5,6,8 \\
			SSC05 & ICME & 20 Mar  13:05 & 21 Mar  14 & 22 Mar  06 & 400 & 580 & 440 & 420 & -0.4 & 3,6,8 \\
			SSC06 & MC & 23 Mar  10:51 & 24 Mar  12 & 25 Mar  20 & 400 & 480 & 440 & 500 & -5.7 & 1,3,4,5,6,7,8 \\
			SSC07 & CIR & 29 Mar  21:40 &  &  & 310 & 390 &  &  & -1.1 & 2 \\
			SSC08 & Shock & 14 Apr  11:47 &  &  & 390 & 420 &  &  & -4.4 & 3 \\
			SSC09 & MC & 17 Apr  10:20 & 18 Apr  00 & 19 Apr  08 & 340 & 550 & 510 & 600 & -9.7 & 1,3,4,5,6,7,8 \\
			SSC10 & MC & 19 Apr  08:01 & 20 Apr  08 & 21 Apr  18 & 430 & 690 & 600 & 450 & -7.2 & 1,3,4,6,7,8 \\
			SSC11 & Misc. & 23 Apr  04:14 &  &  & 430 & 600 &  &  & -4.6 & 3,8 \\
			SSC12 & Misc. & 10 May  10:29 &  &  & 350 & 400 &  &  & -1.3 & 3,8 \\
			SSC13 & ICME & 11 May  09:24 & 11 May  14 & 12 May  00 & 350 & 450 & 440 & 480 & -11.7 & 2,3,6,7,8 \\
			SSC14 & MC & 18 May  19:18 & 19 May  04 & 20 May  03 & 350 & 480 & 450 & 475 & -3.3 & 1,3,4,5,8 \\
			SSC15 & ICME & 20 May  02:56 & 20 May  10 & 21 May  22 & 430 & 520 & 470 & 380 & -1.2 & 3,6,8 \\
			SSC16 & Shock & 21 May  20:58 &  &  & 360 & 400 &  &  & -1.2 & 8 \\
			SSC17 & MC & 23 May  10:14 & 23 May  20 & 25 May  18 & 430 & 800 & 800 & 450 & -6.6 & 1,3,4,6,7 \\
			SSC18 & Shock & 30 May  01:32 &  &  & 460 & 510 &  &  & -0.9 & 8 \\
			SSC19 & CIR & 08 Jun  10:28 &  &  &  300 & 340 &  &  & -2.6 & 2 \\
			SSC20 & ICME & 17 Jul  15:24 & 18 Jul  12 & 19 Jul  09 & 420 & 510 & 450 & 450 & -1.1 & 3,6,8 \\
			SSC21 & ICME & 19 Jul  09:30 & 20 Jul  04 & 22 Jul  06 & 460 & 550 & 900 & 500 & -1.6 & 3,6,8 \\
			SSC22 & Shock & 29 Jul  12:40 &  &  & 390 & 500 &  &  & -0.3 & 3,8 \\
			SSC23 & MC & 01 Aug  04:22 & 01 Aug  09 & 01 Aug  23 & 375 & 500 & 430 & 480 & -6.2 & 3,4,6,7,8 \\
			SSC24 & MC & 01 Aug  22:19 & 02 Aug  04 & 04 Aug  01 & 440 & 460 & 510 & 425 & -10.1 & 1,3,4,5,6,8 \\
			SSC25 & MC & 18 Aug  18:09 & 19 Aug  16 & 21 Aug  14 & 410 & 580 & 510 & 440 & -3.2 & 3,5,6,7,8 \\
			SSC26 & Misc. & 26 Aug  10:45 &  &  & 310 & 400 &  &  & -4.2 & 3,8 \\
			SSC27 & ICME & 07 Sep  16:08 & 08 Sep  04 & 08 Sep  20 & 390 & 560 & 470 & 440 & -12.1 & 3,6,7,8 \\
			SSC28 & MC & 30 Sep  07:20 & 30 Sep  21 & 01 Oct  15 & 300 & 340 & 370 & 380 & -7.1 & 1,2,3,4,5,6,8 \\
			SSC29 & CIR & 09 Nov  16:58 &  &  & 340 & 360 &  &  & -2.0 & 2 \\
			SSC30 & idem & 09 Nov  17:51 &  &  &  & 390 &  &  & -2.2 & 8 \\
			SSC31 & CIR & 11 Nov  11:50 &  &  & 440 & 560 &  &  & -1.5 & 2 \\
			SSC32 & Misc. & 26 Nov  21:08 &  &  & 380 & 580 &  &  & -3.3 & 3,8 \\
		\end{tabular}
		\label{table:Observations_at_L1}
	\end{center}
\end{table}

\begin{landscape}
	\begin{table}[htb]
		\begin{center}
			\caption{Solar-wind properties in the sheath and in the ICME (MC and non-MC) itself. Each ICME is presented using its negative peak value $B_{\rm pz}$.
			For the sheath and the ICME itself, we list the following: total (sheath or ICME) duration, duration between structure start time and time of the minimum value of Dst, Dst minimum value inside the structure,  mean $\beta$, mean $M_{A}$, normalized and integrated $B_{\rm z<0}^*$ value until the time of min(Dst) (n/a means no negative $z$-component of the IMF during the period for which $B_{\rm z<0}^*$ is calculated), field rotation (e.g. SEN stands for south--east--north and indicates the sequence of the field direction; small letters indicate a small component) and chirality (in the case of MC).
			We considered the OMNI data set propagated to the bow-shock nose location.}
			\begin{tabular}{l c   c c cccc| c c ccc c  c c c  }
				SSC & $B_{\rm pz}$  & $\Delta t$ & $\Delta t_{\rm Dst}$ &  Dst & $\beta$ & $M_{A}$ &  $B_{\rm z<0}^*$  &   $\Delta t$ & $\Delta t_{\rm Dst}$ &  Dst & $\beta$ & $M_{A}$ & $B_{\rm z<0}^*$ & Field rot. & Chi. \\
				No. &[nT]&  [h]& [h]& [nT] &  &  & [nT] & [h]& [h]& [nT]&    & & [nT]&&  &   \\
				\hline
				& & \multicolumn{6}{c|}{ \textbf{MC Sheath}} & \multicolumn{9}{c}{\textbf{MC}}  \\
				SSC28&  -24  & 14:00 & 06:00 & -49 & 0.9 & 5.4 & -5 & 16:30 & 18:00 & -176 & 1.1 & 7.4 & -9 &  NES & RH \\
				SSC10&  -20& 28:30 & 24:00 & -149 & 0.6 & 5.8 & -6 & 30:00 & 00:30 & -103 & 0.4 & 6.1 & -7 & SEN & LH \\
				SSC09 & -32  & 17:00 & 06:30 & -98 & 0.7 & 5.6 & -9 & 28:30 & 03:30 & -127 & 0.1 & 2.4 & -12 & SWN & RH \\
				SSC17 & -43  & 11:00 & 06:30 & -109 & 0.4 & 3.6 & -7 & 35:00 & 00:30 & -96 & 0.03 & 1.0 & 0 & SEN & LH \\
				SSC25&  -13   & 24:45 & 13:00 & -53 & 1.6 & 8.2 & -3 & 42:30 & 11:00 & -106 & 0.4 & 5.4 & -4 & WSE & LH \\
				SSC24&  -16 & 08:00 & 06:30 & -102 & 0.4 & 5.0 & -9 & 17:00 & 15:30 & -69 & 0.3 & 4.7 & -2 & NWS & RH \\
				SSC06 & -11 & 25:30 & 22:00 & -100 & 0.8 & 6.5 & -5 & 26:00 & 07:30 & -97 & 0.1 & 2.9 & -5 & SWN & RH \\
				SSC01&  -15  & 28:75 & 18:00 & -33 & 1.1 & 6.4 & -5 & 16:00 & 07:30 & -86 & 1.4 & 7.4 & -10 & SWN & RH \\
				SSC03&  -16 & 13:00 & 02:45 & -18 & 2.7 & 9.7 & -4 & 30:30 & 07:30 & -71 & 0.6 & 5.5 & -8 & ESW & RH \\
				SSC14 & -18  & 08:00 & 08:00 & -15 & 2.1 & 8.5 & -2 &  24:00 & 02:30 & -58 & 0.4 & 4.4 &-5 &  SEN & LH \\
				SSC23 & -14 & 05:00 & 02:30 & -20 & 0.3 & 3.6 & -4 & 13:00 & 03:30 & -51 & 0.5 & 5.8 & -9 & SWN & RH \\
				SSC04 & -17  & 32:30 & 17:00 & -37 & 2.9 & 7.5 & -2 & 17:00 & 12:30 & -18 & 0.2 & 1.6 & 0 & NEs & RH \\
				& &\multicolumn{6}{c|}{\textbf{ICME Sheath}}& \multicolumn{9}{c}{\textbf{ICME}}  \\
				SSC27 & -27 & 12:30 & 08:00 & -181 & 0.4 & 4.2 & -11 & 15:00 & 00:30 & -127 & 0.8 & 6.3 & n/a & ENW &  \\
				SSC13 & -21 & 05:15 & 05:15 & -64 & 1.3 & 7.0 & -9 & 08:30 & 04:00 & -110 & 0.9 & 5.1 & -11 & ESW &  \\
				SSC21&  -12 & 27:30 & 19:00 & -28 & 0.4 & 3.8 & -1 & 44:00 & 10:00 & -36 & 0.8 & 8.5 & -2 & NES &  \\
				SSC15&  -9 & 07:00 & 06:00 & -7 & 1.4 & 7.3 & -1 & 35:30 & 15:30 & -36 & 1.0 & 6.7 & 0 & NES&  \\
				SSC20 & -16 & 21:00 & 20:30 & -17 & 2.2 & 9.0 & 0 & 21:00 & 00:30 & -15 & 0.3 & 3.8 & 0 & ENW &  \\
				SSC05 & -11  & 18:00 & 18:00 & -6 & 2.0 & 9.5 & 0 & 22:00 & 24:30 & -13 & 1.0 & 8.0 & n/a &  NWS &  \\
			\end{tabular}
			\label{table:SW_ICME}
		\end{center}
	\end{table}
\end{landscape}

\begin{table}[htb]
	\begin{center}
		\caption{Solar-wind properties in the Miscellaneous and the Shocks. We list the following: normalized and integrated $B_{\rm z<0}^*$ parameter, event duration, duration between SSC and min(Dst), min(Dst)
			reached, negative peak value $B_{\rm pz}$, mean $\beta$, mean $M_{A}$, mean $P_{\rm SW}$, and maximum jump in solar-wind velocity inside the Miscellaneous structures or across the shock for the Shocks.
			We considered the OMNI data set propagated to the bow-shock nose location.
		}
		\begin{tabular}{l r c rrrrrrr }
			SSC & $B_{\rm z<0}^*$ & $\Delta t $ & $\Delta t_{\rm Dst}$ &  Dst & $B_{\rm pz}$  & $\beta$ & $M_{A}$ &  $P_{\rm SW}$ & $\Delta{V_{\rm SW}}$ \\
			No. & [nT]& [h]  & [h]  & [nT] & [nT]  &  &  &  [nPa] &  \\
			\hline
			\multicolumn{10}{c}{\textbf{Misc.} - after discontinuity}  \\
			SSC32& -3.9  & 21:15 & 06:45 & -64 & -18 & 1.3 & 7.9 & 6.5 & 1.5 \\
			SSC11& -2.4  & 30:15 & 10:45 & -57 & -18 & 2.9 & 9.5 & 4.6 & 1.3 \\
			SSC26& -4.2 & 36:00 & 08:00 & -45 & -14  & 1.0 & 5.8 & 3.6 & 1.5 \\
			SSC12& -1.4  & 22:25 & 12:00 & -14 & -8 & 2.6 & 10.6 & 6.0 & 1.3 \\
			\multicolumn{10}{c}{\textbf{Shock} - after discontinuity}  \\
			SSC08 & -4.2  & 44:30 & 03:00 & -23 & -9 & 1.35 & 6.7 & 2.4 & 1.1 \\
			SSC18& -0.8  & 42:00 & 14:30 & -13 & -4 & 1.35 & 7.7 & 1.4 & 1.1 \\
			SSC16 & -1.2 & 28:00 & 13:30 & -12 & -6 & 2.7 & 9.6 & 3.6 & 1.2 \\
			SSC22 & -0.2 & 25:00 & 16:15 & 0.0 & -10 & 0.9 & 5.8 & 2.5 & 1.2 \\
		\end{tabular}
		\label{table:SW_others}
	\end{center}
\end{table}

\begin{table}[htb]
	\begin{center}
		\caption{Solar-wind properties in the CIR/SIR structures. Each event is presented with its normalized and integrated $B_{\rm z<0}^*$ value. 
		We list the following before and after the SI: event duration (from SSC to SI, and from SI to
		the end), duration between SSC and the time of min(Dst),  min(Dst), negative peak value $B_{\rm pz}$, mean $\beta$, mean $M_{A}$.
		The last two columns show the maximum $P_{\rm SW}$ reached and the maximum jump in solar-wind velocity across the SI. Note that the time between SSC and min(Dst) can be negative before the SI. This is due to the fact
			that the CIR/SIR start time in the literature can be recorded as being before the SSC time.
			We considered the OMNI data set propagated to the location of the bow-shock nose.
		}					
		\begin{tabular}{l r  c rrrrr rr  }
			SSC & $B_{\rm z<0}^*$ & $\Delta t$ & $\Delta t_{\rm Dst}$ &  Dst & $B_{\rm pz}$ & $\beta$ & $M_{A}$ & $P_{\rm SW}$ & $\Delta V_{\rm SW}$  \\
			No. & [nT] & [h] & [h] & [nT] & [nT]& & & [nPa] &  \\
			\hline
			\multicolumn{2}{c}{\textbf{CIR/SIR}}   & \multicolumn{6}{c }{before SI}  &  \multicolumn{2}{c }{across SI}  \\
			SSC07& -1.1  & 03:00 & -01:00 & 13 & - & 2.3 & 8.9 & 18 & 1.8  \\
			SSC31& -0.6  & 02:30 & -02:45 & -32 & - & 1.2& 5.7 &  14 & 1.5 \\
			SSC29\,--\,30& -2.0  & 15:15 & 13:30 & -28 & - & 1.5 & 8.5 &11 & 1.3  \\
			SSC02 & -1.9 & 03:30 & -02:30 & 15 & - & 3.5 & 12.4 & 12 & 1.3\\
			SSC19 & -2.7  & 02:40 & -09:00 & 2 & - & 3.8 & 10.6 &  5 & 1.4 \\
			
			\multicolumn{2}{c}{\textbf{CIR/SIR}}   &\multicolumn{6}{c}{after SI}   & \\
			
			SSC07&& 39:00 & 17:00 & -38 & -14 & 1.3 & 7.5 &  \\
			SSC31&& 05:30 & 02:15 & -32 & -15 & 1.1 & 7.5 &\\
			SSC29\,--\,30&& 22:00 & 02:15 & -25 & -14 & 0.5 & 4.1 &  \\
			SSC02 && 16:00 & 09:15 & -21 & -13 & 2.8 & 7.5 &  \\
			SSC19 && 12:00 & 19:30 & -18 & -13 & 1.1 & 6.1 & \\
		\end{tabular}
		\label{table:SW_CIR}
	\end{center}
\end{table}




\end{article}

\end{document}